\documentclass[aps,showpacs,preprintnumbers,twocolumn,superscriptaddress,floatfix,nofootinbib]{revtex4-2}

\usepackage[linesnumbered,lined,boxed,commentsnumbered,ruled,vlined]{algorithm2e}
\usepackage{algpseudocode}
\usepackage{amsfonts}
\usepackage{amsmath}
\usepackage{amssymb}
\usepackage{amsthm}
\usepackage{bm,bbm}
\usepackage{caption}
\usepackage{comment}
\usepackage{dcolumn}
\usepackage{epstopdf}
\usepackage{float}
\usepackage{graphicx}
\usepackage{hyperref}
\usepackage{cleveref}
\usepackage{mathrsfs}
\usepackage{mathtools}
\usepackage{multirow}
\usepackage{physics}
\usepackage{ragged2e}
\usepackage{scalerel}
\usepackage{subfigure}
\usepackage{tabularx}
\usepackage{tikz}
\usepackage{url}
\usepackage{utfsym}
\usepackage{verbatim}
\usepackage{xcolor}
\usepackage{xkcdcolors}
\usepackage[normalem]{ulem}
\usepackage{lineno}
\DeclareCaptionJustification{justified}{\justifying}
\hypersetup{colorlinks=true, citecolor=orange, urlcolor=blue, linkcolor=magenta}

\definecolor{cel}{rgb}{0.0,0.53,0.74}
\definecolor{green}{rgb}{0.0,0.5,0.0}

\definecolor{colSDRG}{RGB}{51,117,56}
\definecolor{colSDCM}{RGB}{194,106,119}
\definecolor{colRec}{RGB}{230,158,0}
\definecolor{colpp}{RGB}{87,181,232}
\definecolor{colnn}{RGB}{0,158,115}
\definecolor{colpn}{RGB}{204,120,166}

\begin{document}

\title{Multiscale Reconstruction of Weighted Networks from Coarse-Grained Data}


\author{Mattia Marzi}
\email{mattia.marzi@imtlucca.it}
\affiliation{IMT School for Advanced Studies, P.zza San Francesco 19, 55100 Lucca (Italy)}
\affiliation{Lorentz Institute for Theoretical Physics, University of Leiden, Einsteinweg 55, 2333 CC Leiden (The Netherlands)}
\affiliation{Statistics Netherlands, Henri Faasdreef 312, 2492 JP Den Haag (the Netherlands)}
\affiliation{INdAM-GNAMPA Istituto Nazionale di Alta Matematica `Francesco Severi', P.le Aldo Moro 5, 00185 Rome (Italy)}
\author{Frank P. Pijpers}
\affiliation{Statistics Netherlands, Henri Faasdreef 312, 2492 JP Den Haag (the Netherlands)}
\affiliation{Korteweg - de Vries Institute for Mathematics, University of Amsterdam, Amsterdam (the Netherlands)}
\author{Diego Garlaschelli}
\affiliation{IMT School for Advanced Studies, P.zza San Francesco 19, 55100 Lucca (Italy)}
\affiliation{Lorentz Institute for Theoretical Physics, University of Leiden, Einsteinweg 55, 2333 CC Leiden (The Netherlands)}
\affiliation{INdAM-GNAMPA Istituto Nazionale di Alta Matematica `Francesco Severi', P.le Aldo Moro 5, 00185 Rome (Italy)}

\date{\today}

\begin{abstract}
Network reconstruction from partial information is usually performed at the same resolution level at which constraints are observable. This becomes problematic when only coarse-grained information is available, while the relevant process occurs at a finer scale. Here we employ the multiscale model of weighted networks introduced in a companion paper and turn it into a probabilistic framework for reconstructing weighted networks across arbitrary aggregation levels. The model is built to preserve its functional form under coarse-graining, so that global parameters calibrated on an observable aggregate layer can be transferred to finer layers without refitting.
We test the method on two empirical systems with different aggregation mechanisms. 
In the International Trade Network, countries are aggregated into geographic macro-regions and the observed coarse-grained layer is used to infer the underlying country-level network. 
In the Dutch production network, sectoral flows are reconstructed across the hierarchical industrial classification, using coarser sectoral layers to infer finer ones. 
In both geographical and sectoral settings, we benchmark our genuinely multiscale reconstruction method against a state-of-the-art weighted reconstruction model calibrated directly at the target resolution, and therefore using additional information available at the same (finer) scale at which performance is evaluated. 
Despite this informational disadvantage, our method recovers the fine-scale binary structure with high accuracy, improving precision, specificity, accuracy and maximum degree-error diagnostics, while remaining nearly equivalent in sensitivity.
\end{abstract}

\maketitle

\section*{INTRODUCTION}

Economic and financial systems can be naturally represented as networks in which heterogeneous agents interact through weighted relations~\cite{ReconstructionMethods2018,cimini2021reconstructing,bardoscia_physics_2021}. In production systems, links encode relationships between firms or sectors; in financial systems, they represent exposures; in trade systems, they correspond to trade flows.
In many empirically relevant settings, however, bilateral interactions are not directly observable. Confidentiality constraints typically prevent the disclosure of microscopic relations, while only aggregate node-level quantities, such as total activity, total trade or total production volume, are available~\cite{ReconstructionMethods2018,cimini2021reconstructing,Mungo2024SupplyNetworksReview}. The reconstruction problem then consists in inferring a statistically consistent microscopic structure from partial macroscopic information.

Within the maximum-entropy framework~\cite{jaynes1957information,park2004statistical,garlaschelli2008maximum,squartini2011analytical,squartini2017maximum}, this task amounts to identifying the least biased ensemble of graphs reproducing a chosen set of constraints on average. When the empirical degree sequence is enforced, one recovers the Configuration Model~\cite{park2004statistical,squartini2011analytical,squartini2017maximum}, whose dyadic probabilities take the form
\begin{equation}
p_{ij}=\frac{x_i x_j}{1+x_i x_j}.
\end{equation}
The node-specific parameters $\{x_i\}$ are determined so that expected and observed degrees coincide.

We use uppercase letters for adjacency and weight matrices and lowercase letters for their entries; model averages are indicated by explicit expectation operators.

In economic applications, degree sequences are rarely observable, while node strengths $\{s_i\}$ are typically available. A common dimensionality reduction consists in expressing the heterogeneous parameters through a fitness ansatz~\cite{GarlaschelliLoffredo2004WTW,CimiModel2015,Mazzarisi2017LimitedInformation,Anand2018MissingLinks,Lebacher2019LostEdges,Ramadiah2020ReconstructingAndStressTesting,marzi2026reproducing,marzi2026Bayesian}. In its density-corrected gravity formulation~\cite{EnhancedGravityModelTrade2019,DiVece2022GravityMER}, one sets
\begin{equation}
x_i\sim\sqrt{\delta}\,s_i,
\end{equation}
leading to
\begin{equation}
p_{ij}=\frac{\delta\,s_i s_j}{1+\delta\,s_i s_j},
\end{equation}
where a single global parameter $\delta$ controls overall density.
Although this specification preserves the information encoded in node-level aggregates, it implicitly assumes a fixed description scale. If nodes are aggregated into larger units, the functional form of the probability is not preserved under coarse graining and parameters must be re-estimated at each level. In hierarchical economic systems~\cite{FunctionalStructureProductionNetworks2021,ReconstructingFirmLevelInteractionsDutchInputOutputNetwork2022}, such dependence on the chosen aggregation level becomes a structural limitation.

{
A different perspective consists in using a reconstruction model whose functional form is invariant under coarse graining~\cite{Garuccio2023,LalliThesis,IalongoBangmaJansenGarlaschelli2024,Gabrielli2025}. In the binary MultiScale Model, microscopic link probabilities are written as
\begin{equation}
p_{ij}^{\mathrm{MSM}}
=
1-\exp\left(-\delta s_i s_j\right),
\end{equation}
where $s_i$ and $s_j$ are node strengths and $\delta$ is a global density parameter. If microscopic nodes are grouped into disjoint blocks, the same functional form is preserved at the aggregated level:
\begin{equation}
p_{IJ}^{\mathrm{MSM}}
=
1-\exp\left(-\delta s_I s_J\right).
\end{equation}
The block strengths entering the coarse-grained probability are obtained by additivity,
\begin{equation}
s_I
=
\sum_{i\in I}s_i.
\end{equation}
Thus, coarse graining changes only the variables entering the model, from node strengths to block strengths, but not the probability law itself {for distinct blocks}. This closure is the key property exploited below: the density parameter $\delta$ can be calibrated at an observed aggregate layer and transferred to a finer layer without refitting the model family.
}

The difference with standard maximum-entropy reconstructions based on fitness specifications is therefore not merely parametric, but concerns the behavior of the model under aggregation. Both specifications depend on bilinear combinations of node-level quantities; however, only the exponential form remains closed under aggregation with additive parameters. This distinction becomes particularly relevant in empirical settings where certain constraints, for instance the overall link density, are observable only at aggregated resolutions. In such cases, one may wish to calibrate the model at a coarse-grained level and then infer microscopic structure consistently with those constraints. If the functional form is not preserved under aggregation, this procedure becomes ill-defined: parameters must be re-estimated at each resolution, and the resulting microscopic network is no longer guaranteed to reproduce the observed aggregated constraints.


In this paper, we extend the scale-invariant construction from binary topology to integer-valued weighted networks by introducing a compound-Poisson specification that separates the control of link probabilities from the normalization of expected weights. This yields a reconstruction scheme in which the density parameter can be calibrated at an observed aggregate layer and then transferred unchanged to finer layers, while additive strengths fix the weight scale consistently across resolutions. We test this principle in two empirical systems with different aggregation mechanisms, one geographical and one sectoral, and show that the transferred coarse-scale calibration remains informative at the microscopic level. In particular, the method accurately predicts the number of fine-scale links, preserves strength information up to the chosen diagonal convention, and reproduces local structural profiles that are not imposed as constraints. When compared with a fine-scale weighted reconstruction calibrated directly at the target resolution and sharing the same unconditional gravity expectation once strengths are fixed, the scale-invariant approach systematically improves the selectivity of the inferred binary support, increasing precision, specificity and accuracy while keeping sensitivity nearly unchanged. This shows that the main empirical advantage of the approach lies in the multiscale reconstruction of topology, rather than in a different specification of expected weights.

\section*{RESULTS}

\subsection*{The weighted MultiScale Model (wMSM)}

In the companion paper~\cite{Marzi2026wMSM}, we introduce the weighted MultiScale Model as an aggregation-invariant ensemble for integer-valued weighted networks. In the present work, we use this construction as a reconstruction model and recall only the formulas needed for the empirical analysis.  The model extends the binary MSM to integer-valued weights by assigning to each dyad a compound-Poisson distribution~\cite{Feller1971,JohnsonKempKotz2005} with intensity $\lambda_{ij}\coloneq\delta s_i s_j$ and geometrically distributed positive marks. The Poisson count specifies the number of exchange events, and each mark specifies the amount contributed by one event. Each amount starts at one weight unit and increases by one unit with probability $1-\rho$ until stopping with probability $\rho$, giving $P(X=n)=\rho(1-\rho)^{n-1}$ for $n\ge1$. This geometric law maximizes entropy on positive integers at fixed mean. The dyadic weight is the sum of these contributions. This choice gives
\begin{equation}
P^{\mathrm{wMSM}}(w_{ij}=0)
=
\exp\left(-\delta s_i s_j\right),
\end{equation}
and, for $n\ge1$,
\begin{align}
&P^{\mathrm{wMSM}}(w_{ij}=n)=\exp\Big(-\delta\,s_i s_j\Big)\nonumber\\
&\cdot\sum_{k=1}^{n}\frac{\left(\delta\,s_i s_j\right)^k}{k!}\binom{n-1}{k-1}\rho^k(1-\rho)^{n-k}.
\end{align}
The induced link probability is therefore
\begin{equation}
p_{ij}^{\mathrm{wMSM}}
=
P^{\mathrm{wMSM}}(w_{ij}>0)
=
1-\exp\left(-\delta s_i s_j\right),
\end{equation}
while the expected dyadic weight is
\begin{equation}
\left\langle w_{ij}\right\rangle_{\mathrm{wMSM}}
=
\frac{\delta}{\rho}s_i s_j.
\end{equation}
Thus $\delta$ controls the topology, whereas $\rho$ fixes the scale of weights without affecting link probabilities.

The same functional forms are preserved under aggregation {between distinct blocks}. If microscopic nodes are grouped into two distinct blocks $I$ and $J$, weights are aggregated by summation $w_{IJ}=\sum_{i\in I}\sum_{j\in J}w_{ij}$, and strengths are additive, $s_I=\sum_{i\in I}s_i$ and $s_J=\sum_{j\in J}s_j$. Since independent Poisson intensities add, the block intensity is $\lambda_{IJ}=\sum_{i\in I}\sum_{j\in J}\lambda_{ij}=\delta s_I s_J$.
Therefore the coarse-grained weight $w_{IJ}$ is again compound-Poisson with the same geometric mark distribution and with the same functional dependence on the aggregated strengths. In particular,
\begin{equation}
p_{IJ}^{\mathrm{wMSM}}
=
1-\exp\left(-\delta s_I s_J\right),
\end{equation}
and, for $n\ge1$,
\begin{align}
&P^{\mathrm{wMSM}}(w_{IJ}=n)=\exp\Big(-\delta\,s_I s_J\Big)\nonumber\\
&\cdot\sum_{k=1}^{n}\frac{\left(\delta\,s_I s_J\right)^k}{k!}\binom{n-1}{k-1}\rho^k(1-\rho)^{n-k}.
\end{align}
Consequently, the parameters of the model can be calibrated at one aggregation layer and used unchanged at any other layer, since coarse graining preserves the probability law {for inter-block dyads}. In particular, the density parameter $\delta$ is fixed at the chosen calibration layer by matching the expected number of links to the observed one,
\begin{equation}
\left\langle L\right\rangle
=
\sum_{(i,j)\in\mathcal D}
p_{ij}^{\mathrm{wMSM}}
=
L,
\end{equation}
where $\mathcal D$ contains the dyads counted as possible links at the calibration layer: it includes diagonal entries if self-loops are retained, and excludes them otherwise. 

The weight-scale parameter $\rho$ is then fixed by the strength normalization. Let
\begin{equation}
W
=
\sum_i s_i
\end{equation}
be the observed total strength. Moreover, geometric marks have mean $1/\rho$, so the expected strength satisfies
\begin{equation}
\left\langle s_i\right\rangle_{\mathrm{wMSM}}
=
\sum_j
\left\langle w_{ij}\right\rangle_{\mathrm{wMSM}}
=
\frac{\delta}{\rho}s_i W.
\end{equation}
Therefore, choosing
\begin{equation}
\rho
=
\delta W
\end{equation}
gives
\begin{equation}
\left\langle w_{ij}\right\rangle_{\mathrm{wMSM}}
=
\frac{s_i s_j}{W},
\end{equation}
and reproduces node strengths exactly whenever the matrix convention of the layer retains diagonal entries. Since both $\delta$ and $W$ are invariant across layers after calibration, the same is true for $\rho$. If diagonal entries are removed, {the normalization $\rho=\delta W$ gives $\langle s_i\rangle=s_i(1-s_i/W)$, so individual strengths are no longer reproduced exactly}, a standard correction in strength-driven gravity reconstructions~\cite{ReconstructingFirmLevelInteractionsDutchInputOutputNetwork2022,CReM2020}. {Here we apply the same dyadic law to retained diagonals, preserving the gravity expectation and exact strength matching on full support without fitting the self-loop count separately. A fully aggregation-invariant diagonal construction is also possible, although its self-loop calibration need not be feasible. Its derivation and admissibility conditions are discussed in the Supplementary Information of Ref.~\cite{Marzi2026wMSM}.}

\begin{figure*}[t!]
\centering
\begin{minipage}{\linewidth}
\centering
\includegraphics[width=\linewidth]{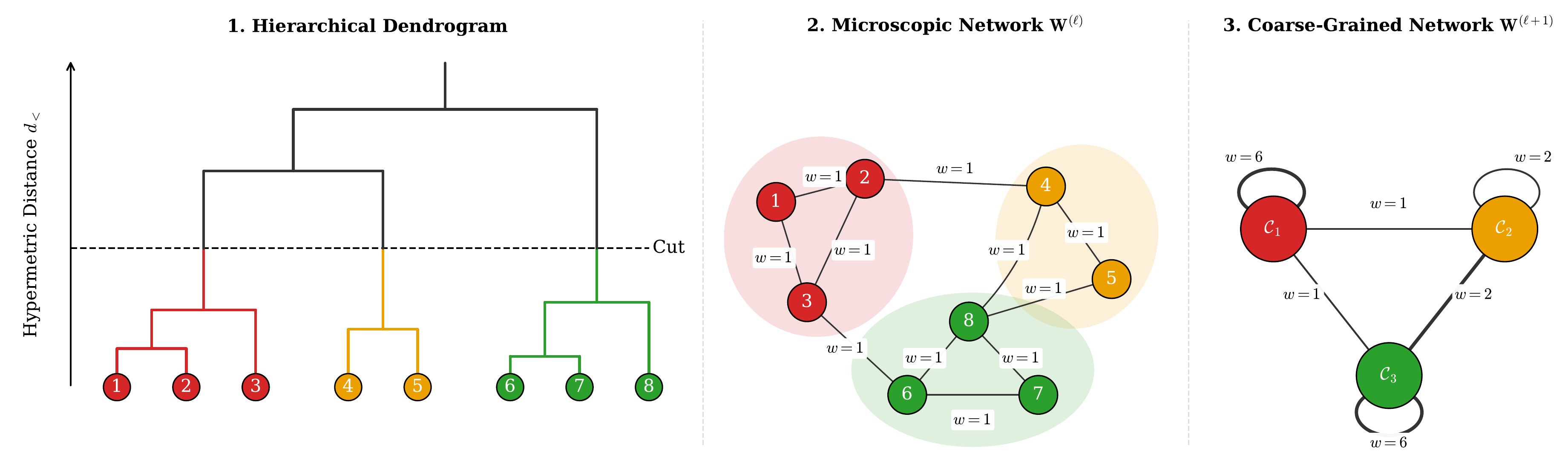}
\par\textbf{(a)}
\end{minipage}
\vspace{0.5em}
\begin{minipage}{0.6\linewidth}
\centering
\includegraphics[width=\linewidth]{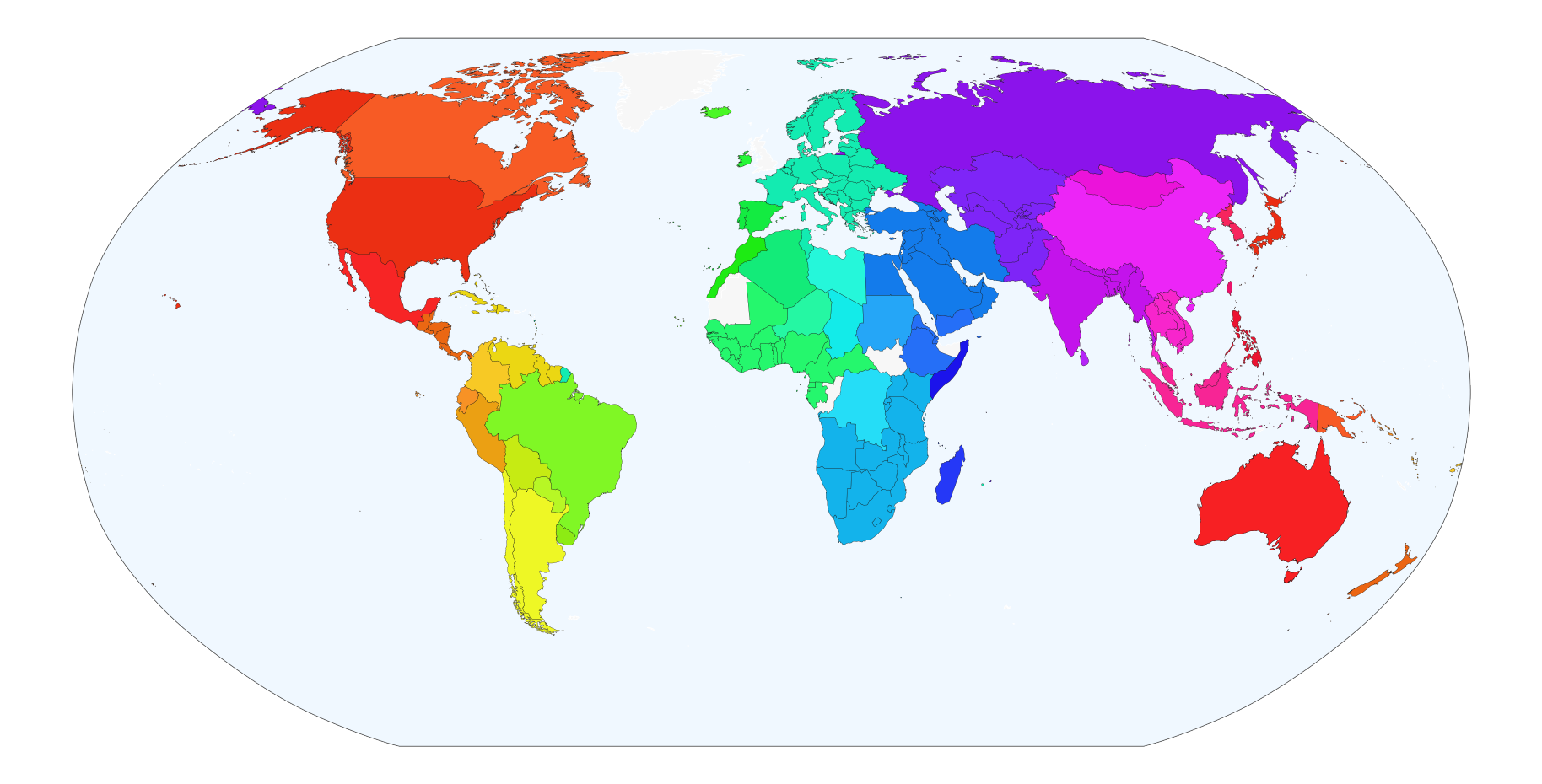}
\par\textbf{(b)}
\end{minipage}
\caption{\textbf{Ultrametric aggregation of the ITN}. Panel~(a) illustrates the construction of the multiscale representation. Countries form the microscopic layer $\Omega_0$, while cutting the single-linkage dendrogram at height $d_c$ defines the macro-regional layer $\Omega_1(d_c)$. Edge weights are aggregated by summation, and self-loops at the macro-regional scale represent total internal weight within a block. Panel~(b) shows the 1993 macro-regional partition obtained with the main cut $d_c=800{\,}\mathrm{km}$, with different colors identifying different macro-regions.}
\label{fig:1}
\end{figure*}

\subsection*{Multiscale reconstruction protocol and diagnostics}

We now test the wMSM in two empirical settings where coarse graining has a different nature. The first application is the International Trade Network (ITN), where countries are aggregated into geographical macro-regions through a distance-based dendrogram. The second application is the Dutch (sectoral) production network (DPN), where economic sectors are aggregated according to the hierarchical structure of the SBI classification, the Dutch standard industrial classification that organizes economic activities into progressively finer sectoral codes. Thus, in one case the observed coarse layer is geographical, while in the other it is sectoral. The same reconstruction principle applies in both cases: the binary density parameter $\delta$ is calibrated at an aggregate scale and then transferred to finer layers, while the weight-scale parameter $\rho$ is fixed by the strength normalization through $\rho=\delta W$, using the additivity of strengths across aggregation levels. The ITN data and preprocessing are described in Appendix~\hyperlink{AppA}{A}, while the DPN data are described in Appendix~\hyperlink{AppB}{B}.

For a generic layer $\Omega_\ell$, let
\begin{equation}
\mathbf{W}^{(\ell)}
=
\left\{
w_{ij}^{(\ell)}
\right\}
\end{equation}
denote the observed weighted matrix, and let
\begin{equation}
a_{ij}^{(\ell)}
=
\mathbbm{1}\!\left(w_{ij}^{(\ell)}>0\right)
\end{equation}
be its binary projection. The dyadic support $\mathcal D^{(\ell)}$ depends on the diagonal convention of the layer. In the ITN application, {the observed country-level data contain no self-loops, and the reconstructed diagonal is excluded from comparison, giving support $i<j$}, while macro-regional self-loops are retained because they represent total trade internal to a macro-region. In the DPN application, diagonal entries are retained at all sectoral resolutions, because they represent within-sector production relations.

The observed number of links at layer $\ell$ is
\begin{equation}
L^{(\ell)}
=
\sum_{(i,j)\in\mathcal D^{(\ell)}}a_{ij}^{(\ell)}.
\end{equation}
Given fitted wMSM probabilities $p_{ij}^{(\ell)}$, the expected number of links is
\begin{equation}
\left\langle L^{(\ell)}\right\rangle
=
\sum_{(i,j)\in\mathcal D^{(\ell)}}p_{ij}^{(\ell)}.
\end{equation}
The corresponding relative error is
\begin{equation}
\mathrm{RE}_L^{(\ell)}
=
\frac{
\left|
\left\langle L^{(\ell)}\right\rangle-L^{(\ell)}
\right|
}{
L^{(\ell)}
}.
\end{equation}

Observed and expected degrees are
\begin{equation}
k_i^{(\ell)}
=
\sum_{{j}}a_{ij}^{(\ell)},
\qquad
\left\langle k_i^{(\ell)}\right\rangle
=
\sum_{{j}}p_{ij}^{(\ell)},
\end{equation}
{In these sums and the local profiles below, excluded diagonal entries are set to zero.} {Degrees are adjacency row sums, so a retained diagonal contributes once; link counts use the upper triangle and count each unordered dyad once}. Their average relative error is
\begin{equation}
\mathrm{ARE}_k^{(\ell)}
=
\frac{1}{N^{(\ell)}}
\sum_{i=1}^{N^{(\ell)}}
\left|
\frac{\left\langle k_i^{(\ell)}\right\rangle}{k_i^{(\ell)}}-1
\right|.
\end{equation}
Analogously, using observed strengths $s_i^{(\ell)}$ and expected strengths
\begin{equation}
\left\langle s_i^{(\ell)}\right\rangle
=
\sum_{{j}}
\left\langle w_{ij}^{(\ell)}\right\rangle,
\end{equation}
we define
\begin{equation}
\mathrm{ARE}_s^{(\ell)}
=
\frac{1}{N^{(\ell)}}
\sum_{i=1}^{N^{(\ell)}}
\left|
\frac{\left\langle s_i^{(\ell)}\right\rangle}{s_i^{(\ell)}}-1
\right|.
\end{equation}

\begin{table*}[t!]
\centering
\scriptsize
\renewcommand{\arraystretch}{1.12}
\setlength{\tabcolsep}{5pt}
\begin{tabular}{|c|c|c|c|c|c|c|}
\hline
Year 
& $\mathrm{RE}_L^{(1)}$ 
& $\mathrm{ARE}_k^{(1)}$ 
& $\mathrm{ARE}_s^{(1)}$ 
& $\mathrm{RE}_L^{(0)}$ 
& $\mathrm{ARE}_k^{(0)}$ 
& $\mathrm{ARE}_s^{(0)}$ \\
\hline
1991 & 0.0000 & 0.0933 & 0.0000 & 0.0411 & 0.2143 & 0.0058 \\
1992 & 0.0000 & 0.0904 & 0.0000 & 0.0158 & 0.2086 & 0.0057 \\
1993 & 0.0000 & 0.0826 & 0.0000 & 0.0544 & 0.2183 & 0.0057 \\
1994 & 0.0000 & 0.0975 & 0.0000 & 0.0596 & 0.2109 & 0.0057 \\
1995 & 0.0000 & 0.0887 & 0.0000 & 0.0653 & 0.2098 & 0.0057 \\
1996 & 0.0000 & 0.0777 & 0.0000 & 0.0714 & 0.2013 & 0.0057 \\
1997 & 0.0000 & 0.0674 & 0.0000 & 0.0656 & 0.1875 & 0.0057 \\
1998 & 0.0000 & 0.0652 & 0.0000 & 0.0575 & 0.1798 & 0.0057 \\
1999 & 0.0000 & 0.0687 & 0.0000 & 0.0575 & 0.1833 & 0.0057 \\
2000 & 0.0000 & 0.0687 & 0.0000 & 0.0429 & 0.1876 & 0.0057 \\
\hline
Mean & 0.0000 & 0.0800 & 0.0000 & 0.0531 & 0.2001 & 0.0057 \\
\hline
\end{tabular}
\caption{\textbf{Macro-regional calibration and country-level reconstruction.} Performance of the wMSM at the macro-regional calibration layer and at the country reconstruction layer, using the main aggregation scale $d_c=800{\,}\mathrm{km}$. Superscript $(1)$ denotes the macro-regional layer and superscript $(0)$ denotes the country layer. The macro-regional link count is reproduced by construction, while the country-level link count is predicted after transferring the parameter calibrated at the coarse scale.}
\label{tab:wmsm_absolute_results}
\end{table*}

We also inspect local structural profiles that are not imposed as constraints. The average nearest-neighbor degree is defined as
\begin{equation}
k_i^{nn}
=
\frac{1}{k_i}
\sum_{{j}}a_{ij}k_j,
\end{equation}
while the {implemented binary clustering profile is}
\begin{equation}
c_i
={\frac{\sum_j\sum_k a_{ij}a_{ik}a_{jk}}{k_i(k_i-1)},\qquad k_i>1}.
\end{equation}
For the weighted layer, {using a matrix extension of} standard weighted-network diagnostics~\cite{Barrat2004WeightedNetworks,Onnela2005WeightedMotifs}, we compute the average nearest-neighbor strength
\begin{equation}
s_i^{nn}
=
\frac{1}{k_i}
\sum_{{j}}a_{ij}s_j,
\end{equation}
and the {implemented weighted clustering profile}
\begin{equation}
{c_i^{w}=\frac{\sum_j\sum_k\left(\widetilde w_{ij}\widetilde w_{jk}\widetilde w_{ki}\right)^{1/3}}{k_i(k_i-1)},\quad k_i>1,}
\end{equation}
{
\begin{equation}
\widetilde w_{ij}=\frac{w_{ij}}{U},\qquad
U=\sum_{(i,j)\in\mathcal D}w_{ij}.
\end{equation}
}
Here $U$ is the observed unordered-dyad weight total, distinct from the row-strength total $W$. Expected nearest-neighbor profiles are computed by replacing the observed adjacency with the fitted probability matrix and observed strengths with expected strengths, with explicit conditioning on link existence for weighted nearest-neighbor strength. Since $c_i$ and $c_i^{w}$ are nonlinear ratios, their expected profiles are estimated by sampling networks from the fitted ensemble and averaging the same node-level estimators used for the observed network.

At a calibration layer $\Omega_{\mathrm{cal}}$, the parameter $\delta$ is fitted by matching the observed number of links,
\begin{equation}
\sum_{(I,J)\in\mathcal D^{(\mathrm{cal})}}
\left[
1-\exp\!\left(-\delta\,s_I^{(\mathrm{cal})}s_J^{(\mathrm{cal})}\right)
\right]
=
L^{(\mathrm{cal})}.
\end{equation}
The fitted value of $\delta$ is then transferred unchanged to the target layer. Therefore, at any reconstructed layer $\Omega_\ell$, probabilities and expected weights are
\begin{equation}
p_{ij}^{(\ell)}
=
1-\exp\!\left(-\delta\,s_i^{(\ell)}s_j^{(\ell)}\right),
\end{equation}
and
\begin{equation}
\left\langle w_{ij}^{(\ell)}\right\rangle
=
\frac{\delta}{\rho}\,s_i^{(\ell)}s_j^{(\ell)},
\end{equation}
with $\rho=\delta W$, where $W$ denotes the observed total strength of the empirical network under consideration. Since strengths are additive, the same $W$ is preserved across aggregation layers.

\begin{figure*}[t!]
\centering
\begin{minipage}{0.48\linewidth}
\centering
\includegraphics[width=\linewidth]{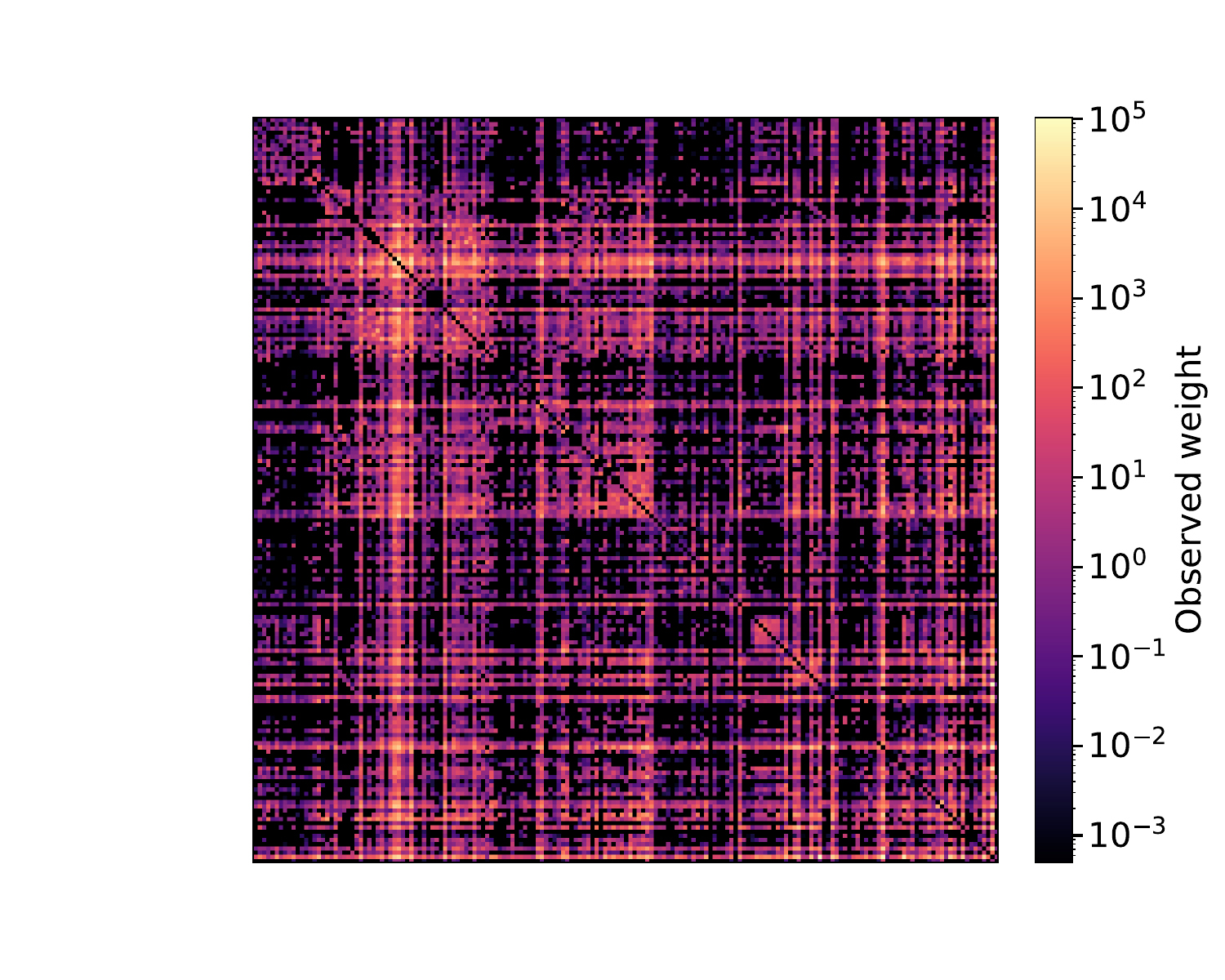}
\par\textbf{(a)}
\end{minipage}
\hfill
\begin{minipage}{0.48\linewidth}
\centering
\includegraphics[width=\linewidth]{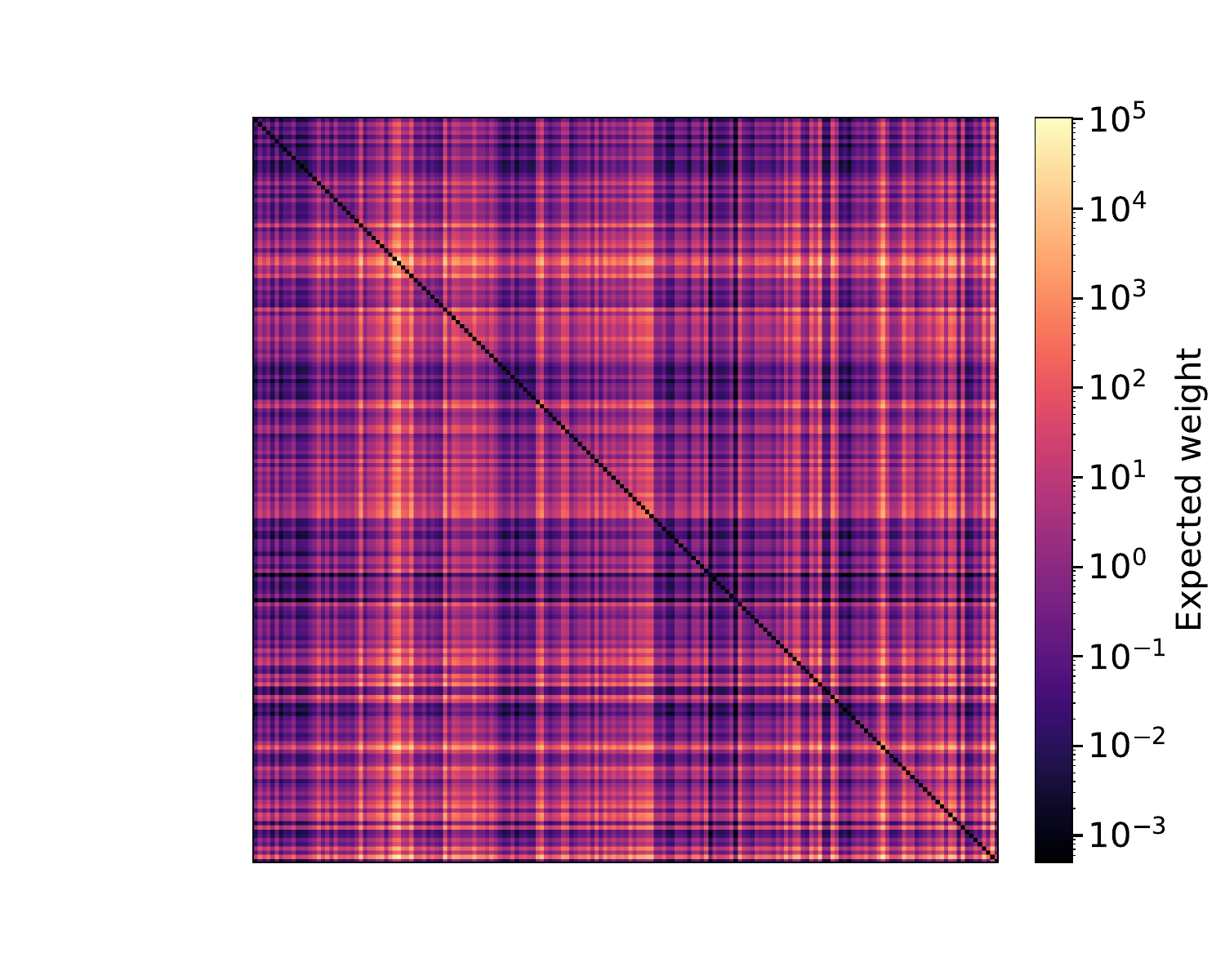}
\par\textbf{(b)}
\end{minipage}
\vspace{0.4em}
\begin{minipage}{0.48\linewidth}
\centering
\includegraphics[width=\linewidth]{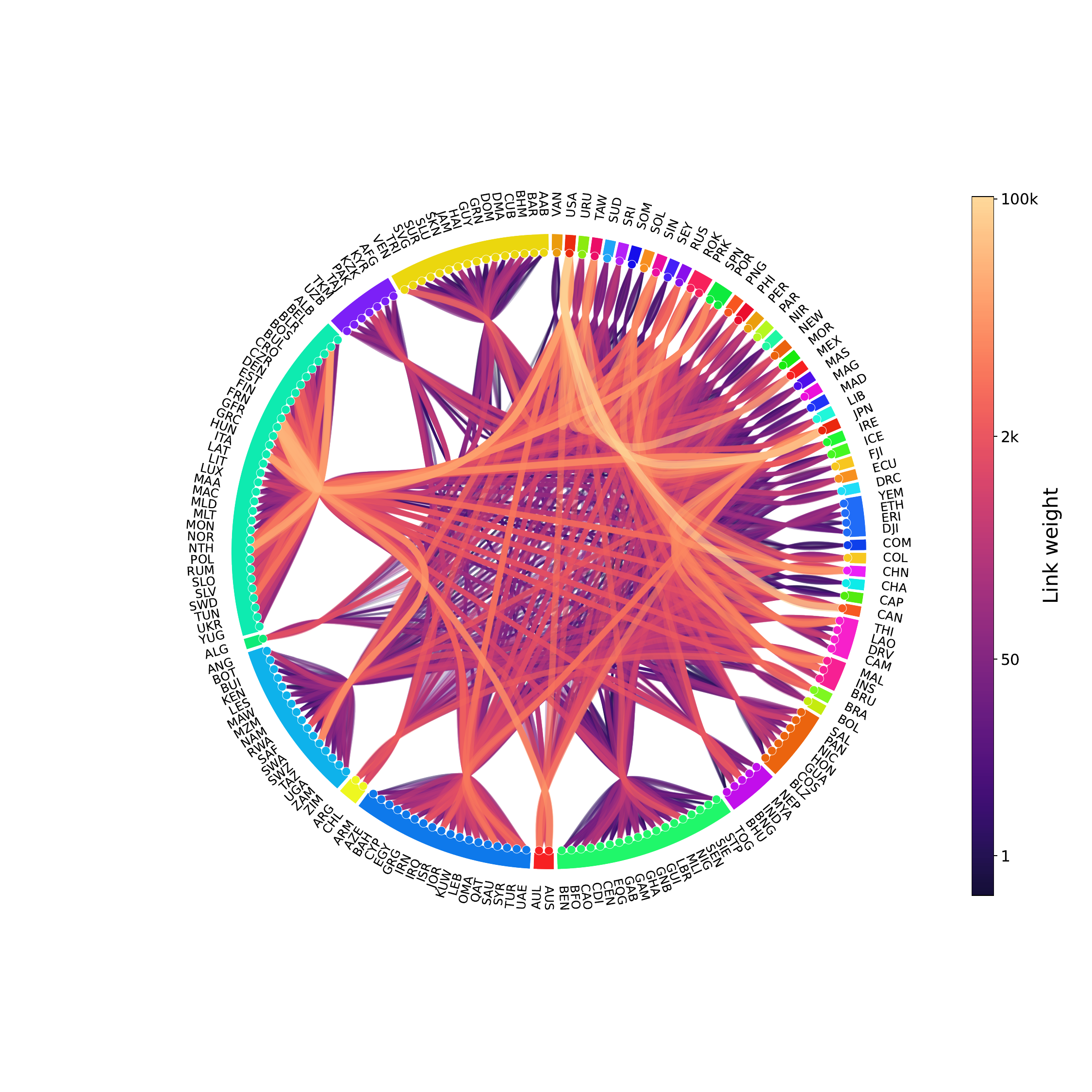}
\par\textbf{(c)}
\end{minipage}
\hfill
\begin{minipage}{0.48\linewidth}
\centering
\includegraphics[width=\linewidth]{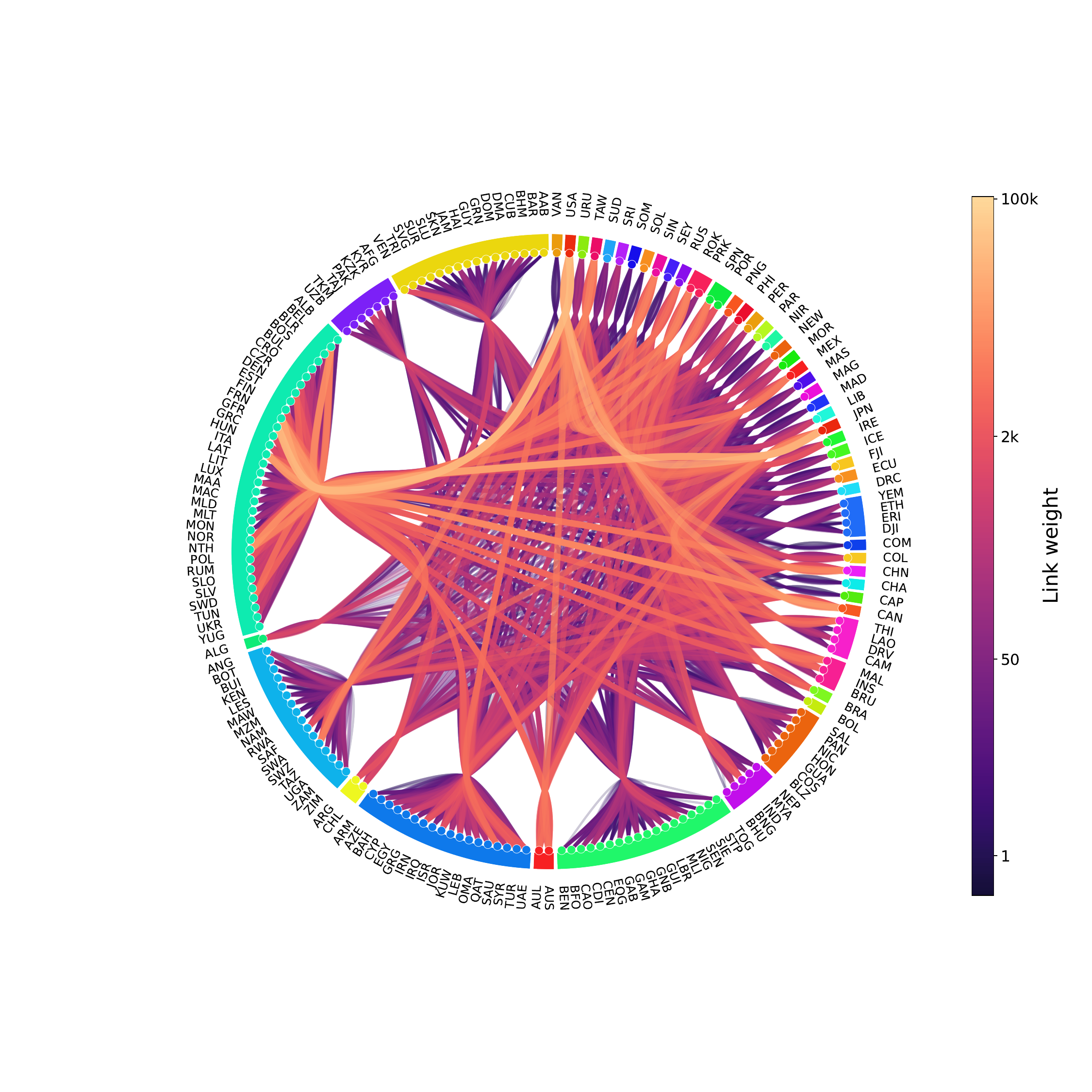}
\par\textbf{(d)}
\end{minipage}
\caption{\textbf{Observed and reconstructed country-level weighted ITN in 1993}. Panels~(a,b): observed country-to-country weights and wMSM expected weights, with countries ordered by macro-region. Black entries correspond to absent links or values outside the plotted range. Panels~(c,d): circular visualization of the same observed and reconstructed weighted networks. In the reconstructed network, the displayed support is obtained by retaining the top-$\langle L^{(0)}\rangle$ dyads according to the fitted probabilities. This gives the most likely binary realization predicted by the model. Edge width encodes the corresponding expected trade volume. Node colors identify macro-regions.}
\label{fig:2}
\end{figure*}

\subsection*{Reconstructing country-level trade from macro-regional data}

We first apply the wMSM to the ITN from 1991 to 2000. Countries form the target microscopic layer $\Omega_0$, while calibration is performed on the macro-regional layer $\Omega_1(d_c)$ obtained by cutting a single-linkage dendrogram of geographical distances. The figures shown in this section refer to the 1993 snapshot, which we use as a representative visual example, while quantitative results are reported for the full period 1991--2000.

\begin{figure*}[t!]
\centering
\begin{minipage}{0.32\linewidth}
\centering
\includegraphics[width=\linewidth]{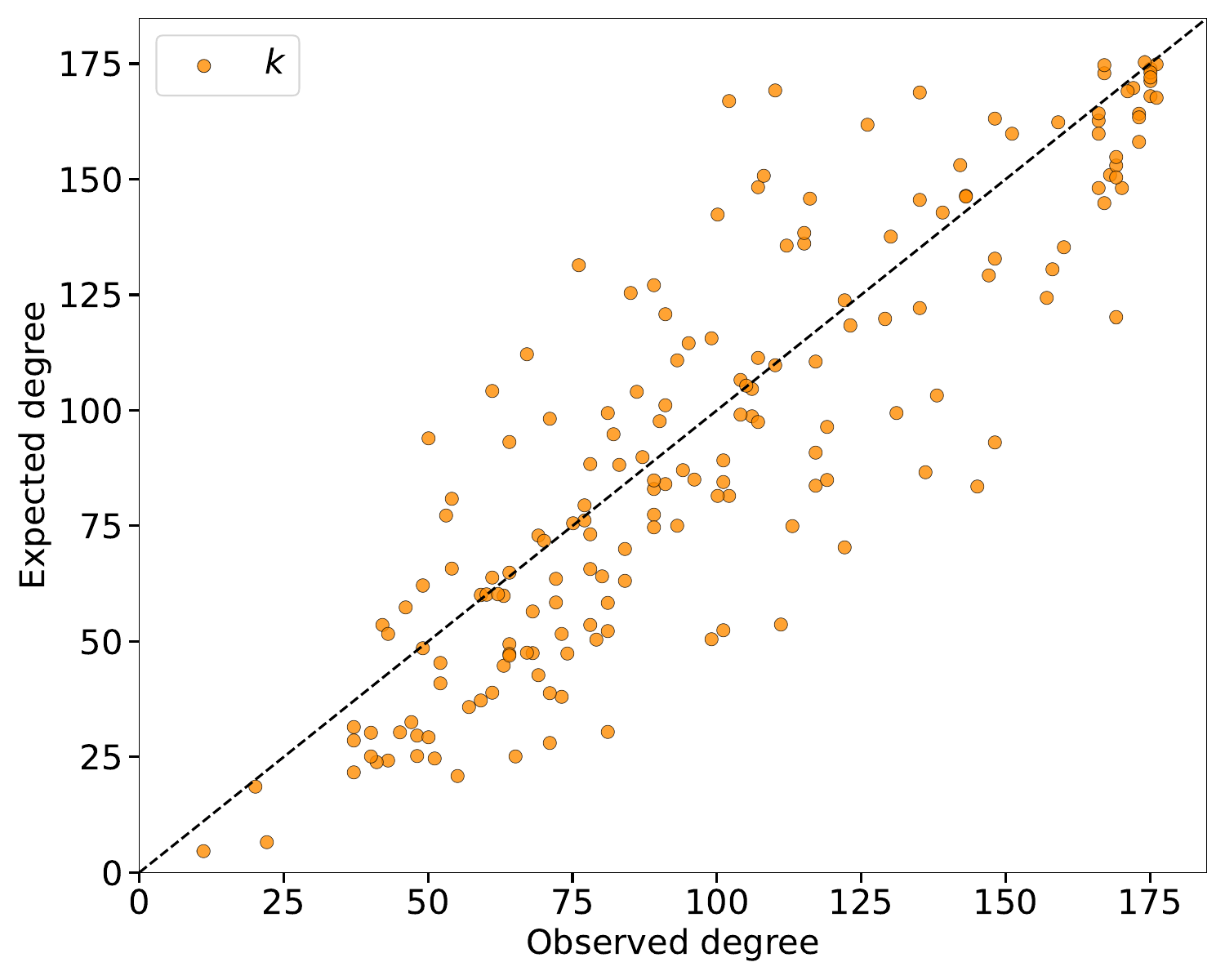}
\par\textbf{(a)}
\end{minipage}
\hfill
\begin{minipage}{0.32\linewidth}
\centering
\includegraphics[width=\linewidth]{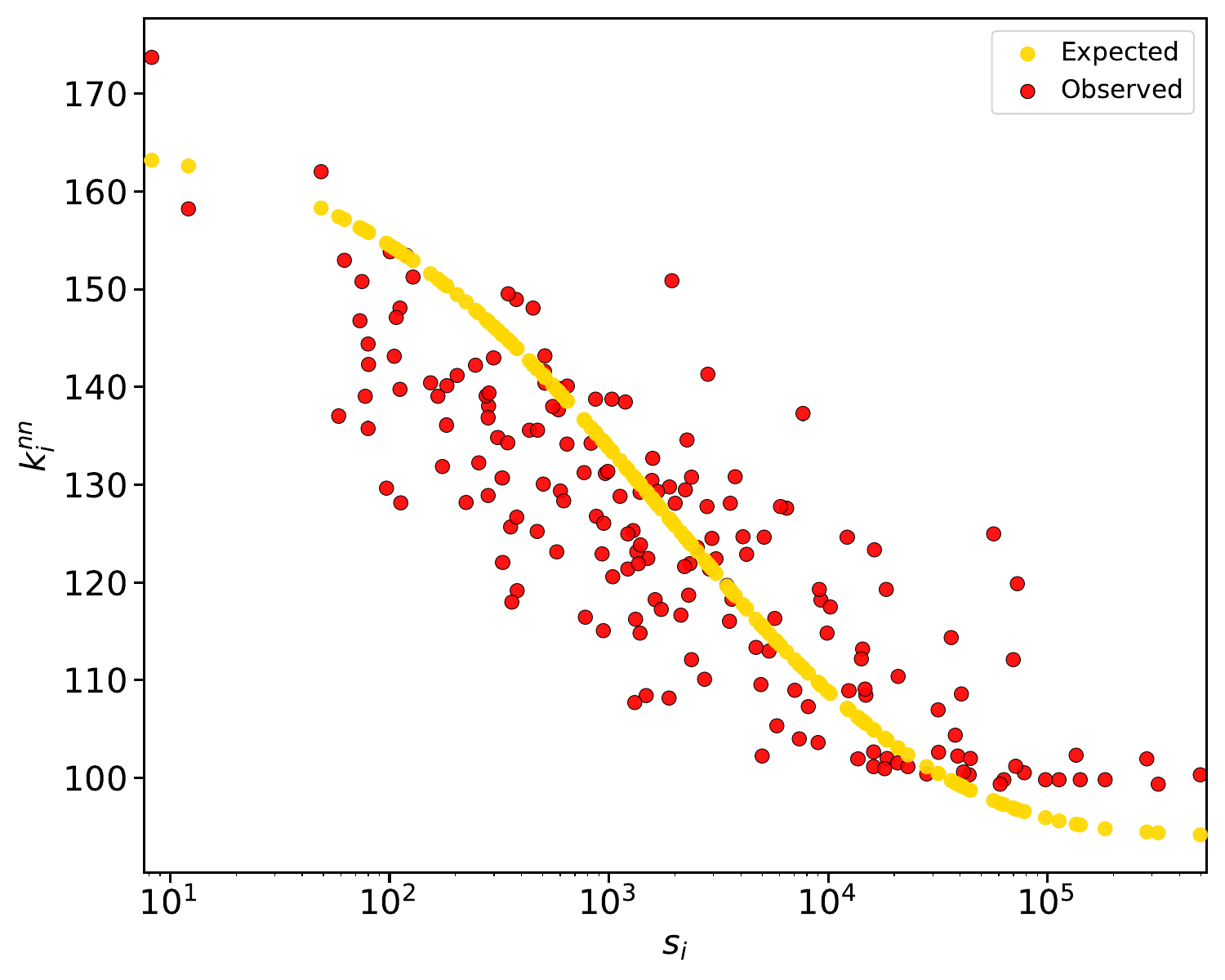}
\par\textbf{(b)}
\end{minipage}
\hfill
\begin{minipage}{0.32\linewidth}
\centering
\includegraphics[width=\linewidth]{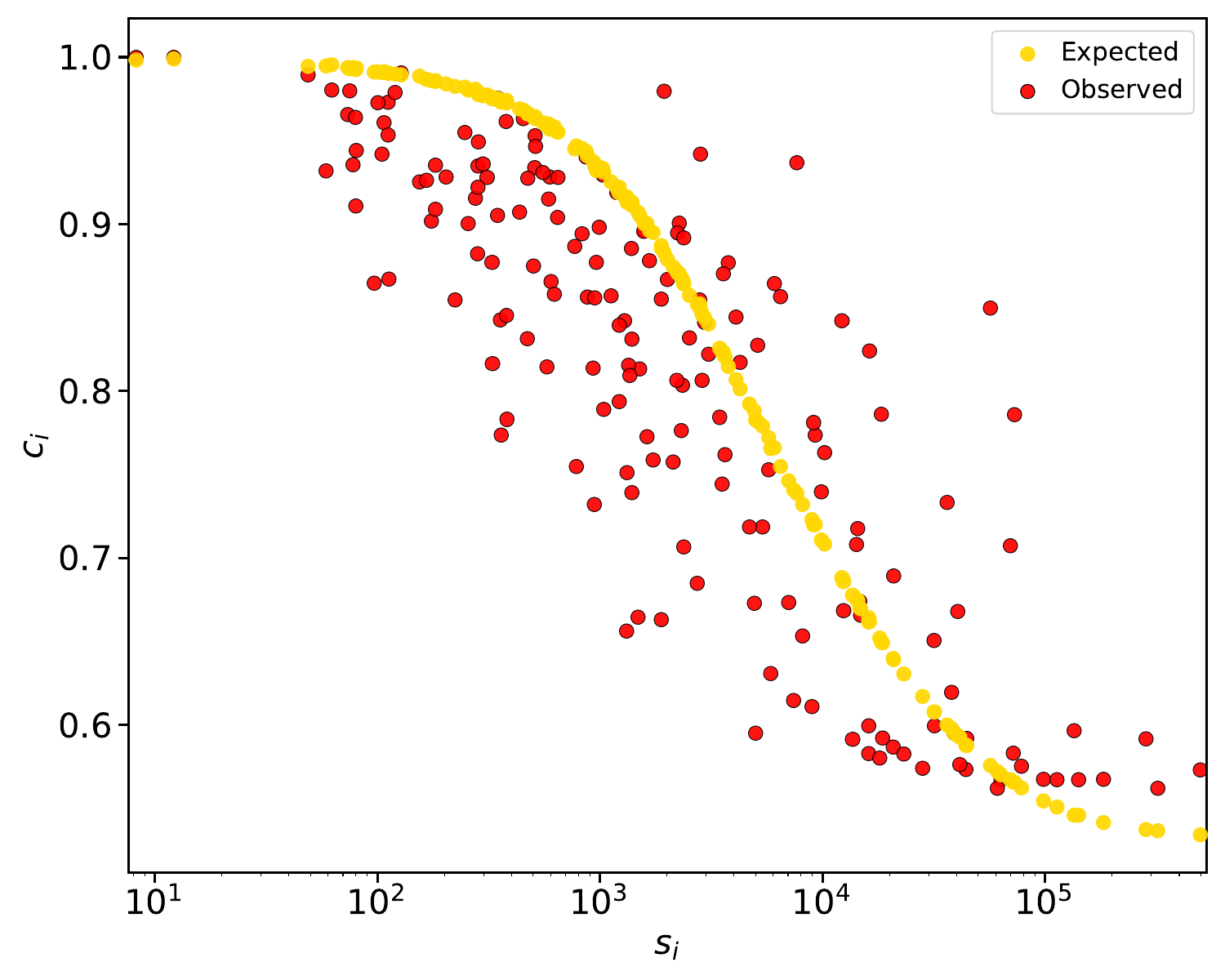}
\par\textbf{(c)}
\end{minipage}
\vspace{0.4em}
\begin{minipage}{0.32\linewidth}
\centering
\includegraphics[width=\linewidth]{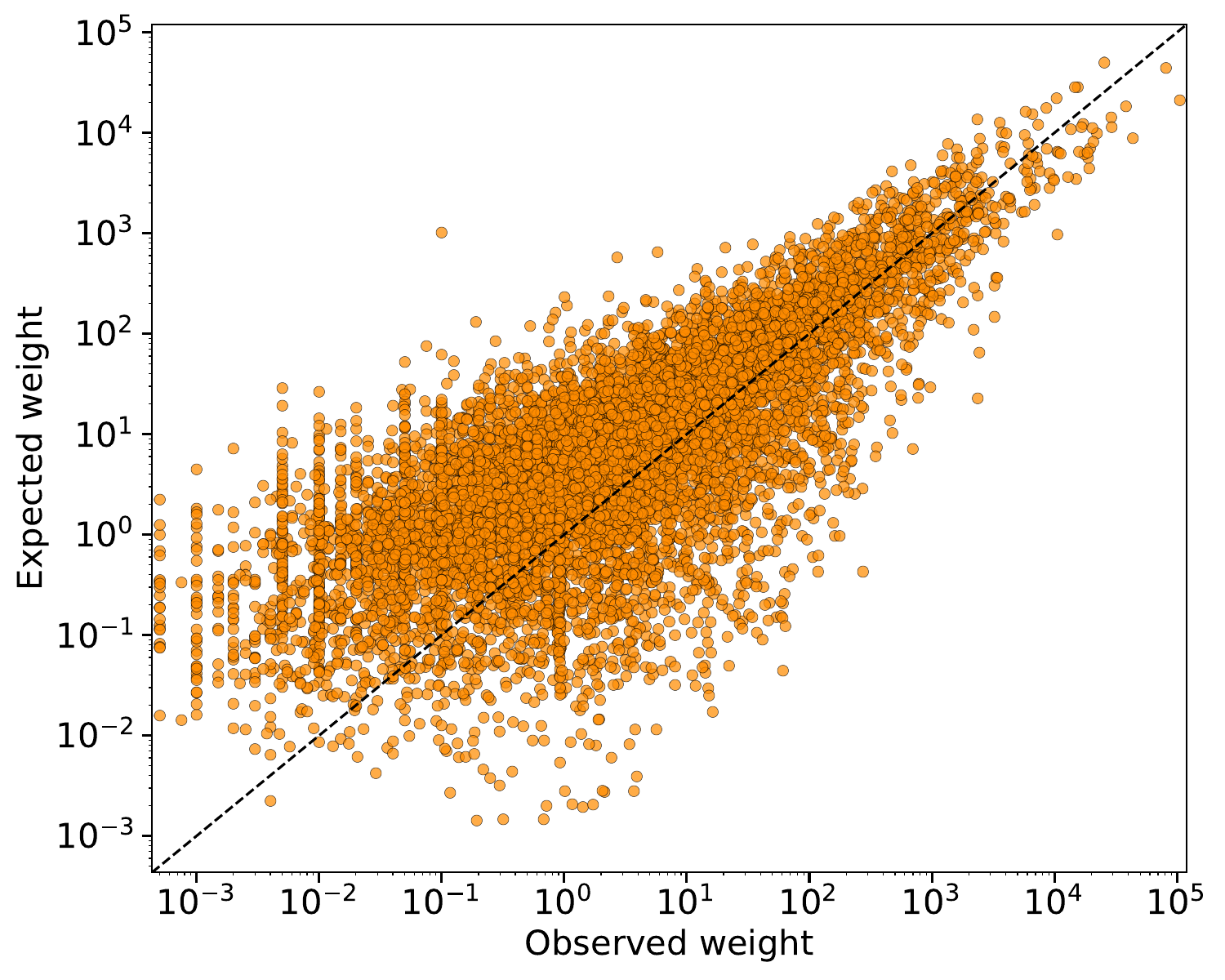}
\par\textbf{(d)}
\end{minipage}
\hfill
\begin{minipage}{0.32\linewidth}
\centering
\includegraphics[width=\linewidth]{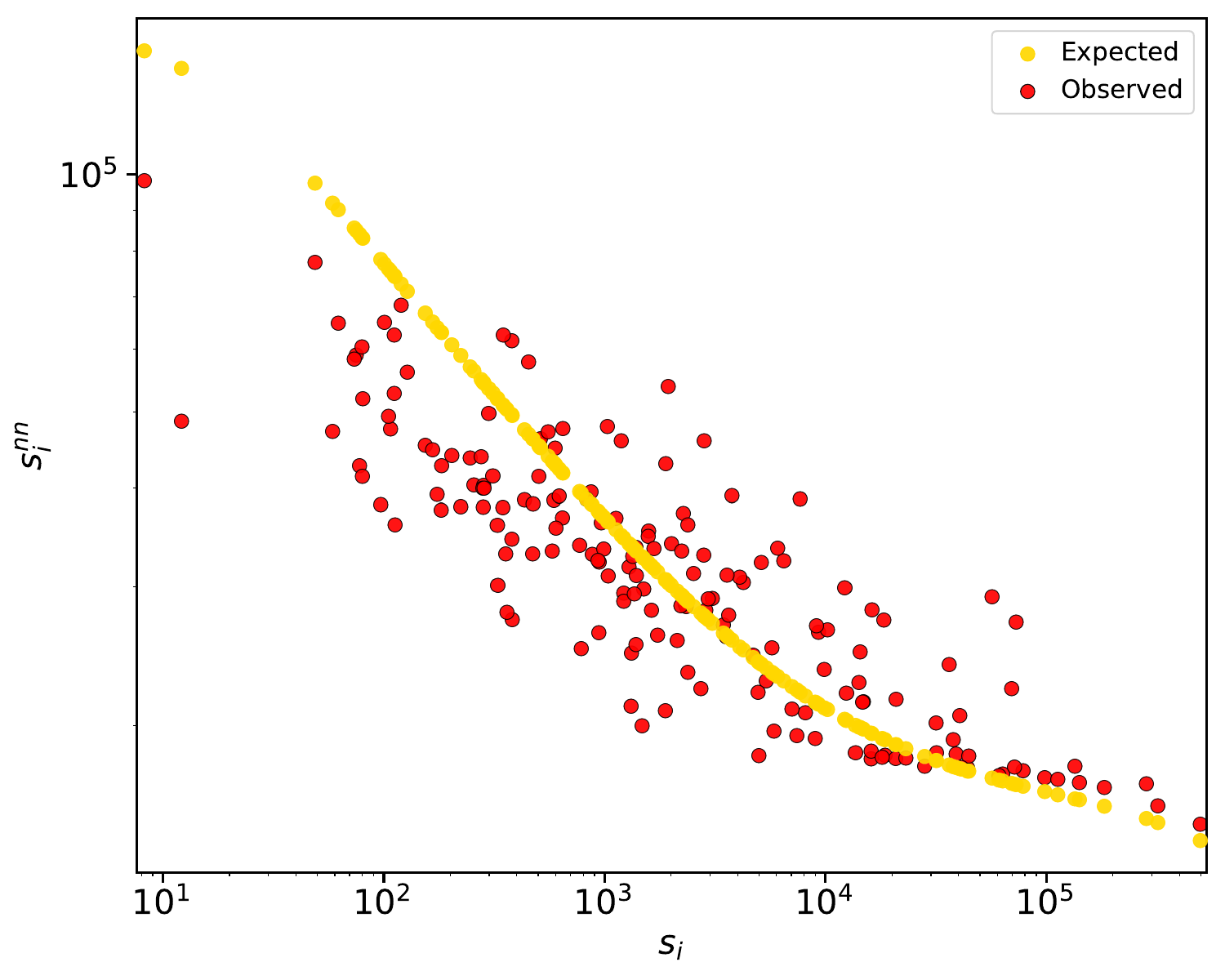}
\par\textbf{(e)}
\end{minipage}
\hfill
\begin{minipage}{0.32\linewidth}
\centering
\includegraphics[width=\linewidth]{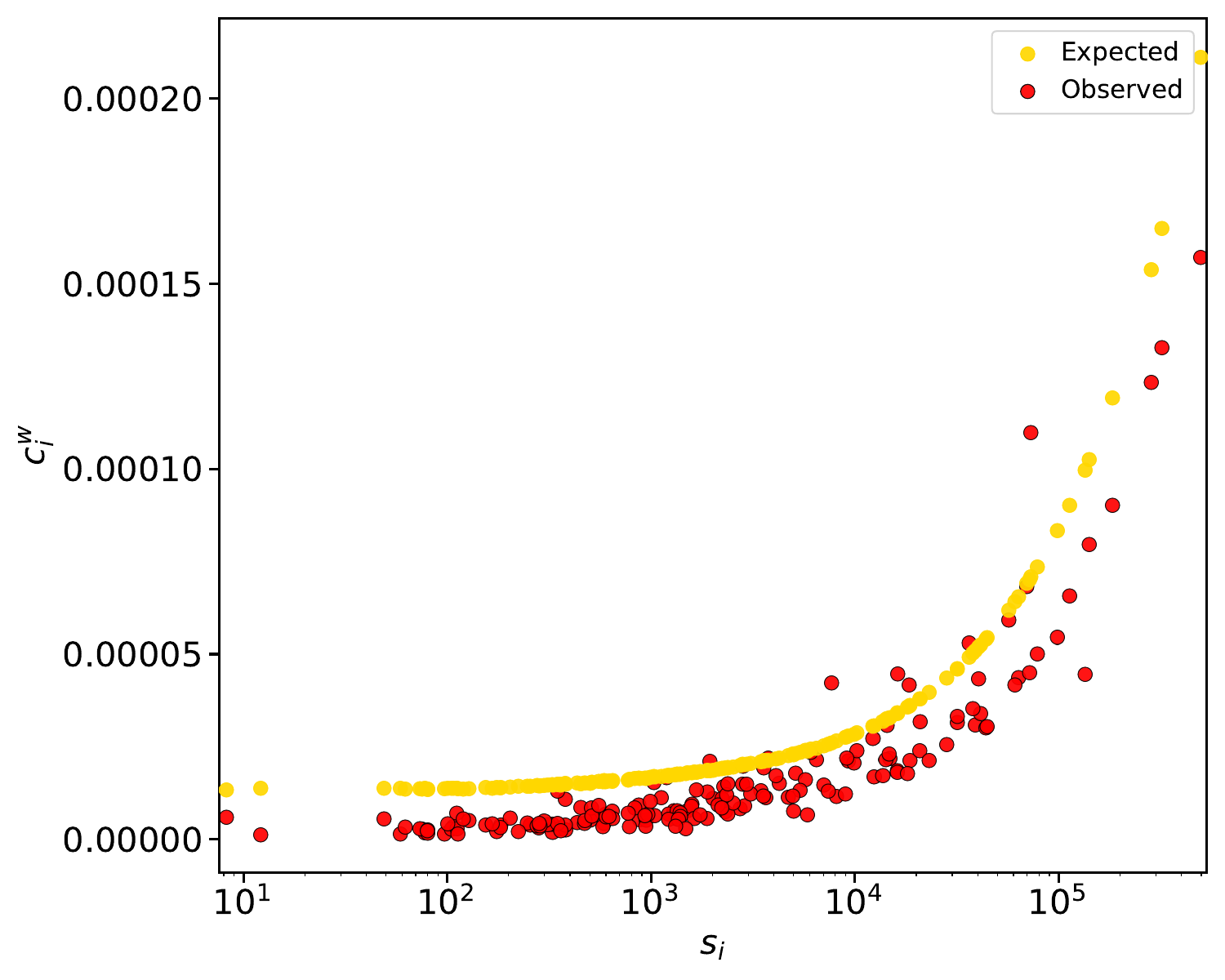}
\par\textbf{(f)}
\end{minipage}
\caption{\textbf{Country-level diagnostics for the 1993 ITN reconstructed by the wMSM}. Panels~(a–c): observed versus expected degrees, average nearest-neighbor degree $k_i^{nn}$, and binary clustering coefficient $c_i$. Panels~(d–f): observed positive dyadic weights versus expected weights, average nearest-neighbor strength $s_i^{nn}$, and weighted clustering coefficient $c_i^w$. In the local-profile panels, points and curves compare observed country-level values with expected values under the fitted ensemble.}
\label{fig:3}
\end{figure*}

To construct the macro-regional layer, we start from the matrix of bilateral geographical distances between countries. We then apply single-linkage hierarchical clustering, which builds a dendrogram by progressively merging the closest countries or groups of countries. At the beginning, each country forms its own cluster. As the distance threshold increases, nearby countries merge first, while more distant clusters merge only at larger heights. The height at which two countries first become part of the same cluster defines their dendrogram, or ultrametric, distance. Cutting the dendrogram at a prescribed height $d_c$ therefore produces a partition of the country set into macro-regions: countries connected below the cut belong to the same macro-region, while countries whose common ancestor lies above the cut remain in different macro-regions. In the main analysis we use $d_c=800\,\mathrm{km}$.

Given a macro-regional partition $\Omega_1(d_c)=\{I_1,\dots,I_{N^{(1)}}\}$, country-level weights are aggregated by summation,
\begin{equation}
w_{IJ}^{(1)}(t)
=
\sum_{i\in I}\sum_{j\in J}w_{ij}^{(0)}(t).
\end{equation}
The corresponding binary macro-regional projection is
\begin{equation}
a_{IJ}^{(1)}(t)
=
\mathbbm{1}\!\left(w_{IJ}^{(1)}(t)>0\right).
\end{equation}
Macro-regional self-loops are retained, since $w_{II}^{(1)}(t)$ represents the total trade internal to macro-region $I$. 

At each year, the parameter $\delta$ is calibrated on the macro-regional layer by matching the observed number of macro-regional links,
\begin{equation}
\sum_{I\le J}
\left[
1-\exp\!\left(-\delta\,s_I^{(1)}(t)s_J^{(1)}(t)\right)
\right]
=
L^{(1)}(t).
\end{equation}
The fitted value of $\delta$ is then transferred unchanged to the country layer. Figure~\ref{fig:1} summarizes the construction and shows the 1993 macro-regional partition. The map is only used to define the calibration layer: after aggregation, no dyadic distance enters the reconstruction model. Thus the country-level prediction depends only on the transferred parameter and on country strengths.

Before evaluating the microscopic reconstruction, we first check the behavior of the model at the macro-regional scale where calibration is performed. Since $\delta$ is fitted by matching the macro-regional link count, the expected number of macro-regional links is reproduced by construction, up to numerical precision. Moreover, because macro-regional self-loops are retained, the additive strength constraints are preserved at the calibration layer. The relevant test is therefore whether the same parameter, once transferred to the country layer, still produces a consistent microscopic reconstruction.

Table~\ref{tab:wmsm_absolute_results} shows that the macro-regional layer is reproduced accurately, as expected from the calibration procedure. The macro-regional link-count error is zero up to numerical precision, while macro-regional strengths are reproduced exactly under the retained-diagonal convention. After transferring the same parameter to the country layer, the average relative error on the number of country links is approximately $5.3\%$. Degree errors remain stable across years. The strength error is instead very small, with a mean value below $0.6\%$, because the weighted part of the model preserves the additive strength information up to the country-level diagonal convention.

The 1993 matrices in Fig.~\ref{fig:2} provide a visual diagnostic of the weighted reconstruction. Countries are ordered by macro-region, so that diagonal blocks correspond to within-macro-region trade and off-diagonal blocks to trade between different macro-regions. The observed and expected heatmaps display similar large-scale organization, especially for the strongest trade flows. The circular visualizations provide the same observed-versus-expected comparison in network form, offering a compact representation of both scales at once. For the expected circular network we use the top-$\langle L^{(0)}\rangle$ projection of the fitted probability matrix. This deterministic projection can be interpreted as the most likely support predicted by the model at the expected density.

{Figure~\ref{fig:3} shows the local binary and weighted profiles for the 1993 snapshot, with similar behavior observed in the other years. wMSM captures the main dependence of these quantities on country strength, even though nearest-neighbor and clustering profiles are not directly constrained. This is a stringent test: the model uses only country strengths and two global parameters, with the binary parameter calibrated at the macro-regional scale rather than on countries, yet it still reproduces nonlinear local observables that are not imposed during calibration. The most visible discrepancy concerns the weighted clustering coefficient $c_i^w$, which is slightly overestimated. This offset is consistent with the way the transferred scale-invariant probability matrix distributes link probability across dyads. Since $p_{ij}^{\mathrm{wMSM}}=1-\exp(-\delta s_i s_j)$ is an increasing and saturating function of the strength product, the model tends to concentrate probability on dyads involving high-strength countries once the coarse-scale density has been fixed. As a result, triangles involving strong countries become slightly more likely in the reconstructed ensemble. Because these dyads also carry the largest expected weights, this effect is amplified in weighted triangular observables such as $c_i^w$. The same mechanism explains the small offset visible in the nearest-neighbor strength profile, especially for low-strength countries, whose expected neighborhoods are biased toward stronger trading partners. Thus the deviations are not due to an error in the weighted normalization, which is controlled by the strength constraints, but to the selective binary support induced by the coarse-calibrated scale-invariant topology. Overall, the agreement remains satisfactory given that these local profiles are not fitted directly.}

\begin{table*}[t!]
\centering
\scriptsize
\renewcommand{\arraystretch}{1.08}
\setlength{\tabcolsep}{3.4pt}
\begin{tabular}{|c|ccc|ccc|ccc|}
\hline
& \multicolumn{3}{c|}{$\Omega_2$ calibration (SBI2)}
& \multicolumn{3}{c|}{$\Omega_1$ reconstruction (SBI3)}
& \multicolumn{3}{c|}{$\Omega_0$ reconstruction (SBI4)} \\
\hline
$g$
& $\mathrm{RE}_L^{(2)}$
& $\mathrm{ARE}_k^{(2)}$
& $\mathrm{ARE}_s^{(2)}$
& $\mathrm{RE}_L^{(1)}$
& $\mathrm{ARE}_k^{(1)}$
& $\mathrm{ARE}_s^{(1)}$
& $\mathrm{RE}_L^{(0)}$
& $\mathrm{ARE}_k^{(0)}$
& $\mathrm{ARE}_s^{(0)}$ \\
\hline
0 & 0.0000 & 0.7160 & 0.0000 & 0.1948 & 0.8402 & 0.0000 & 0.1940 & 0.8831 & 0.0000 \\
1 & 0.0000 & 0.6570 & 0.0000 & 0.1415 & 0.7750 & 0.0000 & 0.1242 & 0.7738 & 0.0000 \\
2 & 0.0000 & 0.5061 & 0.0000 & 0.0204 & 0.4865 & 0.0000 & 0.0192 & 0.5051 & 0.0000 \\
3 & 0.0000 & 0.4911 & 0.0000 & 0.0711 & 0.4867 & 0.0000 & 0.1247 & 0.4777 & 0.0000 \\
4 & 0.0000 & 0.4067 & 0.0000 & 0.1273 & 0.5069 & 0.0000 & 0.1747 & 0.5251 & 0.0000 \\
5 & 0.0000 & 0.4809 & 0.0000 & 0.0914 & 0.5444 & 0.0000 & 0.1243 & 0.5552 & 0.0000 \\
6 & 0.0000 & 0.3884 & 0.0000 & 0.1855 & 0.5108 & 0.0000 & 0.2256 & 0.5232 & 0.0000 \\
7 & 0.0000 & 0.4273 & 0.0000 & 0.1466 & 0.4009 & 0.0000 & 0.1818 & 0.3962 & 0.0000 \\
8 & 0.0000 & 0.5321 & 0.0000 & 0.1551 & 0.4922 & 0.0000 & 0.1920 & 0.5024 & 0.0000 \\
9 & 0.0000 & 0.5935 & 0.0000 & 0.1395 & 0.5746 & 0.0000 & 0.1748 & 0.5922 & 0.0000 \\
\hline
Mean & 0.0000 & 0.5199 & 0.0000 & 0.1273 & 0.5618 & 0.0000 & 0.1535 & 0.5734 & 0.0000 \\
\hline
\end{tabular}
\caption{\textbf{Performance of the wMSM on the DPN}. For each one-digit good group $g$, the model is calibrated at the two-digit SBI layer $\Omega_2$ and transferred to the three-digit and four-digit layers, $\Omega_1$ and $\Omega_0$. Diagonal entries are retained at all sectoral resolutions, because they represent within-sector production relations. Consequently, strengths are reproduced exactly up to numerical precision at every layer.}
\label{tab:dutch_wmsm_absolute_results}
\end{table*}

The same analysis was repeated for the alternative cuts $d_c=400,600,1000\,\mathrm{km}$. The qualitative behavior is stable across cuts: finer partitions retain more information but are closer to the country layer, while coarser partitions produce denser macro-regional networks and reduce the amount of binary information available for calibration. A systematic robustness analysis across ultrametric aggregation scales is reported in Appendix~\hyperlink{AppC}{C}.

\subsection*{Reconstructing Dutch production networks across sectoral resolutions}

We now apply the same procedure to the DPN. In this case the multiscale structure is not geographical but sectoral. The finest layer $\Omega_0$ corresponds to four-digit SBI sectors, the intermediate layer $\Omega_1$ to three-digit SBI sectors, and the coarsest layer $\Omega_2$ to two-digit SBI sectors. The data contain ten one-digit good groups, indexed by $g=0,\dots,9$. For each good group, we calibrate the wMSM on the coarse $\Omega_2$ layer and transfer the fitted parameter to both $\Omega_1$ and $\Omega_0$.

The diagonal convention differs from the country-level ITN case. Sectoral self-loops are retained at all resolutions, because they represent within-sector production relations and therefore have empirical meaning. As a consequence, the strength constraints are matched exactly up to numerical precision at all three sectoral scales.

At each good group $g$, the parameter $\delta$ is calibrated at the two-digit SBI layer by solving
\begin{equation}
\sum_{I\le J}
\left[
1-\exp\!\left(-\delta\,s_I^{(2,g)}s_J^{(2,g)}\right)
\right]
=
L^{(2,g)}.
\end{equation}
The same value of $\delta$ is then used to reconstruct the three-digit and four-digit layers,
\begin{equation}
p_{ij}^{(\ell,g)}
=
1-\exp\!\left(-\delta\,s_i^{(\ell,g)}s_j^{(\ell,g)}\right),
\qquad
\ell\in\{0,1\},
\end{equation}
with expected weights
\begin{equation}
\left\langle w_{ij}^{(\ell,g)}\right\rangle
=
\frac{\delta}{\rho}\,s_i^{(\ell,g)}s_j^{(\ell,g)}.
\end{equation}

\begin{figure*}[t!]
\centering
\begin{minipage}{0.48\linewidth}
\centering
\includegraphics[width=\linewidth]{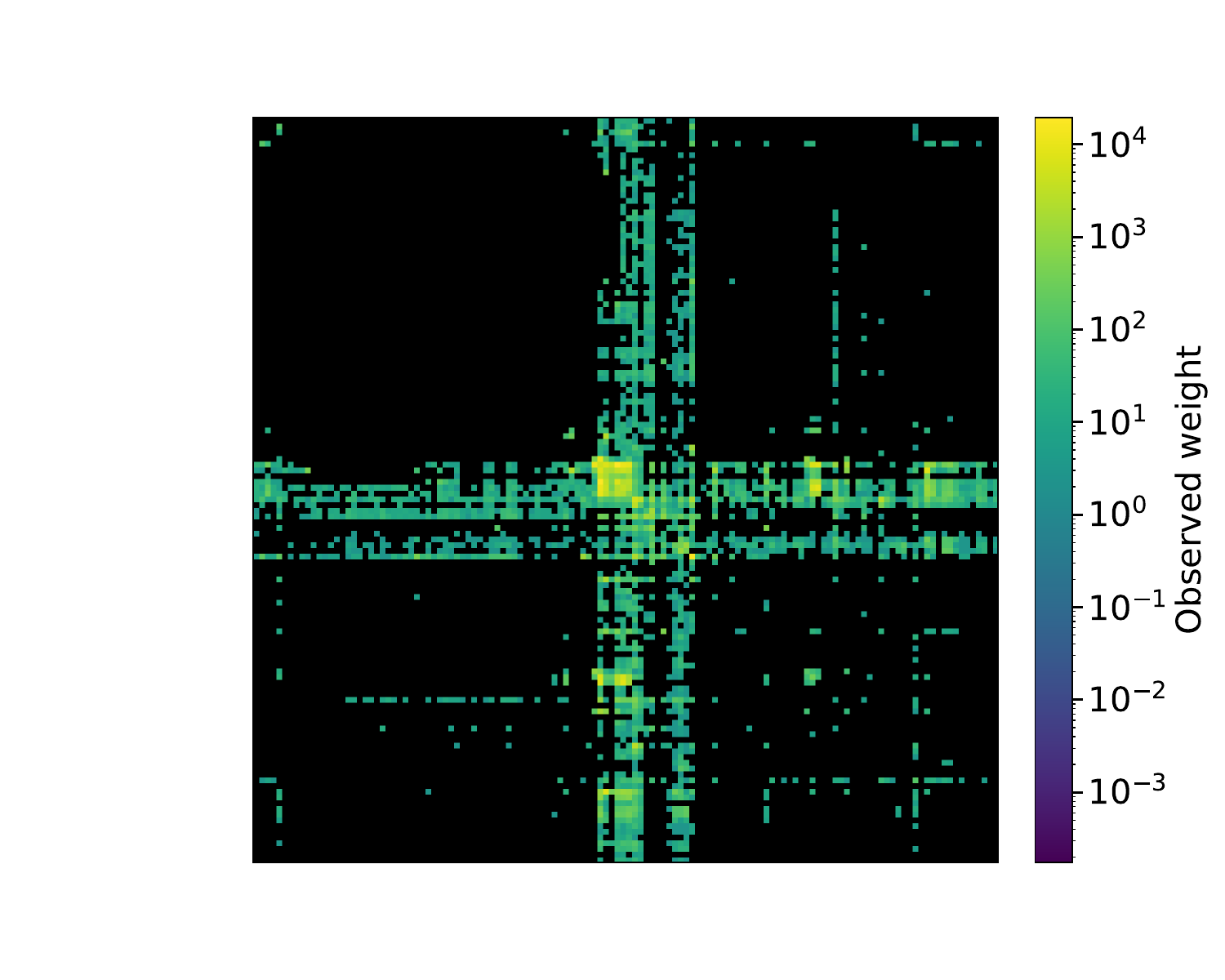}
\par\textbf{(a)}
\end{minipage}
\hfill
\begin{minipage}{0.48\linewidth}
\centering
\includegraphics[width=\linewidth]{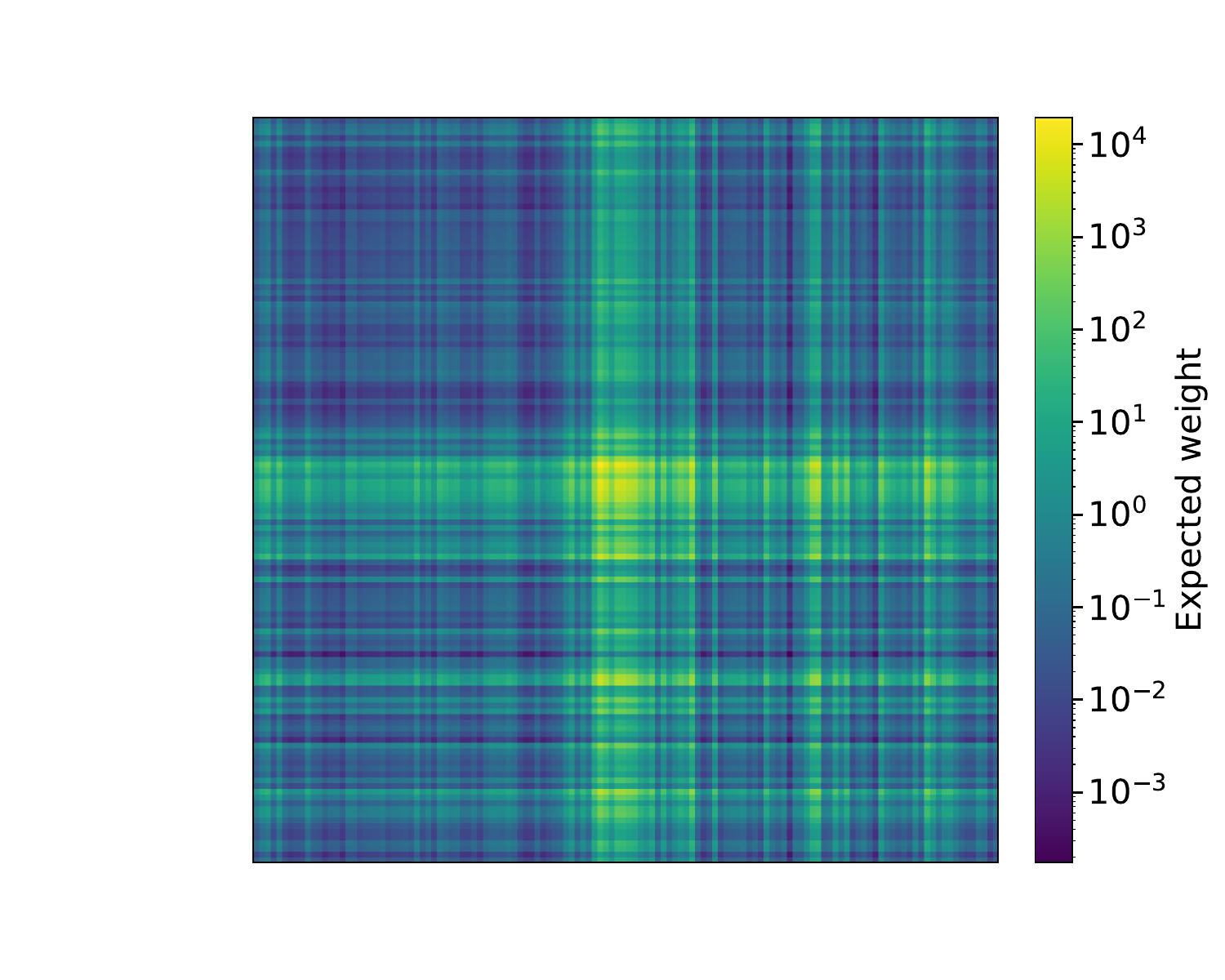}
\par\textbf{(b)}
\end{minipage}
\vspace{0.4em}
\begin{minipage}{0.48\linewidth}
\centering
\includegraphics[width=\linewidth]{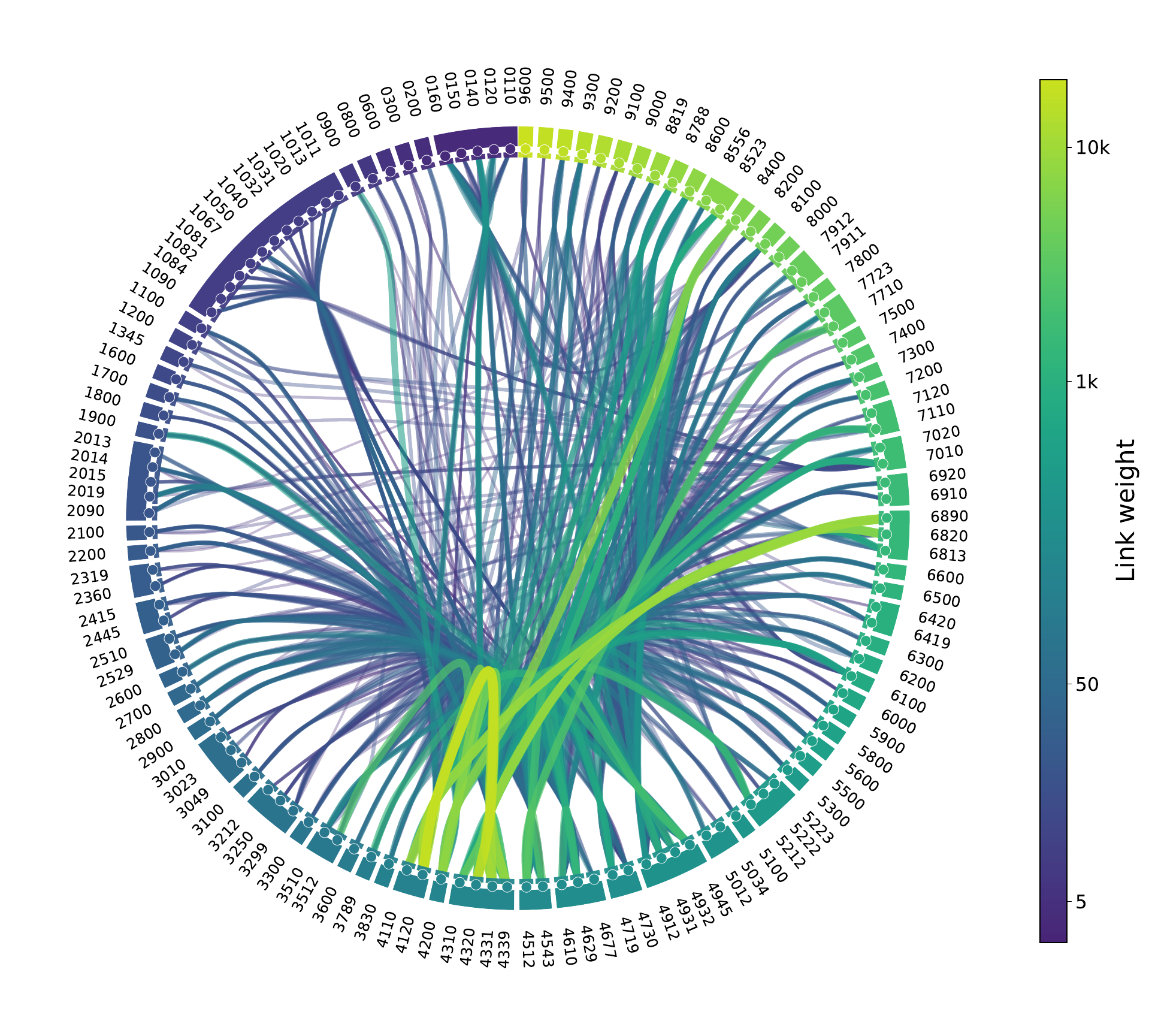}
\par\textbf{(c)}
\end{minipage}
\hfill
\begin{minipage}{0.48\linewidth}
\centering
\includegraphics[width=\linewidth]{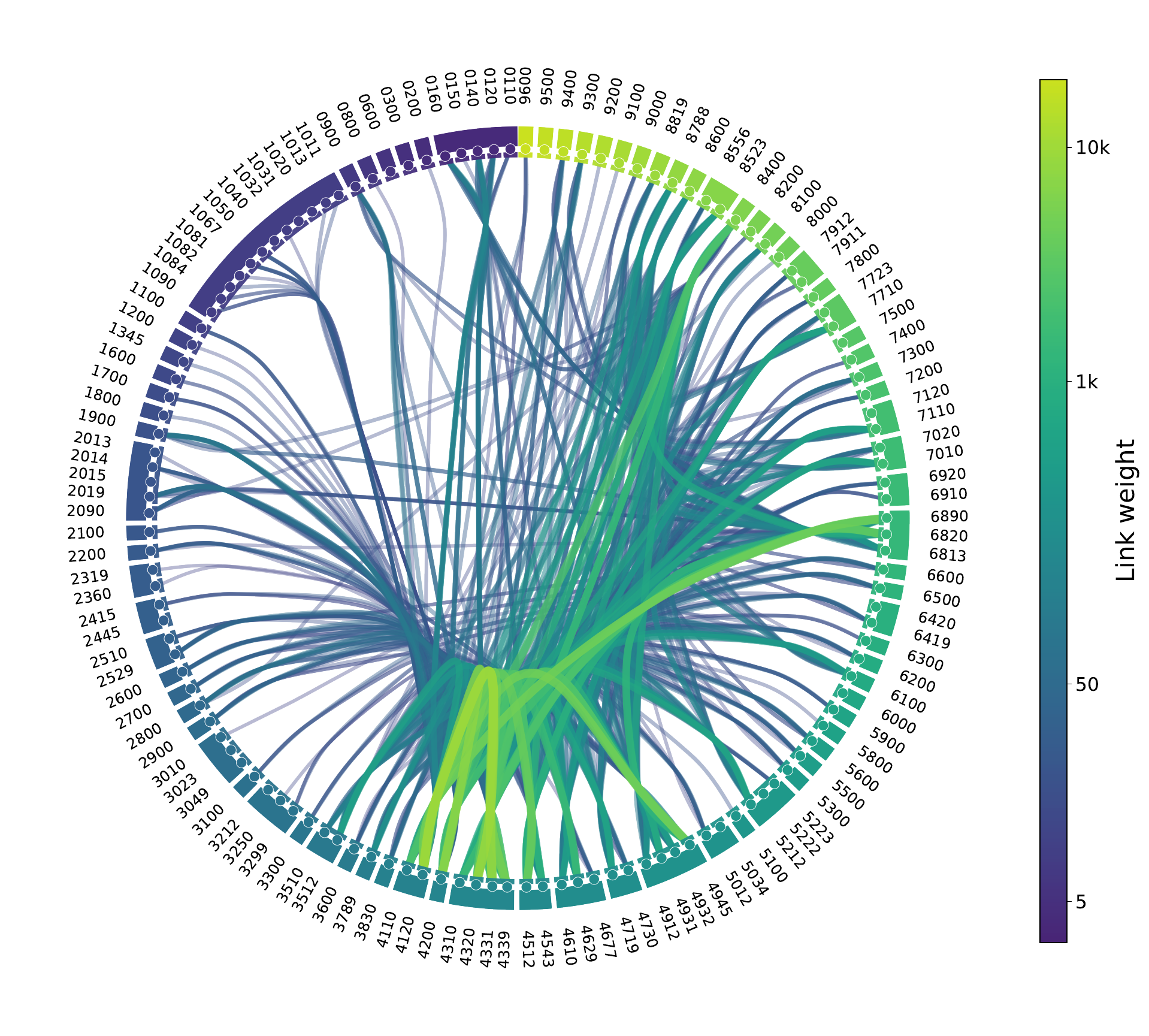}
\par\textbf{(d)}
\end{minipage}
\caption{\textbf{Observed and reconstructed four-digit SBI production network for good group $g=4$}. Panels~(a,b): observed sector-to-sector weights and wMSM expected weights, with sectors ordered according to the SBI hierarchy. Black entries correspond to absent links or values outside the plotted range. Panels~(c,d): circular visualization of the same observed and reconstructed weighted networks. In the reconstructed network, the displayed support is obtained by retaining the top-$\langle L^{(0)}\rangle$ dyads according to the fitted probabilities. This gives the most likely binary realization predicted by the model. Edge width encodes the corresponding expected production volume. Node colors identify sectoral groups.}
\label{fig:4}
\end{figure*}

Table~\ref{tab:dutch_wmsm_absolute_results} shows that the fitted model reproduces the coarse SBI2 calibration layer by construction. When transferred to finer sectoral resolutions, the expected number of links remains reasonably close to the observed one across all good groups. The mean link-count error is about $12.7\%$ at the three-digit layer and about $15.4\%$ at the four-digit layer. This deterioration is expected, because the same parameter is transferred from a substantially coarser sectoral representation. Nevertheless, the reconstruction remains informative across all ten product layers, and the strength error remains zero up to numerical precision because sectoral diagonal entries are retained.

For visual diagnostics, Fig.~\ref{fig:4} focuses on the four-digit SBI layer for good group $g=4$. The observed and expected heatmaps display a similar block organization induced by the SBI hierarchy. As in the ITN case, the expected circular network is shown through the top-$\langle L^{(0)}\rangle$ projection of the fitted probability matrix. The agreement is notable because the four-digit sectoral network is reconstructed using only four-digit strengths and two global parameters, with the binary parameter calibrated at the two-digit layer.

\begin{figure*}[t!]
\centering
\begin{minipage}{0.32\linewidth}
\centering
\includegraphics[width=\linewidth]{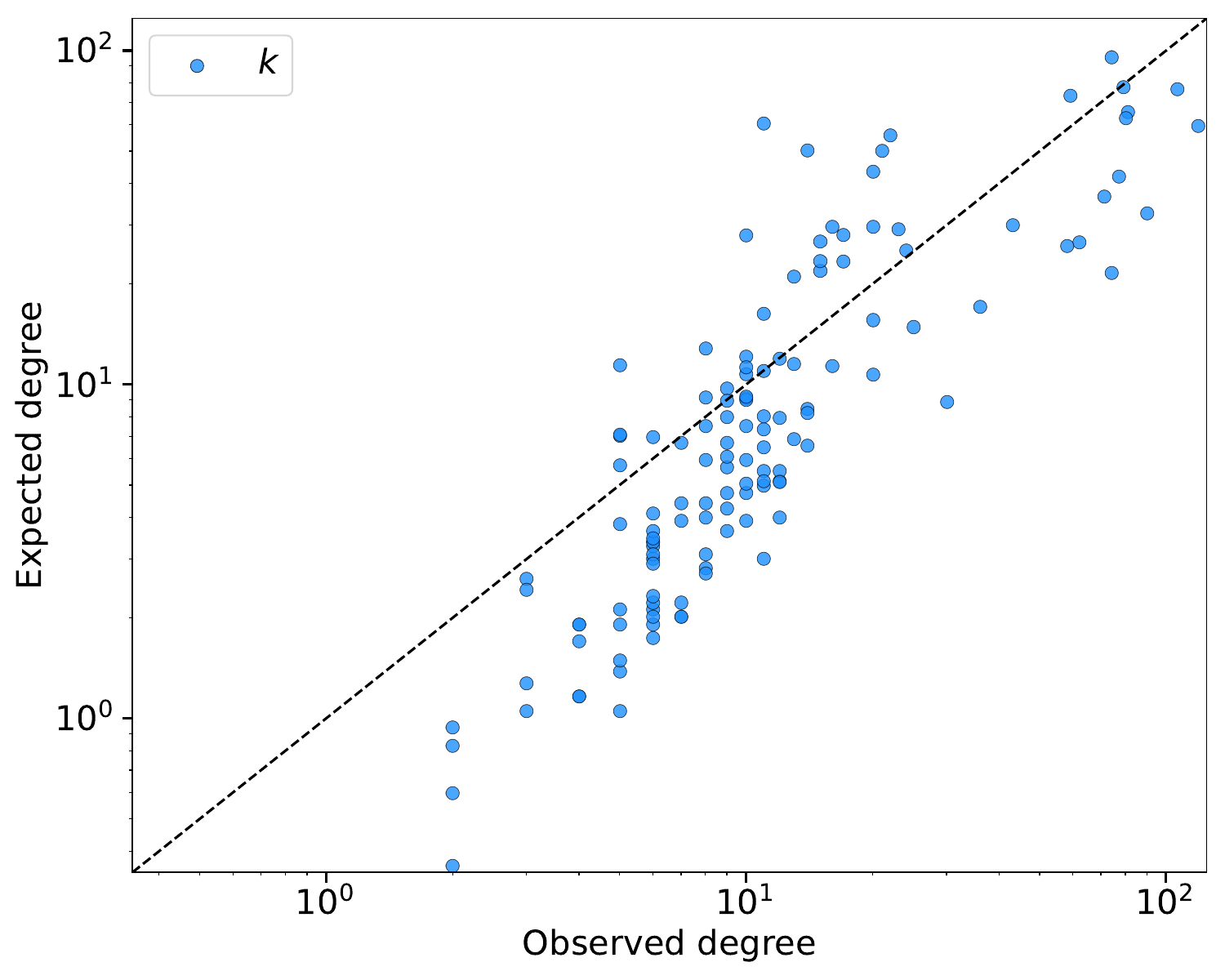}
\par\textbf{(a)}
\end{minipage}
\hfill
\begin{minipage}{0.32\linewidth}
\centering
\includegraphics[width=\linewidth]{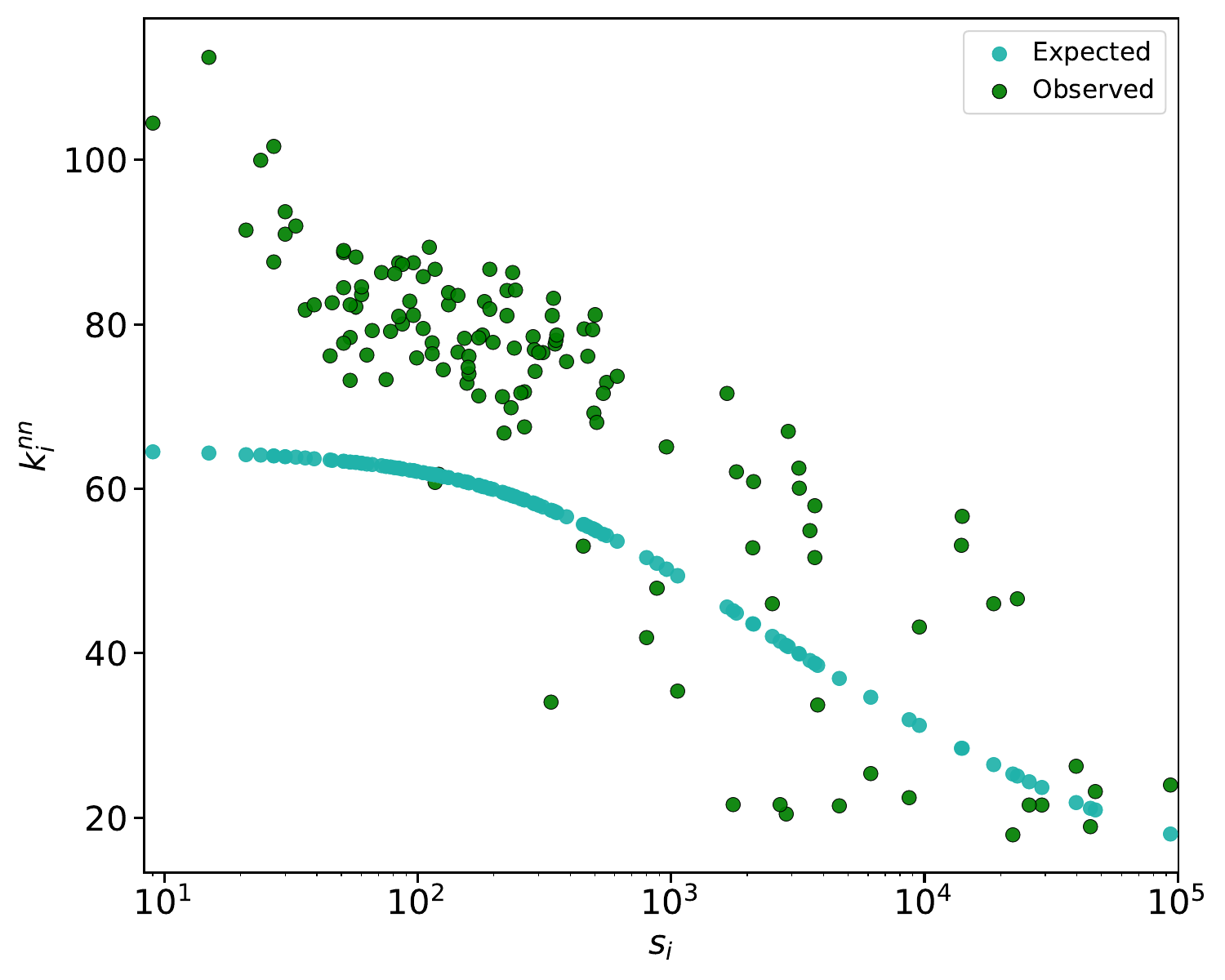}
\par\textbf{(b)}
\end{minipage}
\hfill
\begin{minipage}{0.32\linewidth}
\centering
\includegraphics[width=\linewidth]{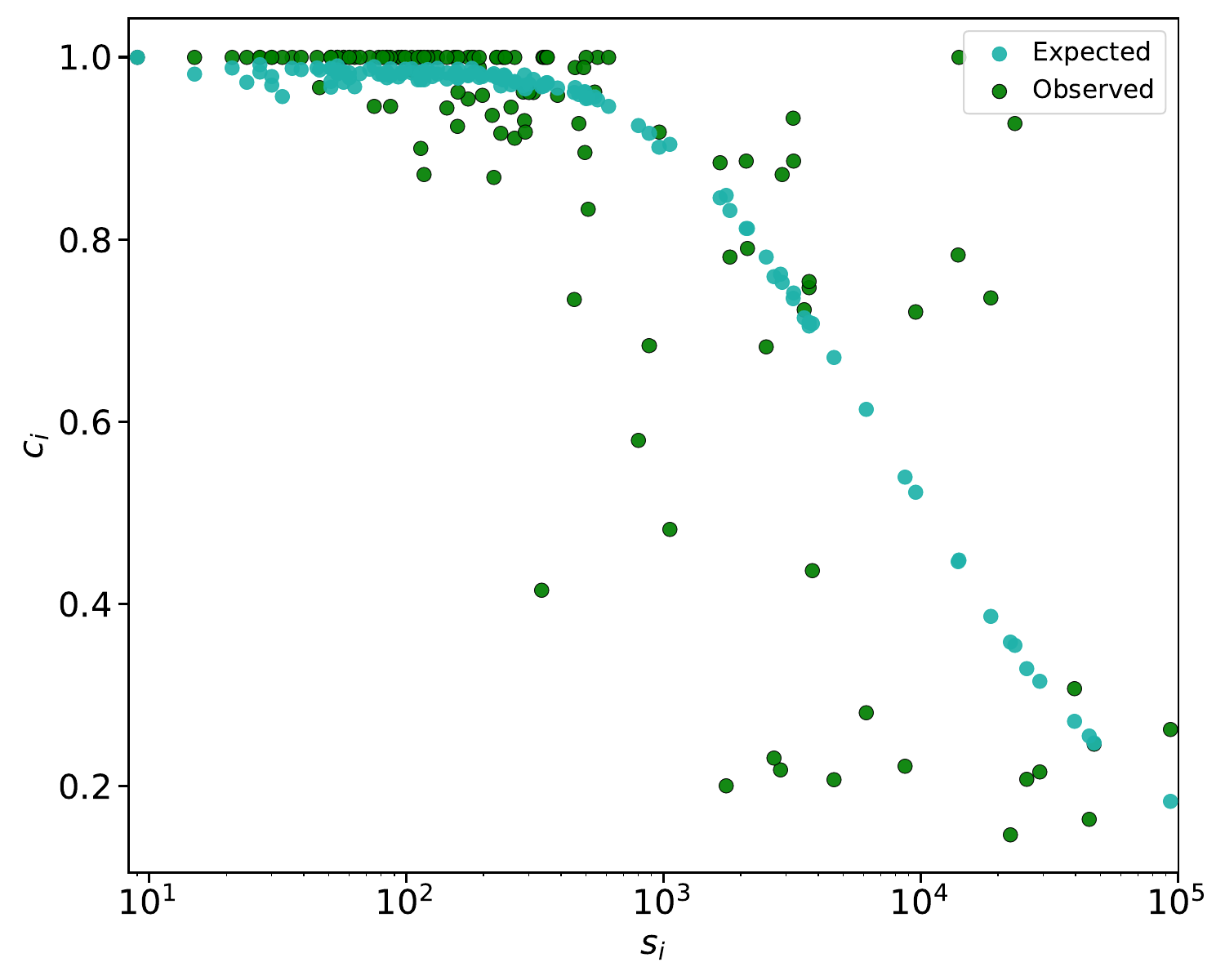}
\par\textbf{(c)}
\end{minipage}
\vspace{0.4em}
\begin{minipage}{0.32\linewidth}
\centering
\includegraphics[width=\linewidth]{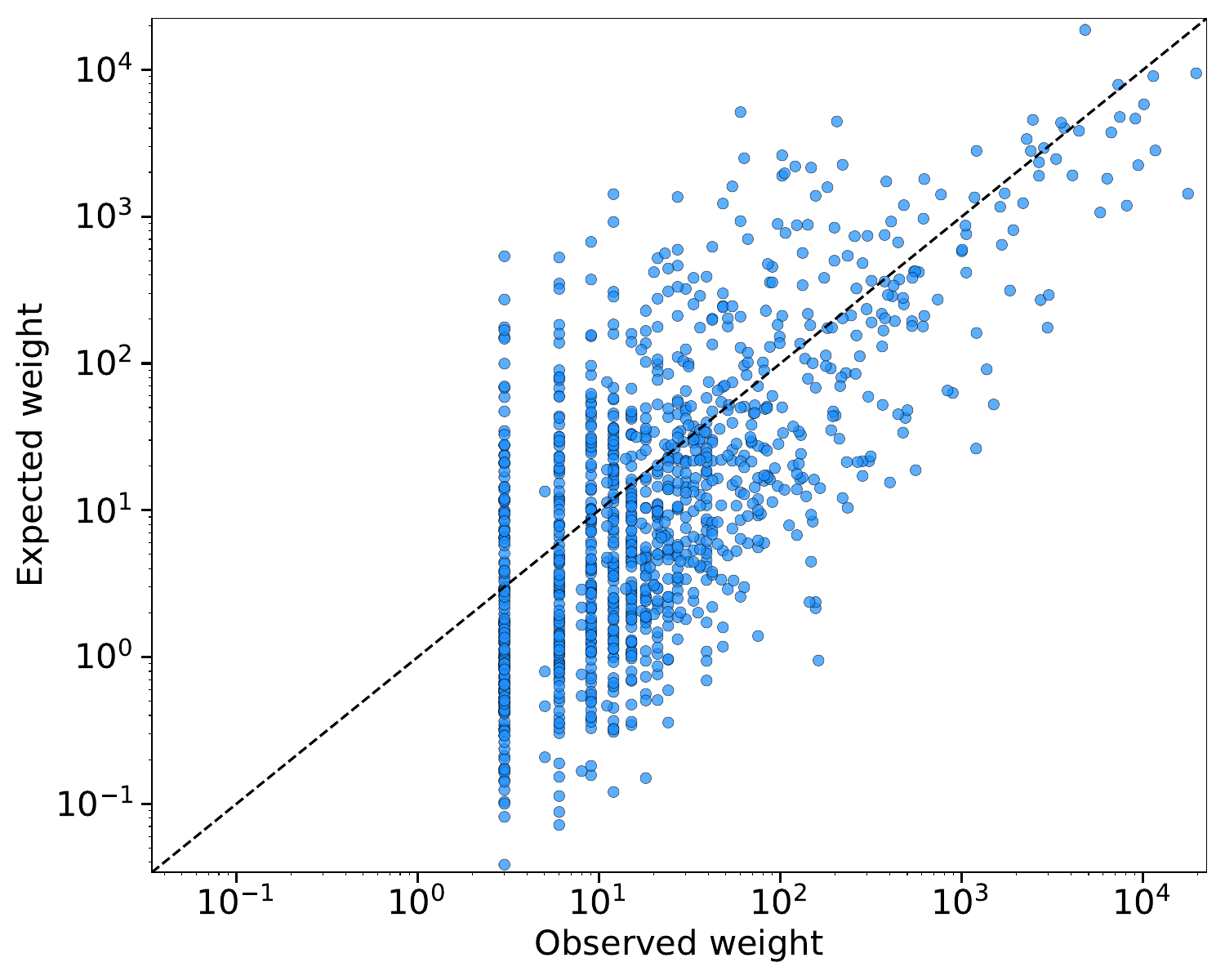}
\par\textbf{(d)}
\end{minipage}
\hfill
\begin{minipage}{0.32\linewidth}
\centering
\includegraphics[width=\linewidth]{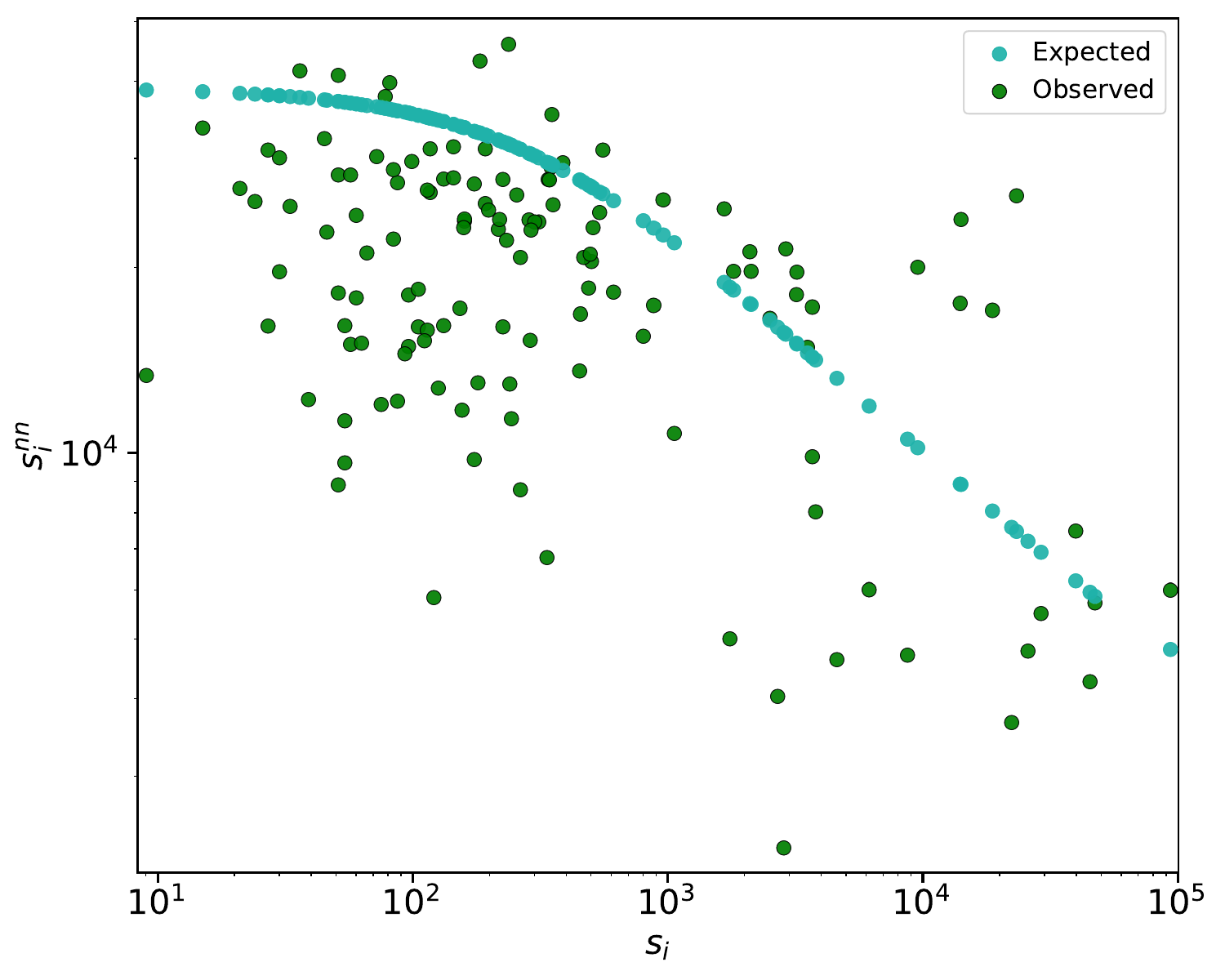}
\par\textbf{(e)}
\end{minipage}
\hfill
\begin{minipage}{0.32\linewidth}
\centering
\includegraphics[width=\linewidth]{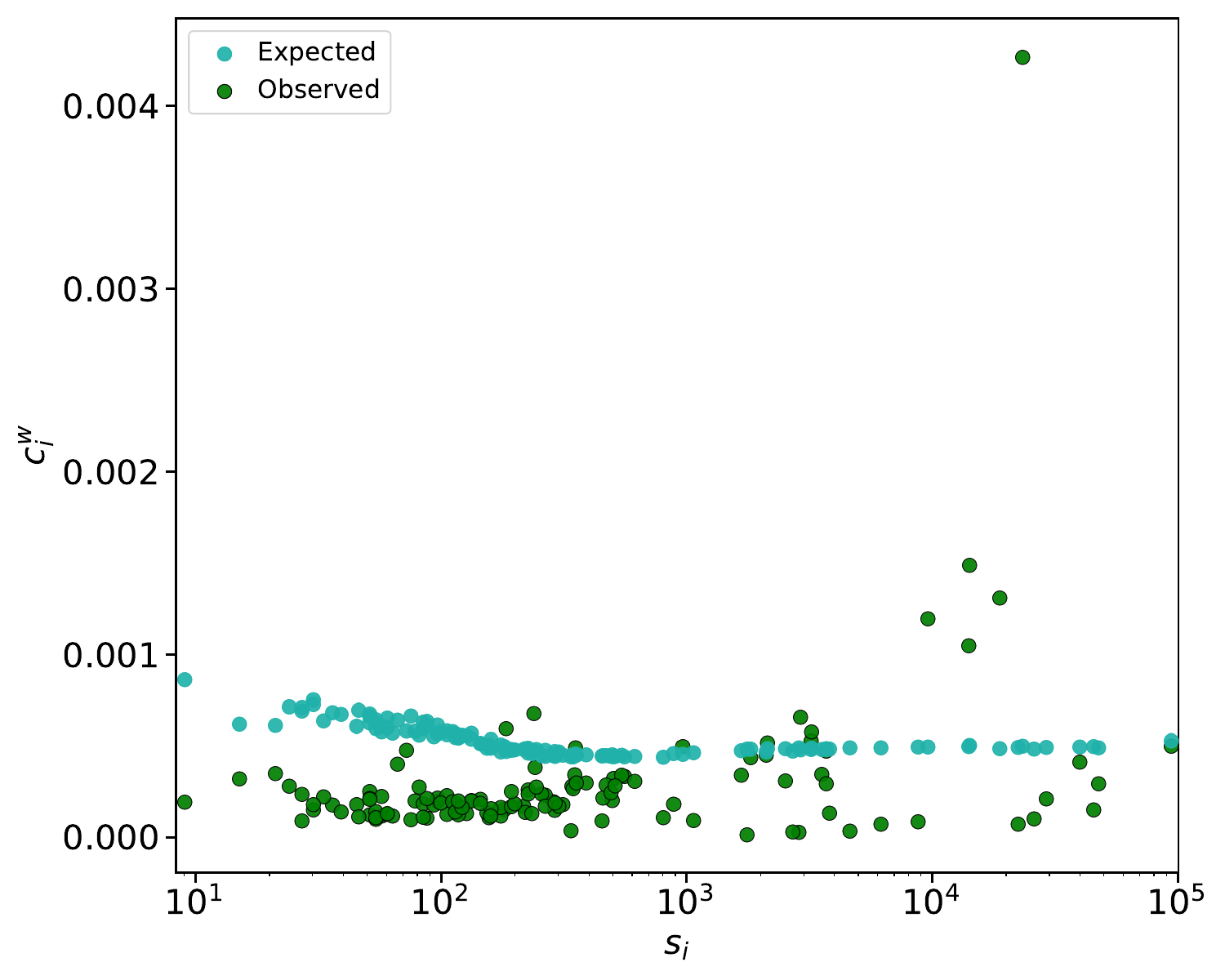}
\par\textbf{(f)}
\end{minipage}
\caption{\textbf{Four-digit SBI diagnostics for the DPN reconstructed by the wMSM, shown for good group $g=4$}. Panels~(a–c): observed versus expected degrees, average nearest-neighbor degree $k_i^{nn}$, and binary clustering coefficient $c_i$. Panels~(d–f): observed positive dyadic weights versus expected weights, average nearest-neighbor strength $s_i^{nn}$, and weighted clustering coefficient $c_i^w$. In the local-profile panels, points and curves compare observed sector-level values with expected values under the fitted ensemble.}
\label{fig:5}
\end{figure*}

Figure~\ref{fig:5} shows that the reconstructed four-digit layer captures the main structural organization of the observed DPN beyond the quantities imposed during calibration. The nearest-neighbor and clustering profiles are not constrained, and yet their dependence on sector strength is reproduced with the correct qualitative behavior. This confirms that the same scale-invariant mechanism can be transferred from a geographical aggregation problem to a sectoral aggregation problem without changing the model.

\subsection*{Benchmark against fine-scale weighted reconstruction}

The previous sections show that the wMSM reconstructs both the ITN and the DPN visually and through several local diagnostics. To assess whether this performance is competitive, we compare it with the CReMB, a state-of-the-art weighted reconstruction model based on the Conditional Reconstruction Method~\cite{CReM2020}, whose construction builds on maximum-entropy approaches for economic and financial networks~\cite{Mastrandrea2014EnhancedReconstruction,EstimatingTopologicalProperties2015,CimiModel2015,NetworkReconstructionDensitySampling2017,ReconstructionMethods2018,cimini2021reconstructing}. A detailed description of the benchmark is provided in Appendix~\hyperlink{AppD}{D}.

The comparison is deliberately unfavorable to the wMSM. The CReMB is calibrated directly at the same fine-grained layer at which it is evaluated. In the ITN application, this means that the CReMB is calibrated on the country layer. In the Dutch application, it is calibrated separately on the SBI3 and SBI4 layers. By contrast, the wMSM calibrates its binary density parameter only at a coarser layer and then transfers it to the target layer without using the target link count. Thus, if the wMSM matches or improves the CReMB at the fine scale, it does so despite having access to strictly coarser topological information.

On a generic target support $\mathcal D$, the CReMB uses the density-corrected Gravity Model~\cite{EnhancedGravityModelTrade2019,DiVece2022GravityMER},
\begin{equation}
p_{ij}^{\mathrm{CReMB}}
=
\frac{z\,s_is_j}{1+z\,s_is_j},
\end{equation}
where the parameter $z$ is fixed by imposing the fine-scale link-count constraint
\begin{equation}
\sum_{(i,j)\in\mathcal D}p_{ij}^{\mathrm{CReMB}}=L.
\end{equation}
The weighted layer is then assigned through the gravity expectation
\begin{equation}
\left\langle w_{ij}\right\rangle_{\mathrm{CReMB}}
=
\frac{s_is_j}{W},
\end{equation}
with positive weights generated conditionally on link existence.

\begin{figure*}[t!]
\centering
\begin{minipage}{\linewidth}
\centering
\includegraphics[width=\linewidth]{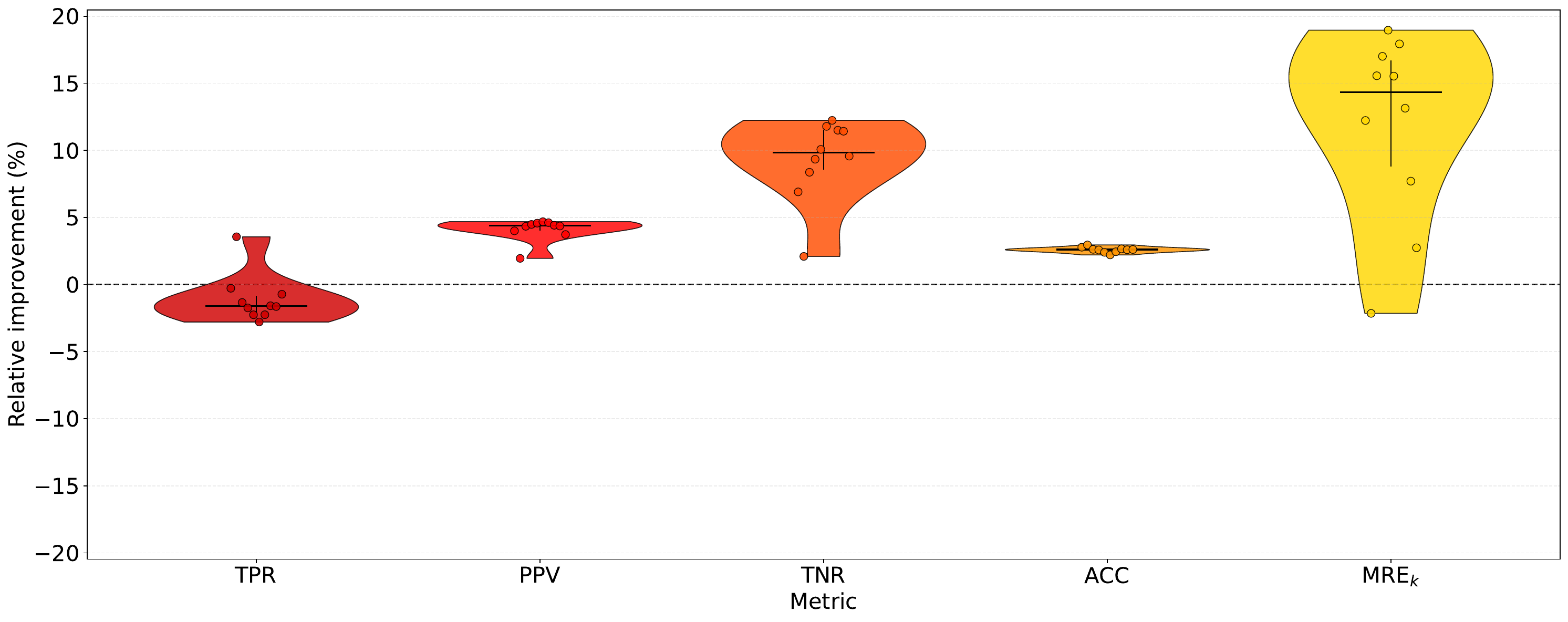}
\par\textbf{(a)}
\end{minipage}
\vspace{0.4em}
\begin{minipage}{\linewidth}
\centering
\includegraphics[width=\linewidth]{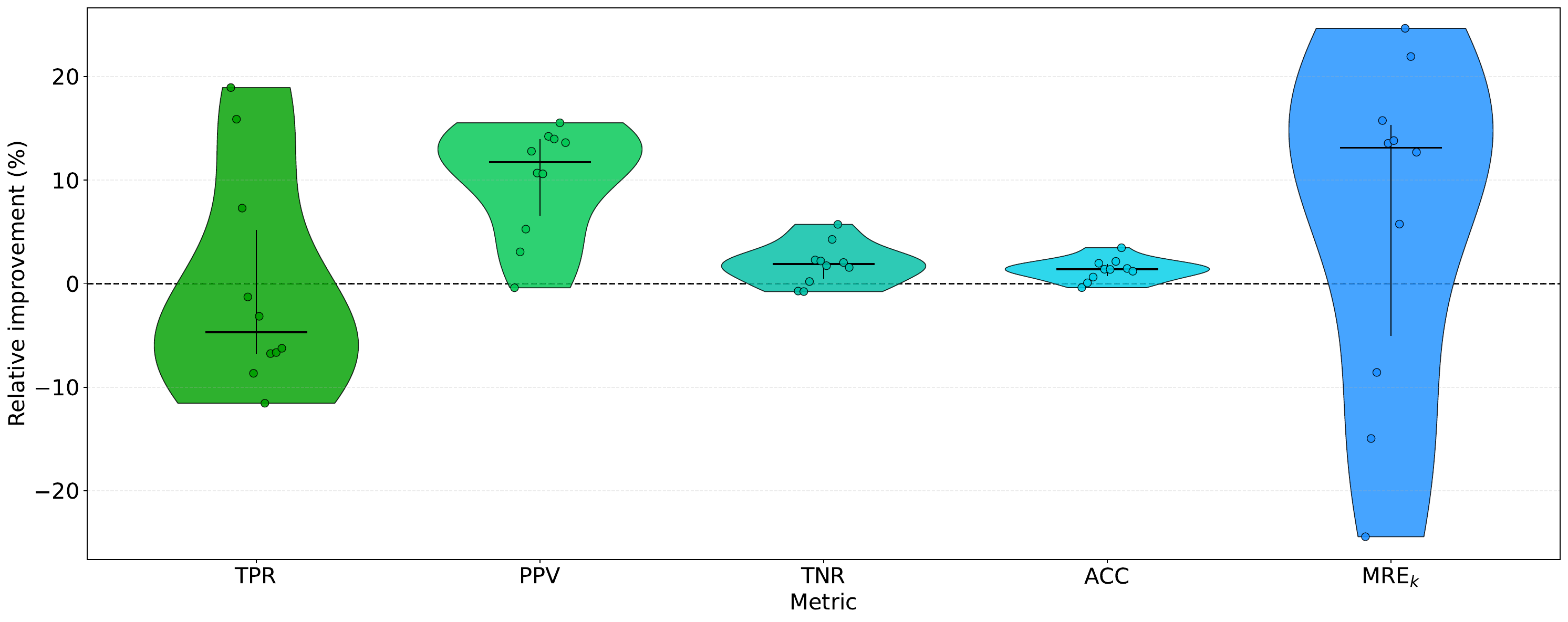}
\par\textbf{(b)}
\end{minipage}
\caption{\textbf{Comparison between the coarse-calibrated wMSM and the fine-calibrated CReMB}. For each metric $m$, the signed relative improvement $\Delta_m$ is expressed in percentage and defined so that positive values favor the wMSM and negative values favor the CReMB. Panel~(a): ITN comparison over the ten yearly snapshots 1991–2000, with the wMSM calibrated at the macro-regional layer and the CReMB calibrated at the country layer. In each ITN violin, dots represent individual years, ordered from left to right from 1991 to 2000. Panel~(b): DPN comparison over the ten one-digit good groups at the four-digit SBI layer, with the wMSM calibrated at the two-digit SBI layer and the CReMB calibrated directly at the four-digit SBI layer. In each Dutch violin, dots represent individual good groups, ordered from left to right from $g=0$ to $g=9$. The dashed horizontal line marks $\Delta_m=0$. The displayed diagnostics are $\mathrm{TPR}$, $\mathrm{PPV}$, $\mathrm{TNR}$, $\mathrm{ACC}$ and $\mathrm{MRE}_k$, distinguished by color.}
\label{fig:6}
\end{figure*}

Before comparing the two models, it is useful to isolate which quantities can actually distinguish them. Once strengths are fixed, the wMSM expected weights reduce to
\begin{equation}
\left\langle w_{ij}\right\rangle_{\mathrm{wMSM}}
=
\frac{\delta}{\rho}s_is_j
=
\frac{s_is_j}{W},
\end{equation}
because $\rho=\delta W$. This is exactly the same unconditional gravity expectation used by the CReMB. Consequently, any metric depending only on $\left\langle w_{ij}\right\rangle$, such as the weighted Jaccard similarity, the root mean squared error on dyadic weights, the Pearson correlation coefficient, or the coefficient of determination $R_w^2$, gives identical results for the two models. These metrics are therefore not informative for the present head-to-head comparison. The relevant difference lies instead in the binary probability matrix: the wMSM assigns probabilities through the scale-invariant exponential form transferred from the coarse layer, whereas the CReMB assigns probabilities through the density-corrected gravity form calibrated directly on the target layer.

We therefore focus on five binary diagnostics. On the target support $\mathcal D$, the expected confusion-matrix entries are
\begin{align}
\mathrm{TP}
&=
\sum_{(i,j)\in\mathcal D}a_{ij}p_{ij},
\\
\mathrm{FP}
&=
\sum_{(i,j)\in\mathcal D}\left(1-a_{ij}\right)p_{ij},
\\
\mathrm{TN}
&=
\sum_{(i,j)\in\mathcal D}\left(1-a_{ij}\right)\left(1-p_{ij}\right),
\\
\mathrm{FN}
&=
\sum_{(i,j)\in\mathcal D}a_{ij}\left(1-p_{ij}\right).
\end{align}
Here, $\mathrm{TP}$ and $\mathrm{TN}$ quantify the expected number of correctly recovered links and non-links, respectively, while $\mathrm{FP}$ and $\mathrm{FN}$ quantify the expected number of spurious links and missed links.

Sensitivity, or true positive rate, measures the fraction of observed links recovered in expectation,
\begin{equation}
\mathrm{TPR}
=
\frac{\mathrm{TP}}{\mathrm{TP}+\mathrm{FN}}.
\end{equation}
Precision, or positive predictive value, measures the fraction of expected links that correspond to observed links,
\begin{equation}
\mathrm{PPV}
=
\frac{\mathrm{TP}}{\mathrm{TP}+\mathrm{FP}}.
\end{equation}
Specificity, or true negative rate, measures the fraction of observed non-links correctly identified as absent,
\begin{equation}
\mathrm{TNR}
=
\frac{\mathrm{TN}}{\mathrm{TN}+\mathrm{FP}}.
\end{equation}
Accuracy measures the overall expected fraction of correctly classified dyads,
\begin{equation}
\mathrm{ACC}
=
\frac{\mathrm{TP}+\mathrm{TN}}{\mathrm{TP}+\mathrm{FP}+\mathrm{TN}+\mathrm{FN}}.
\end{equation}
We also report the maximum relative degree error,
\begin{equation}
\mathrm{MRE}_k
=
\max_{1\le i\le N}
\left|
\frac{\left\langle k_i\right\rangle}{k_i}-1
\right|,
\end{equation}
which measures the worst node-level discrepancy in the expected degree sequence.

For each metric $m$ and realization index $r$, we define a relative signed improvement $\Delta_m(r)$, in percentage, so that positive values always favor the wMSM. For metrics to be maximized, namely $\mathrm{TPR}$, $\mathrm{PPV}$, $\mathrm{TNR}$ and $\mathrm{ACC}$, we set
\begin{equation}
\Delta_m(r)
=
100\,
\frac{
m_{\mathrm{wMSM}}(r)-m_{\mathrm{CReMB}}(r)
}{
\left|m_{\mathrm{CReMB}}(r)\right|
}.
\end{equation}
For $\mathrm{MRE}_k$, which must be minimized, we set
\begin{equation}
\Delta_{\mathrm{MRE}_k}(r)
=
100\,
\frac{
\mathrm{MRE}_{k,\mathrm{CReMB}}(r)
-
\mathrm{MRE}_{k,\mathrm{wMSM}}(r)
}{
\left|\mathrm{MRE}_{k,\mathrm{CReMB}}(r)\right|
}.
\end{equation}
Thus $\Delta_m(r)>0$ means that the wMSM performs better for realization $r$.

In the ITN application, the wMSM improves most binary diagnostics despite its informational disadvantage. At the main aggregation scale $d_c=800\,\mathrm{km}$, precision improves on average by about $4.1\%$, specificity by about $9.3\%$, accuracy by about $2.6\%$, and the maximum relative degree error by about $11.9\%$. These improvements occur in all ten yearly snapshots for precision, specificity and accuracy, and in nine out of ten yearly snapshots for $\mathrm{MRE}_k$. The CReMB performs better only in sensitivity, with an average advantage of about $1.1\%$. This indicates that the two models recover a comparable fraction of observed links, but the wMSM is more selective: it assigns less probability mass to absent dyads, thereby improving precision and specificity while preserving essentially the same ability to recover the observed support.

In the DPN application, the comparison is repeated over ten one-digit good groups and for both reconstructed sectoral layers. At the three-digit SBI layer, the wMSM improves precision by about $8.9\%$ on average, specificity by about $1.8\%$, accuracy by about $1.4\%$, and $\mathrm{MRE}_k$ by about $5.0\%$. At the four-digit SBI layer, shown in the lower panel of Fig.~\ref{fig:6}, the wMSM improves precision by about $9.9\%$, specificity by about $1.9\%$, accuracy by about $1.3\%$, and $\mathrm{MRE}_k$ by about $6.0\%$. Sensitivity is essentially comparable: the wMSM is slightly better on average at the three-digit layer and slightly worse at the four-digit layer. Across the twenty Dutch target-layer reconstructions, the wMSM improves precision in eighteen cases, specificity in sixteen cases, accuracy in eighteen cases, and $\mathrm{MRE}_k$ in fifteen cases. The gain in precision and specificity is accompanied by fewer expected links in nine of the ten ITN snapshots and fifteen of the twenty Dutch reconstructions. False positives decrease on average by $19.7\%$ in the ITN and $14.7\%$--$17.6\%$ in the DPN, while true positives change by only $-1.1\%$ and $-0.2\%$ to $+1.9\%$, respectively. Scale invariance may help compensate for the coarser calibration information by imposing additional structure on the reconstruction.

Figure~\ref{fig:6} displays the same comparison realization by realization. The top row summarizes the ITN results across years, while the bottom row summarizes the Dutch results across good groups at the four-digit SBI layer. Values above the dashed horizontal line favor the wMSM, whereas values below it favor the CReMB. The violins make clear that the advantage of the wMSM is not limited to a single empirical system. In both the geographical aggregation of the ITN and the sectoral aggregation of DPN, the scale-invariant transfer of a coarse-grained calibration is competitive with a fine-scale reconstruction that uses the target density directly, and it systematically improves the selectivity of the reconstructed binary support.

\section*{DISCUSSION}

{
The present work uses the wMSM introduced in the theoretical companion  paper~\cite{Marzi2026wMSM} as an empirical reconstruction framework. The relevant point for reconstruction is not only that the model assigns probabilities and expected weights from node strengths, but that these assignments are consistent across aggregation levels. In particular, aggregation preserves the functional form of both the binary projection and the weighted distribution {between distinct blocks}, while strengths add deterministically from microscopic nodes to aggregate blocks. 

This property changes the usual reconstruction setting. Instead of fitting a separate ensemble at each resolution, the same parameters can be estimated at the layer where information is available and then used at another layer where the network is unobserved. In our applications, the density parameter $\delta$ is calibrated from coarse-grained topology, while the weight-scale parameter $\rho$ is fixed from the total strength, which is preserved under aggregation. The coarse-grained topology therefore constrains the fine-scale ensemble through the transferred density parameter.
}

The empirical applications show that this multiscale constraint is informative. In the ITN, the wMSM transfers a density parameter calibrated on distance-based macro-regions to the country layer. In the DPN, the same procedure transfers information from SBI2 to SBI3 and SBI4 sectoral resolutions. These two settings involve different aggregation mechanisms, geographical in one case and sectoral in the other, yet the same reconstruction principle remains effective. This suggests that the central requirement is not the specific nature of the blocks, but the additivity of strengths and the closure of the probability law under aggregation {between distinct blocks}.

The comparison with the CReMB clarifies the origin of the improvement. Once strengths are fixed, the wMSM and the CReMB have the same unconditional gravity expectation for dyadic weights. Therefore, their difference lies in the binary probability matrix. The CReMB is calibrated directly at the target scale, whereas the wMSM uses a density parameter learned at a coarser scale. Despite this informational disadvantage, the wMSM improves precision, specificity, accuracy and maximum degree-error diagnostics in both empirical systems, while remaining nearly equivalent in sensitivity. The advantage is therefore not a generic improvement of weighted expectations, but a more selective reconstruction of the binary support.

There are also limitations. The present specification is deliberately parsimonious and uses only strengths and one global density parameter. Dyadic covariates such as distance, sectoral similarity or technological compatibility are not included, although they could in principle be incorporated through additional scale-consistent dyadic factors. Moreover, the model assumes conditional dyadic independence, so higher-order structure is not imposed directly but can only emerge from heterogeneity in strengths and probabilities.  Finally, the weighted model is formulated for integer weights, while a full likelihood treatment of continuous economic flows would require a dedicated extension. {The empirical importers retain the supplied numerical weights without integer rounding; the discrete mark law is therefore a modeling approximation for these monetary flows, while the reported moment predictions use their original numerical values}.

Overall, the wMSM shows that coarse-grained observations can contain usable information about fine-scale topology when the reconstruction model is built to commute with aggregation {for inter-block distributions}. This provides a route toward reconstruction methods that remain meaningful across resolutions and can be applied when microscopic interactions are hidden but aggregate network information is available.

\section*{METHODS}

\subsection*{Implementation}

All analyses were carried out in Python using the accompanying \texttt{wMSM} package. The routines support both directed and undirected networks, but all empirical results reported here use the undirected specification.

The wMSM density parameter was estimated by solving the one-dimensional link-count equation at the calibration layer. The solver uses a safeguarded Newton scheme: starting from the interval $[0,1]$, the upper bound is doubled until the expected number of links exceeds the observed one, after which Newton updates are accepted only if they remain inside the current bracket; otherwise, bisection is used. The relative tolerance on the link-count equation was set to $10^{-8}$, with a maximum of $200$ iterations. 

For the CReMB, the binary parameter was fitted directly at the target layer. Its calibration was performed in log-space, $\theta=\log z$, by damped Newton iterations with Armijo backtracking. The initial value was given by the sparse approximation $z_0=L/\sum_{(i,j)\in\mathcal D}s_is_j$. The stopping tolerance was again $10^{-8}$ times the target link count, with a maximum of $200$ iterations. The same dyadic support and diagonal convention used for the wMSM were used for the CReMB.

\subsection*{ITN preprocessing}

The ITN analysis is based on yearly bilateral trade flows between countries from $1991$ to $2000$. For each year, we constructed an undirected weighted network by retaining positive bilateral trade volumes and collapsing each country pair into a single unordered dyad. 

The geographic calibration layer was obtained from bilateral country distances. Countries were clustered by single-linkage hierarchical clustering, and macro-regions were defined by cutting the resulting dendrogram at a prescribed distance threshold. The main analysis uses $d_c=800\,\mathrm{km}$, while robustness checks were performed at $d_c=200,400,600,1000\,\mathrm{km}$. Macro-regional self-loops were retained because they represent total trade internal to a macro-region.

\subsection*{DPN preprocessing}

The DPN was constructed from sector-to-sector production flows resolved by product group. We aggregated product groups to their first digit, obtaining ten product layers $g=0,\dots,9$.

For each product layer, seller and buyer sector codes were mapped to their first two, three or four digits, defining the SBI2, SBI3 and SBI4 resolutions. The network was treated as undirected: reciprocal seller-buyer flows were collapsed to unordered sector pairs and the corresponding weights were summed. Diagonal entries were retained at every SBI resolution, because they represent within-sector production relations.

\subsection*{Evaluation protocol}

For each fitted model, expected link counts, degrees and strengths were computed analytically from the fitted probability and expected-weight matrices.
Expected confusion-matrix entries were computed probabilistically from the fitted link probabilities and the observed adjacency matrix. The reported binary scores, sensitivity, precision, specificity and accuracy, are therefore ensemble expectations. The maximum relative degree error was computed from the expected degree sequence.
Local nearest-neighbor profiles were computed by plug-in expectations. For weighted nearest-neighbor strength, the expected strength of a neighbor was conditioned on the existence of the focal link, so that the contribution of the focal dyad was not treated as independent of the conditioning event. Clustering profiles were instead estimated by Monte Carlo sampling, because both binary and weighted clustering coefficients are nonlinear ratios. For each fitted ensemble we sampled $300$ networks in batches of $64$ realizations. The random seed was fixed to $12345$ for single-snapshot diagnostics and to $54321$ for the multi-year ITN workflow.

\section*{DATA AVAILABILITY}

The empirical analysis combines publicly available international trade data with restricted DPN data. Bilateral trade flows are obtained from the Expanded Trade and GDP Data by Gleditsch~\cite{Gleditsch2002ExpandedTradeGDP}. Country-to-country distances used to construct the macro-regional aggregation are obtained from the CEPII GeoDist database~\cite{MayerZignago2011GeoDist}. The country-level networks, distance-based macro-regional partitions and aggregated trade matrices used in the ITN application are generated from these sources through the preprocessing procedure described in Appendix~\hyperlink{AppA}{A}.

The DPN data are restricted microdata made available through a scientific collaboration with Centraal Bureau voor de Statistiek (CBS). These data cannot be publicly redistributed because of confidentiality restrictions. The sectoral production networks used in the analysis, their aggregation levels and the corresponding summary statistics are described in Appendix~\hyperlink{AppB}{B}. Access to the underlying Dutch microdata is subject to CBS data-access procedures and confidentiality requirements.

\section*{CODE AVAILABILITY}

The Python package named \texttt{wMSM} (\textit{weighted MultiScale Model}), implementing the algorithms described in the main text, is available on PyPI and at the URL \url{https://github.com/mattiamarzi/wMSM}.

\bibliography{references}

\begin{thebibliography}{41}%
\makeatletter
\providecommand \@ifxundefined [1]{%
 \@ifx{#1\undefined}
}%
\providecommand \@ifnum [1]{%
 \ifnum #1\expandafter \@firstoftwo
 \else \expandafter \@secondoftwo
 \fi
}%
\providecommand \@ifx [1]{%
 \ifx #1\expandafter \@firstoftwo
 \else \expandafter \@secondoftwo
 \fi
}%
\providecommand \natexlab [1]{#1}%
\providecommand \enquote  [1]{``#1''}%
\providecommand \bibnamefont  [1]{#1}%
\providecommand \bibfnamefont [1]{#1}%
\providecommand \citenamefont [1]{#1}%
\providecommand \href@noop [0]{\@secondoftwo}%
\providecommand \href [0]{\begingroup \@sanitize@url \@href}%
\providecommand \@href[1]{\@@startlink{#1}\@@href}%
\providecommand \@@href[1]{\endgroup#1\@@endlink}%
\providecommand \@sanitize@url [0]{\catcode `\\12\catcode `\$12\catcode `\&12\catcode `\#12\catcode `\^12\catcode `\_12\catcode `\%12\relax}%
\providecommand \@@startlink[1]{}%
\providecommand \@@endlink[0]{}%
\providecommand \url  [0]{\begingroup\@sanitize@url \@url }%
\providecommand \@url [1]{\endgroup\@href {#1}{\urlprefix }}%
\providecommand \urlprefix  [0]{URL }%
\providecommand \Eprint [0]{\href }%
\providecommand \doibase [0]{https://doi.org/}%
\providecommand \selectlanguage [0]{\@gobble}%
\providecommand \bibinfo  [0]{\@secondoftwo}%
\providecommand \bibfield  [0]{\@secondoftwo}%
\providecommand \translation [1]{[#1]}%
\providecommand \BibitemOpen [0]{}%
\providecommand \bibitemStop [0]{}%
\providecommand \bibitemNoStop [0]{.\EOS\space}%
\providecommand \EOS [0]{\spacefactor3000\relax}%
\providecommand \BibitemShut  [1]{\csname bibitem#1\endcsname}%
\let\auto@bib@innerbib\@empty
\bibitem [{\citenamefont {Squartini}\ \emph {et~al.}(2018)\citenamefont {Squartini}, \citenamefont {Caldarelli}, \citenamefont {Cimini}, \citenamefont {Gabrielli},\ and\ \citenamefont {Garlaschelli}}]{ReconstructionMethods2018}%
  \BibitemOpen
  \bibfield  {author} {\bibinfo {author} {\bibfnamefont {T.}~\bibnamefont {Squartini}}, \bibinfo {author} {\bibfnamefont {G.}~\bibnamefont {Caldarelli}}, \bibinfo {author} {\bibfnamefont {G.}~\bibnamefont {Cimini}}, \bibinfo {author} {\bibfnamefont {A.}~\bibnamefont {Gabrielli}},\ and\ \bibinfo {author} {\bibfnamefont {D.}~\bibnamefont {Garlaschelli}},\ }\bibfield  {title} {\bibinfo {title} {Reconstruction methods for networks: The case of economic and financial systems},\ }\href {https://doi.org/10.1016/j.physrep.2018.06.008} {\bibfield  {journal} {\bibinfo  {journal} {Phys. Rep.}\ }\textbf {\bibinfo {volume} {757}},\ \bibinfo {pages} {1} (\bibinfo {year} {2018})}\BibitemShut {NoStop}%
\bibitem [{\citenamefont {Cimini}\ \emph {et~al.}(2021)\citenamefont {Cimini}, \citenamefont {Mastrandrea},\ and\ \citenamefont {Squartini}}]{cimini2021reconstructing}%
  \BibitemOpen
  \bibfield  {author} {\bibinfo {author} {\bibfnamefont {G.}~\bibnamefont {Cimini}}, \bibinfo {author} {\bibfnamefont {R.}~\bibnamefont {Mastrandrea}},\ and\ \bibinfo {author} {\bibfnamefont {T.}~\bibnamefont {Squartini}},\ }\href {https://doi.org/10.1017/9781108771030} {\emph {\bibinfo {title} {Reconstructing Networks}}},\ Elements in the Structure and Dynamics of Complex Networks\ (\bibinfo  {publisher} {Cambridge Univ. Press},\ \bibinfo {year} {2021})\BibitemShut {NoStop}%
\bibitem [{\citenamefont {Bardoscia}\ \emph {et~al.}(2021)\citenamefont {Bardoscia}, \citenamefont {Barucca}, \citenamefont {Battiston}, \citenamefont {Caccioli}, \citenamefont {Cimini}, \citenamefont {Garlaschelli}, \citenamefont {Saracco}, \citenamefont {Squartini},\ and\ \citenamefont {Caldarelli}}]{bardoscia_physics_2021}%
  \BibitemOpen
  \bibfield  {author} {\bibinfo {author} {\bibfnamefont {M.}~\bibnamefont {Bardoscia}}, \bibinfo {author} {\bibfnamefont {P.}~\bibnamefont {Barucca}}, \bibinfo {author} {\bibfnamefont {S.}~\bibnamefont {Battiston}}, \bibinfo {author} {\bibfnamefont {F.}~\bibnamefont {Caccioli}}, \bibinfo {author} {\bibfnamefont {G.}~\bibnamefont {Cimini}}, \bibinfo {author} {\bibfnamefont {D.}~\bibnamefont {Garlaschelli}}, \bibinfo {author} {\bibfnamefont {F.}~\bibnamefont {Saracco}}, \bibinfo {author} {\bibfnamefont {T.}~\bibnamefont {Squartini}},\ and\ \bibinfo {author} {\bibfnamefont {G.}~\bibnamefont {Caldarelli}},\ }\bibfield  {title} {\bibinfo {title} {The physics of financial networks},\ }\href {https://doi.org/10.1038/s42254-021-00322-5} {\bibfield  {journal} {\bibinfo  {journal} {Nature Reviews Physics}\ }\textbf {\bibinfo {volume} {3}},\ \bibinfo {pages} {490} (\bibinfo {year} {2021})}\BibitemShut {NoStop}%
\bibitem [{\citenamefont {Mungo}\ \emph {et~al.}(2024)\citenamefont {Mungo}, \citenamefont {Brintrup}, \citenamefont {Garlaschelli},\ and\ \citenamefont {Lafond}}]{Mungo2024SupplyNetworksReview}%
  \BibitemOpen
  \bibfield  {author} {\bibinfo {author} {\bibfnamefont {L.}~\bibnamefont {Mungo}}, \bibinfo {author} {\bibfnamefont {A.}~\bibnamefont {Brintrup}}, \bibinfo {author} {\bibfnamefont {D.}~\bibnamefont {Garlaschelli}},\ and\ \bibinfo {author} {\bibfnamefont {F.}~\bibnamefont {Lafond}},\ }\bibfield  {title} {\bibinfo {title} {Reconstructing supply networks},\ }\href {https://doi.org/10.1088/2632-072X/ad30bf} {\bibfield  {journal} {\bibinfo  {journal} {J. Phys. Complex.}\ }\textbf {\bibinfo {volume} {5}},\ \bibinfo {pages} {012001} (\bibinfo {year} {2024})}\BibitemShut {NoStop}%
\bibitem [{\citenamefont {Jaynes}(1957)}]{jaynes1957information}%
  \BibitemOpen
  \bibfield  {author} {\bibinfo {author} {\bibfnamefont {E.~T.}\ \bibnamefont {Jaynes}},\ }\bibfield  {title} {\bibinfo {title} {Information theory and statistical mechanics},\ }\href {https://doi.org/10.1103/PhysRev.106.620} {\bibfield  {journal} {\bibinfo  {journal} {Phys. Rev.}\ }\textbf {\bibinfo {volume} {106}},\ \bibinfo {pages} {620} (\bibinfo {year} {1957})}\BibitemShut {NoStop}%
\bibitem [{\citenamefont {Park}\ and\ \citenamefont {Newman}(2004)}]{park2004statistical}%
  \BibitemOpen
  \bibfield  {author} {\bibinfo {author} {\bibfnamefont {J.}~\bibnamefont {Park}}\ and\ \bibinfo {author} {\bibfnamefont {M.~E.~J.}\ \bibnamefont {Newman}},\ }\bibfield  {title} {\bibinfo {title} {Statistical mechanics of networks},\ }\href {https://doi.org/10.1103/PhysRevE.70.066117} {\bibfield  {journal} {\bibinfo  {journal} {Phys. Rev. E}\ }\textbf {\bibinfo {volume} {70}},\ \bibinfo {pages} {066117} (\bibinfo {year} {2004})}\BibitemShut {NoStop}%
\bibitem [{\citenamefont {Garlaschelli}\ and\ \citenamefont {Loffredo}(2008)}]{garlaschelli2008maximum}%
  \BibitemOpen
  \bibfield  {author} {\bibinfo {author} {\bibfnamefont {D.}~\bibnamefont {Garlaschelli}}\ and\ \bibinfo {author} {\bibfnamefont {M.~I.}\ \bibnamefont {Loffredo}},\ }\bibfield  {title} {\bibinfo {title} {Maximum likelihood: Extracting unbiased information from complex networks},\ }\href {https://doi.org/10.1103/PhysRevE.78.015101} {\bibfield  {journal} {\bibinfo  {journal} {Phys. Rev. E}\ }\textbf {\bibinfo {volume} {78}},\ \bibinfo {pages} {015101} (\bibinfo {year} {2008})}\BibitemShut {NoStop}%
\bibitem [{\citenamefont {Squartini}\ and\ \citenamefont {Garlaschelli}(2011)}]{squartini2011analytical}%
  \BibitemOpen
  \bibfield  {author} {\bibinfo {author} {\bibfnamefont {T.}~\bibnamefont {Squartini}}\ and\ \bibinfo {author} {\bibfnamefont {D.}~\bibnamefont {Garlaschelli}},\ }\bibfield  {title} {\bibinfo {title} {Analytical maximum-likelihood method to detect patterns in real networks},\ }\href {https://doi.org/10.1088/1367-2630/13/8/083001} {\bibfield  {journal} {\bibinfo  {journal} {New J. Phys.}\ }\textbf {\bibinfo {volume} {13}},\ \bibinfo {pages} {083001} (\bibinfo {year} {2011})}\BibitemShut {NoStop}%
\bibitem [{\citenamefont {Squartini}\ and\ \citenamefont {Garlaschelli}(2017)}]{squartini2017maximum}%
  \BibitemOpen
  \bibfield  {author} {\bibinfo {author} {\bibfnamefont {T.}~\bibnamefont {Squartini}}\ and\ \bibinfo {author} {\bibfnamefont {D.}~\bibnamefont {Garlaschelli}},\ }\href {https://doi.org/10.1007/978-3-319-69438-2} {\emph {\bibinfo {title} {Maximum-Entropy Networks}}},\ SpringerBriefs in Complexity\ (\bibinfo  {publisher} {Springer International Publishing},\ \bibinfo {address} {Cham},\ \bibinfo {year} {2017})\BibitemShut {NoStop}%
\bibitem [{\citenamefont {Garlaschelli}\ and\ \citenamefont {Loffredo}(2004)}]{GarlaschelliLoffredo2004WTW}%
  \BibitemOpen
  \bibfield  {author} {\bibinfo {author} {\bibfnamefont {D.}~\bibnamefont {Garlaschelli}}\ and\ \bibinfo {author} {\bibfnamefont {M.~I.}\ \bibnamefont {Loffredo}},\ }\bibfield  {title} {\bibinfo {title} {Fitness-dependent topological properties of the world trade web},\ }\href {https://doi.org/10.1103/PhysRevLett.93.188701} {\bibfield  {journal} {\bibinfo  {journal} {Phys. Rev. Lett.}\ }\textbf {\bibinfo {volume} {93}},\ \bibinfo {pages} {188701} (\bibinfo {year} {2004})}\BibitemShut {NoStop}%
\bibitem [{\citenamefont {Cimini}\ \emph {et~al.}(2015{\natexlab{a}})\citenamefont {Cimini}, \citenamefont {Squartini}, \citenamefont {Garlaschelli},\ and\ \citenamefont {Gabrielli}}]{CimiModel2015}%
  \BibitemOpen
  \bibfield  {author} {\bibinfo {author} {\bibfnamefont {G.}~\bibnamefont {Cimini}}, \bibinfo {author} {\bibfnamefont {T.}~\bibnamefont {Squartini}}, \bibinfo {author} {\bibfnamefont {D.}~\bibnamefont {Garlaschelli}},\ and\ \bibinfo {author} {\bibfnamefont {A.}~\bibnamefont {Gabrielli}},\ }\bibfield  {title} {\bibinfo {title} {Systemic risk analysis on reconstructed economic and financial networks},\ }\href {https://doi.org/10.1038/srep15758} {\bibfield  {journal} {\bibinfo  {journal} {Sci. Rep.}\ }\textbf {\bibinfo {volume} {5}},\ \bibinfo {pages} {15758} (\bibinfo {year} {2015}{\natexlab{a}})}\BibitemShut {NoStop}%
\bibitem [{\citenamefont {Mazzarisi}\ and\ \citenamefont {Lillo}(2017)}]{Mazzarisi2017LimitedInformation}%
  \BibitemOpen
  \bibfield  {author} {\bibinfo {author} {\bibfnamefont {P.}~\bibnamefont {Mazzarisi}}\ and\ \bibinfo {author} {\bibfnamefont {F.}~\bibnamefont {Lillo}},\ }\bibfield  {title} {\bibinfo {title} {Methods for reconstructing interbank networks from limited information: A comparison},\ }in\ \href {https://doi.org/10.1007/978-3-319-47705-3_15} {\emph {\bibinfo {booktitle} {Econophysics and Sociophysics: Recent Progress and Future Directions}}},\ \bibinfo {series and number} {New Economic Windows},\ \bibinfo {editor} {edited by\ \bibinfo {editor} {\bibfnamefont {F.}~\bibnamefont {Abergel}}, \bibinfo {editor} {\bibfnamefont {B.~K.}\ \bibnamefont {Chakrabarti}}, \bibinfo {editor} {\bibfnamefont {A.}~\bibnamefont {Ghosh}}, \bibinfo {editor} {\bibfnamefont {M.}~\bibnamefont {Mitra}}, \bibinfo {editor} {\bibfnamefont {M.}~\bibnamefont {Patriarca}},\ and\ \bibinfo {editor} {\bibfnamefont {A.}~\bibnamefont {Vespignani}}}\ (\bibinfo  {publisher} {Springer, Cham},\ \bibinfo {year} {2017})\ pp.\ \bibinfo {pages}
  {201--215}\BibitemShut {NoStop}%
\bibitem [{\citenamefont {Anand}\ \emph {et~al.}(2018)\citenamefont {Anand}, \citenamefont {van Lelyveld}, \citenamefont {Banai}, \citenamefont {Friedrich}, \citenamefont {Garratt}, \citenamefont {Ha{\l}aj}, \citenamefont {Fique}, \citenamefont {Hansen}, \citenamefont {Jaramillo}, \citenamefont {Lee}, \citenamefont {Molina-Borboa}, \citenamefont {Nobili}, \citenamefont {Rajan}, \citenamefont {Salakhova}, \citenamefont {Silva}, \citenamefont {Silvestri},\ and\ \citenamefont {de~Souza}}]{Anand2018MissingLinks}%
  \BibitemOpen
  \bibfield  {author} {\bibinfo {author} {\bibfnamefont {K.}~\bibnamefont {Anand}}, \bibinfo {author} {\bibfnamefont {I.}~\bibnamefont {van Lelyveld}}, \bibinfo {author} {\bibfnamefont {{\'{A}}.}~\bibnamefont {Banai}}, \bibinfo {author} {\bibfnamefont {S.}~\bibnamefont {Friedrich}}, \bibinfo {author} {\bibfnamefont {R.}~\bibnamefont {Garratt}}, \bibinfo {author} {\bibfnamefont {G.}~\bibnamefont {Ha{\l}aj}}, \bibinfo {author} {\bibfnamefont {J.}~\bibnamefont {Fique}}, \bibinfo {author} {\bibfnamefont {I.}~\bibnamefont {Hansen}}, \bibinfo {author} {\bibfnamefont {S.~M.}\ \bibnamefont {Jaramillo}}, \bibinfo {author} {\bibfnamefont {H.}~\bibnamefont {Lee}}, \bibinfo {author} {\bibfnamefont {J.~L.}\ \bibnamefont {Molina-Borboa}}, \bibinfo {author} {\bibfnamefont {S.}~\bibnamefont {Nobili}}, \bibinfo {author} {\bibfnamefont {S.}~\bibnamefont {Rajan}}, \bibinfo {author} {\bibfnamefont {D.}~\bibnamefont {Salakhova}}, \bibinfo {author} {\bibfnamefont {T.~C.}\ \bibnamefont {Silva}}, \bibinfo {author} {\bibfnamefont
  {L.}~\bibnamefont {Silvestri}},\ and\ \bibinfo {author} {\bibfnamefont {S.~R.~S.}\ \bibnamefont {de~Souza}},\ }\bibfield  {title} {\bibinfo {title} {The missing links: A global study on uncovering financial network structures from partial data},\ }\href {https://doi.org/10.1016/j.jfs.2017.05.012} {\bibfield  {journal} {\bibinfo  {journal} {J. Financ. Stability}\ }\textbf {\bibinfo {volume} {35}},\ \bibinfo {pages} {107} (\bibinfo {year} {2018})}\BibitemShut {NoStop}%
\bibitem [{\citenamefont {Lebacher}\ \emph {et~al.}(2019)\citenamefont {Lebacher}, \citenamefont {Cook}, \citenamefont {Klein},\ and\ \citenamefont {Kauermann}}]{Lebacher2019LostEdges}%
  \BibitemOpen
  \bibfield  {author} {\bibinfo {author} {\bibfnamefont {M.}~\bibnamefont {Lebacher}}, \bibinfo {author} {\bibfnamefont {S.}~\bibnamefont {Cook}}, \bibinfo {author} {\bibfnamefont {N.}~\bibnamefont {Klein}},\ and\ \bibinfo {author} {\bibfnamefont {G.}~\bibnamefont {Kauermann}},\ }\bibfield  {title} {\bibinfo {title} {In search of lost edges: A case study on reconstructing financial networks},\ }\href {https://doi.org/10.21314/JNTF.2019.058} {\bibfield  {journal} {\bibinfo  {journal} {J. Network Theory in Finance}\ }\textbf {\bibinfo {volume} {5}},\ \bibinfo {pages} {29} (\bibinfo {year} {2019})}\BibitemShut {NoStop}%
\bibitem [{\citenamefont {Ramadiah}\ \emph {et~al.}(2020)\citenamefont {Ramadiah}, \citenamefont {Caccioli},\ and\ \citenamefont {Fricke}}]{Ramadiah2020ReconstructingAndStressTesting}%
  \BibitemOpen
  \bibfield  {author} {\bibinfo {author} {\bibfnamefont {A.}~\bibnamefont {Ramadiah}}, \bibinfo {author} {\bibfnamefont {F.}~\bibnamefont {Caccioli}},\ and\ \bibinfo {author} {\bibfnamefont {D.}~\bibnamefont {Fricke}},\ }\bibfield  {title} {\bibinfo {title} {Reconstructing and stress testing credit networks},\ }\href {https://doi.org/10.1016/j.jedc.2019.103817} {\bibfield  {journal} {\bibinfo  {journal} {J. Econ. Dyn. Control}\ }\textbf {\bibinfo {volume} {111}},\ \bibinfo {pages} {103817} (\bibinfo {year} {2020})}\BibitemShut {NoStop}%
\bibitem [{\citenamefont {Marzi}\ \emph {et~al.}(2026{\natexlab{a}})\citenamefont {Marzi}, \citenamefont {Giuffrida}, \citenamefont {Garlaschelli},\ and\ \citenamefont {Squartini}}]{marzi2026reproducing}%
  \BibitemOpen
  \bibfield  {author} {\bibinfo {author} {\bibfnamefont {M.}~\bibnamefont {Marzi}}, \bibinfo {author} {\bibfnamefont {F.}~\bibnamefont {Giuffrida}}, \bibinfo {author} {\bibfnamefont {D.}~\bibnamefont {Garlaschelli}},\ and\ \bibinfo {author} {\bibfnamefont {T.}~\bibnamefont {Squartini}},\ }\bibfield  {title} {\bibinfo {title} {Reproducing the first and second moments of empirical degree distributions},\ }\href {https://doi.org/10.1103/3vtj-5nlt} {\bibfield  {journal} {\bibinfo  {journal} {Phys.\ Rev.\ Res.}\ }\textbf {\bibinfo {volume} {8}},\ \bibinfo {pages} {013141} (\bibinfo {year} {2026}{\natexlab{a}})}\BibitemShut {NoStop}%
\bibitem [{\citenamefont {Marzi}\ and\ \citenamefont {Squartini}(2026)}]{marzi2026Bayesian}%
  \BibitemOpen
  \bibfield  {author} {\bibinfo {author} {\bibfnamefont {M.}~\bibnamefont {Marzi}}\ and\ \bibinfo {author} {\bibfnamefont {T.}~\bibnamefont {Squartini}},\ }\bibfield  {title} {\bibinfo {title} {A bayesian approach to out-of-sample network reconstruction},\ }\href {https://doi.org/10.1038/s42005-026-02725-1} {\bibfield  {journal} {\bibinfo  {journal} {Commun. Phys.}\ }\textbf {\bibinfo {volume} {9}},\ \bibinfo {pages} {316} (\bibinfo {year} {2026})}\BibitemShut {NoStop}%
\bibitem [{\citenamefont {Almog}\ \emph {et~al.}(2019)\citenamefont {Almog}, \citenamefont {Bird},\ and\ \citenamefont {Garlaschelli}}]{EnhancedGravityModelTrade2019}%
  \BibitemOpen
  \bibfield  {author} {\bibinfo {author} {\bibfnamefont {A.}~\bibnamefont {Almog}}, \bibinfo {author} {\bibfnamefont {R.}~\bibnamefont {Bird}},\ and\ \bibinfo {author} {\bibfnamefont {D.}~\bibnamefont {Garlaschelli}},\ }\bibfield  {title} {\bibinfo {title} {Enhanced gravity model of trade: Reconciling macroeconomic and network models},\ }\bibfield  {journal} {\bibinfo  {journal} {Front. Phys.}\ }\textbf {\bibinfo {volume} {7}},\ \href {https://doi.org/10.3389/fphy.2019.00055} {10.3389/fphy.2019.00055} (\bibinfo {year} {2019})\BibitemShut {NoStop}%
\bibitem [{\citenamefont {Di~Vece}\ \emph {et~al.}(2022)\citenamefont {Di~Vece}, \citenamefont {Garlaschelli},\ and\ \citenamefont {Squartini}}]{DiVece2022GravityMER}%
  \BibitemOpen
  \bibfield  {author} {\bibinfo {author} {\bibfnamefont {M.}~\bibnamefont {Di~Vece}}, \bibinfo {author} {\bibfnamefont {D.}~\bibnamefont {Garlaschelli}},\ and\ \bibinfo {author} {\bibfnamefont {T.}~\bibnamefont {Squartini}},\ }\bibfield  {title} {\bibinfo {title} {Gravity models of networks: Integrating maximum-entropy and econometric approaches},\ }\href {https://doi.org/10.1103/PhysRevResearch.4.033105} {\bibfield  {journal} {\bibinfo  {journal} {Phys. Rev. Res.}\ }\textbf {\bibinfo {volume} {4}},\ \bibinfo {pages} {033105} (\bibinfo {year} {2022})}\BibitemShut {NoStop}%
\bibitem [{\citenamefont {Mattsson}\ \emph {et~al.}(2021)\citenamefont {Mattsson}, \citenamefont {Takes}, \citenamefont {Heemskerk}, \citenamefont {Diks}, \citenamefont {Buiten}, \citenamefont {Faber},\ and\ \citenamefont {Sloot}}]{FunctionalStructureProductionNetworks2021}%
  \BibitemOpen
  \bibfield  {author} {\bibinfo {author} {\bibfnamefont {C.~E.~S.}\ \bibnamefont {Mattsson}}, \bibinfo {author} {\bibfnamefont {F.~W.}\ \bibnamefont {Takes}}, \bibinfo {author} {\bibfnamefont {E.~M.}\ \bibnamefont {Heemskerk}}, \bibinfo {author} {\bibfnamefont {C.}~\bibnamefont {Diks}}, \bibinfo {author} {\bibfnamefont {G.}~\bibnamefont {Buiten}}, \bibinfo {author} {\bibfnamefont {A.}~\bibnamefont {Faber}},\ and\ \bibinfo {author} {\bibfnamefont {P.~M.~A.}\ \bibnamefont {Sloot}},\ }\bibfield  {title} {\bibinfo {title} {Functional structure in production networks},\ }\bibfield  {journal} {\bibinfo  {journal} {Front. Big Data}\ }\textbf {\bibinfo {volume} {4}},\ \href {https://doi.org/10.3389/fdata.2021.666712} {10.3389/fdata.2021.666712} (\bibinfo {year} {2021})\BibitemShut {NoStop}%
\bibitem [{\citenamefont {Ialongo}\ \emph {et~al.}(2022)\citenamefont {Ialongo}, \citenamefont {de~Valk}, \citenamefont {Marchese}, \citenamefont {Jansen}, \citenamefont {Zmarrou}, \citenamefont {Squartini},\ and\ \citenamefont {Garlaschelli}}]{ReconstructingFirmLevelInteractionsDutchInputOutputNetwork2022}%
  \BibitemOpen
  \bibfield  {author} {\bibinfo {author} {\bibfnamefont {L.~N.}\ \bibnamefont {Ialongo}}, \bibinfo {author} {\bibfnamefont {C.}~\bibnamefont {de~Valk}}, \bibinfo {author} {\bibfnamefont {E.}~\bibnamefont {Marchese}}, \bibinfo {author} {\bibfnamefont {F.}~\bibnamefont {Jansen}}, \bibinfo {author} {\bibfnamefont {H.}~\bibnamefont {Zmarrou}}, \bibinfo {author} {\bibfnamefont {T.}~\bibnamefont {Squartini}},\ and\ \bibinfo {author} {\bibfnamefont {D.}~\bibnamefont {Garlaschelli}},\ }\bibfield  {title} {\bibinfo {title} {Reconstructing firm-level interactions in the dutch input--output network from production constraints},\ }\href {https://doi.org/10.1038/s41598-022-15714-4} {\bibfield  {journal} {\bibinfo  {journal} {Scientific Reports}\ }\textbf {\bibinfo {volume} {12}},\ \bibinfo {pages} {11847} (\bibinfo {year} {2022})}\BibitemShut {NoStop}%
\bibitem [{\citenamefont {Garuccio}\ \emph {et~al.}(2023)\citenamefont {Garuccio}, \citenamefont {Lalli},\ and\ \citenamefont {Garlaschelli}}]{Garuccio2023}%
  \BibitemOpen
  \bibfield  {author} {\bibinfo {author} {\bibfnamefont {E.}~\bibnamefont {Garuccio}}, \bibinfo {author} {\bibfnamefont {M.}~\bibnamefont {Lalli}},\ and\ \bibinfo {author} {\bibfnamefont {D.}~\bibnamefont {Garlaschelli}},\ }\bibfield  {title} {\bibinfo {title} {Multiscale network renormalization: Scale-invariance without geometry},\ }\href {https://doi.org/10.1103/PhysRevResearch.5.043101} {\bibfield  {journal} {\bibinfo  {journal} {Phys. Rev. Res.}\ }\textbf {\bibinfo {volume} {5}},\ \bibinfo {pages} {043101} (\bibinfo {year} {2023})}\BibitemShut {NoStop}%
\bibitem [{\citenamefont {Lalli}(2024)}]{LalliThesis}%
  \BibitemOpen
  \bibfield  {author} {\bibinfo {author} {\bibfnamefont {M.}~\bibnamefont {Lalli}},\ }\emph {\bibinfo {title} {Scale-Invariant Random Graphs: a multiscale approach to network modeling}},\ \href {https://e-theses.imtlucca.it/414/1/Lalli_final%20version.pdf} {Ph.D. thesis},\ \bibinfo  {school} {IMT School for Advanced Studies Lucca} (\bibinfo {year} {2024})\BibitemShut {NoStop}%
\bibitem [{\citenamefont {Ialongo}\ \emph {et~al.}(2024)\citenamefont {Ialongo}, \citenamefont {Bangma}, \citenamefont {Jansen},\ and\ \citenamefont {Garlaschelli}}]{IalongoBangmaJansenGarlaschelli2024}%
  \BibitemOpen
  \bibfield  {author} {\bibinfo {author} {\bibfnamefont {L.~N.}\ \bibnamefont {Ialongo}}, \bibinfo {author} {\bibfnamefont {S.}~\bibnamefont {Bangma}}, \bibinfo {author} {\bibfnamefont {F.}~\bibnamefont {Jansen}},\ and\ \bibinfo {author} {\bibfnamefont {D.}~\bibnamefont {Garlaschelli}},\ }\bibfield  {title} {\bibinfo {title} {Multi-scale reconstruction of large supply networks},\ }\bibfield  {journal} {\bibinfo  {journal} {arXiv}\ }\href {https://doi.org/10.48550/arXiv.2412.16122} {10.48550/arXiv.2412.16122} (\bibinfo {year} {2024})\BibitemShut {NoStop}%
\bibitem [{\citenamefont {Gabrielli}\ \emph {et~al.}(2025)\citenamefont {Gabrielli}, \citenamefont {Garlaschelli}, \citenamefont {Patil},\ and\ \citenamefont {Serrano}}]{Gabrielli2025}%
  \BibitemOpen
  \bibfield  {author} {\bibinfo {author} {\bibfnamefont {A.}~\bibnamefont {Gabrielli}}, \bibinfo {author} {\bibfnamefont {D.}~\bibnamefont {Garlaschelli}}, \bibinfo {author} {\bibfnamefont {S.~P.}\ \bibnamefont {Patil}},\ and\ \bibinfo {author} {\bibfnamefont {M.~{\'A}.}\ \bibnamefont {Serrano}},\ }\bibfield  {title} {\bibinfo {title} {Network renormalization},\ }\href {https://doi.org/10.1038/s42254-025-00817-5} {\bibfield  {journal} {\bibinfo  {journal} {Nat. Rev. Phys.}\ }\textbf {\bibinfo {volume} {7}},\ \bibinfo {pages} {203} (\bibinfo {year} {2025})}\BibitemShut {NoStop}%
\bibitem [{\citenamefont {Marzi}\ \emph {et~al.}(2026{\natexlab{b}})\citenamefont {Marzi}, \citenamefont {Pijpers},\ and\ \citenamefont {Garlaschelli}}]{Marzi2026wMSM}%
  \BibitemOpen
  \bibfield  {author} {\bibinfo {author} {\bibfnamefont {M.}~\bibnamefont {Marzi}}, \bibinfo {author} {\bibfnamefont {F.}~\bibnamefont {Pijpers}},\ and\ \bibinfo {author} {\bibfnamefont {D.}~\bibnamefont {Garlaschelli}},\ }\bibfield  {title} {\bibinfo {title} {Multiscale renormalization of weighted networks},\ }\bibfield  {journal} {\bibinfo  {journal} {arXiv}\ }\href {https://doi.org/10.48550/arXiv.2609.38283} {10.48550/arXiv.2609.38283} (\bibinfo {year} {2026}{\natexlab{b}})\BibitemShut {NoStop}%
\bibitem [{\citenamefont {Feller}(1971)}]{Feller1971}%
  \BibitemOpen
  \bibfield  {author} {\bibinfo {author} {\bibfnamefont {W.}~\bibnamefont {Feller}},\ }\href@noop {} {\emph {\bibinfo {title} {An Introduction to Probability Theory and Its Applications, Volume II}}},\ \bibinfo {edition} {2nd}\ ed.\ (\bibinfo  {publisher} {John Wiley \& Sons},\ \bibinfo {address} {New York},\ \bibinfo {year} {1971})\BibitemShut {NoStop}%
\bibitem [{\citenamefont {Johnson}\ \emph {et~al.}(2005)\citenamefont {Johnson}, \citenamefont {Kemp},\ and\ \citenamefont {Kotz}}]{JohnsonKempKotz2005}%
  \BibitemOpen
  \bibfield  {author} {\bibinfo {author} {\bibfnamefont {N.~L.}\ \bibnamefont {Johnson}}, \bibinfo {author} {\bibfnamefont {A.~W.}\ \bibnamefont {Kemp}},\ and\ \bibinfo {author} {\bibfnamefont {S.}~\bibnamefont {Kotz}},\ }\href {https://doi.org/10.1002/0471715816} {\emph {\bibinfo {title} {Univariate Discrete Distributions}}},\ \bibinfo {edition} {3rd}\ ed.\ (\bibinfo  {publisher} {John Wiley \& Sons},\ \bibinfo {address} {Hoboken, NJ},\ \bibinfo {year} {2005})\BibitemShut {NoStop}%
\bibitem [{\citenamefont {Parisi}\ \emph {et~al.}(2020)\citenamefont {Parisi}, \citenamefont {Squartini},\ and\ \citenamefont {Garlaschelli}}]{CReM2020}%
  \BibitemOpen
  \bibfield  {author} {\bibinfo {author} {\bibfnamefont {F.}~\bibnamefont {Parisi}}, \bibinfo {author} {\bibfnamefont {T.}~\bibnamefont {Squartini}},\ and\ \bibinfo {author} {\bibfnamefont {D.}~\bibnamefont {Garlaschelli}},\ }\bibfield  {title} {\bibinfo {title} {A faster horse on a safer trail: Generalised inference for the efficient reconstruction of weighted networks},\ }\href {https://doi.org/10.1088/1367-2630/ab74a7} {\bibfield  {journal} {\bibinfo  {journal} {New Journal of Physics}\ }\textbf {\bibinfo {volume} {22}},\ \bibinfo {pages} {053053} (\bibinfo {year} {2020})}\BibitemShut {NoStop}%
\bibitem [{\citenamefont {Barrat}\ \emph {et~al.}(2004)\citenamefont {Barrat}, \citenamefont {Barthelemy}, \citenamefont {Pastor-Satorras},\ and\ \citenamefont {Vespignani}}]{Barrat2004WeightedNetworks}%
  \BibitemOpen
  \bibfield  {author} {\bibinfo {author} {\bibfnamefont {A.}~\bibnamefont {Barrat}}, \bibinfo {author} {\bibfnamefont {M.}~\bibnamefont {Barthelemy}}, \bibinfo {author} {\bibfnamefont {R.}~\bibnamefont {Pastor-Satorras}},\ and\ \bibinfo {author} {\bibfnamefont {A.}~\bibnamefont {Vespignani}},\ }\bibfield  {title} {\bibinfo {title} {The architecture of complex weighted networks},\ }\href {https://doi.org/10.1073/pnas.0400087101} {\bibfield  {journal} {\bibinfo  {journal} {Proc. Natl. Acad. Sci. U.S.A.}\ }\textbf {\bibinfo {volume} {101}},\ \bibinfo {pages} {3747} (\bibinfo {year} {2004})}\BibitemShut {NoStop}%
\bibitem [{\citenamefont {Onnela}\ \emph {et~al.}(2005)\citenamefont {Onnela}, \citenamefont {Saram{\"a}ki}, \citenamefont {Kert{\'e}sz},\ and\ \citenamefont {Kaski}}]{Onnela2005WeightedMotifs}%
  \BibitemOpen
  \bibfield  {author} {\bibinfo {author} {\bibfnamefont {J.-P.}\ \bibnamefont {Onnela}}, \bibinfo {author} {\bibfnamefont {J.}~\bibnamefont {Saram{\"a}ki}}, \bibinfo {author} {\bibfnamefont {J.}~\bibnamefont {Kert{\'e}sz}},\ and\ \bibinfo {author} {\bibfnamefont {K.}~\bibnamefont {Kaski}},\ }\bibfield  {title} {\bibinfo {title} {Intensity and coherence of motifs in weighted complex networks},\ }\href {https://doi.org/10.1103/PhysRevE.71.065103} {\bibfield  {journal} {\bibinfo  {journal} {Phys. Rev. E}\ }\textbf {\bibinfo {volume} {71}},\ \bibinfo {pages} {065103} (\bibinfo {year} {2005})}\BibitemShut {NoStop}%
\bibitem [{\citenamefont {Mastrandrea}\ \emph {et~al.}(2014)\citenamefont {Mastrandrea}, \citenamefont {Squartini}, \citenamefont {Fagiolo},\ and\ \citenamefont {Garlaschelli}}]{Mastrandrea2014EnhancedReconstruction}%
  \BibitemOpen
  \bibfield  {author} {\bibinfo {author} {\bibfnamefont {R.}~\bibnamefont {Mastrandrea}}, \bibinfo {author} {\bibfnamefont {T.}~\bibnamefont {Squartini}}, \bibinfo {author} {\bibfnamefont {G.}~\bibnamefont {Fagiolo}},\ and\ \bibinfo {author} {\bibfnamefont {D.}~\bibnamefont {Garlaschelli}},\ }\bibfield  {title} {\bibinfo {title} {Enhanced reconstruction of weighted networks from strengths and degrees},\ }\href {https://doi.org/10.1088/1367-2630/16/4/043022} {\bibfield  {journal} {\bibinfo  {journal} {New J. Phys.}\ }\textbf {\bibinfo {volume} {16}},\ \bibinfo {pages} {043022} (\bibinfo {year} {2014})}\BibitemShut {NoStop}%
\bibitem [{\citenamefont {Cimini}\ \emph {et~al.}(2015{\natexlab{b}})\citenamefont {Cimini}, \citenamefont {Squartini}, \citenamefont {Gabrielli},\ and\ \citenamefont {Garlaschelli}}]{EstimatingTopologicalProperties2015}%
  \BibitemOpen
  \bibfield  {author} {\bibinfo {author} {\bibfnamefont {G.}~\bibnamefont {Cimini}}, \bibinfo {author} {\bibfnamefont {T.}~\bibnamefont {Squartini}}, \bibinfo {author} {\bibfnamefont {A.}~\bibnamefont {Gabrielli}},\ and\ \bibinfo {author} {\bibfnamefont {D.}~\bibnamefont {Garlaschelli}},\ }\bibfield  {title} {\bibinfo {title} {Estimating topological properties of weighted networks from limited information},\ }\href {https://doi.org/10.1103/PhysRevE.92.040802} {\bibfield  {journal} {\bibinfo  {journal} {Phys. Rev. E}\ }\textbf {\bibinfo {volume} {92}},\ \bibinfo {pages} {040802} (\bibinfo {year} {2015}{\natexlab{b}})}\BibitemShut {NoStop}%
\bibitem [{\citenamefont {Squartini}\ \emph {et~al.}(2017)\citenamefont {Squartini}, \citenamefont {Cimini}, \citenamefont {Gabrielli},\ and\ \citenamefont {Garlaschelli}}]{NetworkReconstructionDensitySampling2017}%
  \BibitemOpen
  \bibfield  {author} {\bibinfo {author} {\bibfnamefont {T.}~\bibnamefont {Squartini}}, \bibinfo {author} {\bibfnamefont {G.}~\bibnamefont {Cimini}}, \bibinfo {author} {\bibfnamefont {A.}~\bibnamefont {Gabrielli}},\ and\ \bibinfo {author} {\bibfnamefont {D.}~\bibnamefont {Garlaschelli}},\ }\bibfield  {title} {\bibinfo {title} {Network reconstruction via density sampling},\ }\href {https://doi.org/10.1007/s41109-017-0021-8} {\bibfield  {journal} {\bibinfo  {journal} {Appl. Netw. Sci.}\ }\textbf {\bibinfo {volume} {2}},\ \bibinfo {pages} {3} (\bibinfo {year} {2017})}\BibitemShut {NoStop}%
\bibitem [{\citenamefont {Gleditsch}(2002)}]{Gleditsch2002ExpandedTradeGDP}%
  \BibitemOpen
  \bibfield  {author} {\bibinfo {author} {\bibfnamefont {K.~S.}\ \bibnamefont {Gleditsch}},\ }\bibfield  {title} {\bibinfo {title} {Expanded trade and {GDP} data},\ }\href {https://doi.org/10.1177/0022002702046005006} {\bibfield  {journal} {\bibinfo  {journal} {Journal of Conflict Resolution}\ }\textbf {\bibinfo {volume} {46}},\ \bibinfo {pages} {712} (\bibinfo {year} {2002})}\BibitemShut {NoStop}%
\bibitem [{\citenamefont {Mayer}\ and\ \citenamefont {Zignago}(2011)}]{MayerZignago2011GeoDist}%
  \BibitemOpen
  \bibfield  {author} {\bibinfo {author} {\bibfnamefont {T.}~\bibnamefont {Mayer}}\ and\ \bibinfo {author} {\bibfnamefont {S.}~\bibnamefont {Zignago}},\ }\href@noop {} {\emph {\bibinfo {title} {Notes on {CEPII}'s Distances Measures: The {GeoDist} Database}}},\ \bibinfo {type} {Working Paper}\ \bibinfo {number} {2011-25}\ (\bibinfo  {institution} {CEPII},\ \bibinfo {year} {2011})\BibitemShut {NoStop}%
\bibitem [{\citenamefont {Fagiolo}\ \emph {et~al.}(2010)\citenamefont {Fagiolo}, \citenamefont {Reyes},\ and\ \citenamefont {Schiavo}}]{FagioloReyesSchiavo2010}%
  \BibitemOpen
  \bibfield  {author} {\bibinfo {author} {\bibfnamefont {G.}~\bibnamefont {Fagiolo}}, \bibinfo {author} {\bibfnamefont {J.}~\bibnamefont {Reyes}},\ and\ \bibinfo {author} {\bibfnamefont {S.}~\bibnamefont {Schiavo}},\ }\bibfield  {title} {\bibinfo {title} {The evolution of the world trade web: A weighted-network analysis},\ }\href {https://doi.org/10.1007/s00191-009-0160-x} {\bibfield  {journal} {\bibinfo  {journal} {J. Evol. Econ.}\ }\textbf {\bibinfo {volume} {20}},\ \bibinfo {pages} {479} (\bibinfo {year} {2010})}\BibitemShut {NoStop}%
\bibitem [{\citenamefont {De~Benedictis}\ and\ \citenamefont {Tajoli}(2011)}]{DeBenedictisTajoli2011}%
  \BibitemOpen
  \bibfield  {author} {\bibinfo {author} {\bibfnamefont {L.}~\bibnamefont {De~Benedictis}}\ and\ \bibinfo {author} {\bibfnamefont {L.}~\bibnamefont {Tajoli}},\ }\bibfield  {title} {\bibinfo {title} {The world trade network},\ }\href {https://doi.org/10.1111/j.1467-9701.2011.01360.x} {\bibfield  {journal} {\bibinfo  {journal} {World Econ.}\ }\textbf {\bibinfo {volume} {34}},\ \bibinfo {pages} {1417} (\bibinfo {year} {2011})}\BibitemShut {NoStop}%
\bibitem [{\citenamefont {Hooijmaaijers}\ and\ \citenamefont {Buiten}(2019)}]{MethodologyEstimatingDutchInterfirmTradeNetwork2019}%
  \BibitemOpen
  \bibfield  {author} {\bibinfo {author} {\bibfnamefont {S.}~\bibnamefont {Hooijmaaijers}}\ and\ \bibinfo {author} {\bibfnamefont {G.}~\bibnamefont {Buiten}},\ }\href@noop {} {\emph {\bibinfo {title} {Methodology for estimating the Dutch interfirm trade network}}},\ \bibinfo {type} {Tech. Rep.}\ (\bibinfo  {institution} {Statistics Netherlands},\ \bibinfo {year} {2019})\ \bibinfo {note} {technical Report}\BibitemShut {NoStop}%
\bibitem [{\citenamefont {Buiten}\ \emph {et~al.}(2021)\citenamefont {Buiten}, \citenamefont {de~Jonge}, \citenamefont {Mooijen}, \citenamefont {Hooijmaaijers},\ and\ \citenamefont {Bogaart}}]{Buiten2021Reconstruction}%
  \BibitemOpen
  \bibfield  {author} {\bibinfo {author} {\bibfnamefont {G.}~\bibnamefont {Buiten}}, \bibinfo {author} {\bibfnamefont {E.}~\bibnamefont {de~Jonge}}, \bibinfo {author} {\bibfnamefont {G.}~\bibnamefont {Mooijen}}, \bibinfo {author} {\bibfnamefont {S.}~\bibnamefont {Hooijmaaijers}},\ and\ \bibinfo {author} {\bibfnamefont {P.}~\bibnamefont {Bogaart}},\ }\href {https://doi.org/10.13140/RG.2.2.16685.77286} {\emph {\bibinfo {title} {Reconstruction method for the Dutch interfirm network including a breakdown by commodity for 2018 and 2019 (v1.0)}}},\ \bibinfo {type} {Tech. Rep.}\ (\bibinfo  {institution} {Statistics Netherlands (CBS)},\ \bibinfo {address} {The Hague},\ \bibinfo {year} {2021})\ \bibinfo {note} {technical report, Version 1.0}\BibitemShut {NoStop}%
\bibitem [{\citenamefont {Freeman}\ \emph {et~al.}(2024)\citenamefont {Freeman}, \citenamefont {van~de Plaat},\ and\ \citenamefont {Wache}}]{Freeman2024ExportNetworks}%
  \BibitemOpen
  \bibfield  {author} {\bibinfo {author} {\bibfnamefont {D.}~\bibnamefont {Freeman}}, \bibinfo {author} {\bibfnamefont {M.}~\bibnamefont {van~de Plaat}},\ and\ \bibinfo {author} {\bibfnamefont {B.}~\bibnamefont {Wache}},\ }\href@noop {} {\emph {\bibinfo {title} {Productiviteitsvoordelen van exportnetwerken}}},\ \bibinfo {type} {Tech. Rep.}\ (\bibinfo  {institution} {Centraal Planbureau (CPB)},\ \bibinfo {address} {The Hague},\ \bibinfo {year} {2024})\ \bibinfo {note} {cPB publication}\BibitemShut {NoStop}%
\end{thebibliography}%

\section*{ACKNOWLEDGMENTS}

This publication is part of the projects ``Network renormalization: from theoretical physics to the resilience of societies’’ with file number NWA.1418.24.029 of the research programme NWA L3 - Innovative projects within routes 2024, which is (partly) financed by the Dutch Research Council (NWO) under the grant \url{https://doi.org/10.61686/AOIJP05368}, and ``Redefining renormalization for complex networks’’ with file number OCENW.M.24.039 of the research programme Open Competition Domain Science Package 24-1, which is (partly) financed by the Dutch Research Council (NWO) under the grant \url{https://doi.org/10.61686/PBSEC42210}.

\section*{AUTHOR CONTRIBUTIONS}

Study conception and design: M.M., F.P.P, D.G. Analysis and interpretation of results: M.M., F.P.P, D.G. Draft manuscript preparation: M.M., F.P.P, D.G.

\section*{COMPETING INTERESTS} 

The authors declare no competing interests.

\clearpage

\onecolumngrid

\appendix

\counterwithin*{figure}{section}
\stepcounter{section}
\renewcommand{\thefigure}{C.\arabic{figure}}

\section*{APPENDIX A.\\International Trade Network data and preprocessing}
\hypertarget{AppA}{}

\begin{table*}[!t]
\centering
\begin{minipage}[t]{0.48\textwidth}
\centering
\scriptsize
\renewcommand{\arraystretch}{1.12}
\setlength{\tabcolsep}{5pt}
\begin{tabular}{|c|c|c|c|c|}
\hline
Year & $N^{(0)}$ & $L^{(0)}$ & $d^{(0)}$ & $W$ \\
\hline
1991 & 173 & 8249 & 0.554443 & $3.002\times10^{6}$ \\
1992 & 175 & 8552 & 0.561708 & $3.182\times10^{6}$ \\
1993 & 177 & 8837 & 0.567347 & $3.106\times10^{6}$ \\
1994 & 177 & 9052 & 0.581150 & $3.534\times10^{6}$ \\
1995 & 177 & 9199 & 0.590588 & $4.226\times10^{6}$ \\
1996 & 177 & 9623 & 0.617809 & $4.444\times10^{6}$ \\
1997 & 177 & 9864 & 0.633282 & $4.798\times10^{6}$ \\
1998 & 177 & 9866 & 0.633410 & $4.724\times10^{6}$ \\
1999 & 177 & 9864 & 0.633282 & $4.974\times10^{6}$ \\
2000 & 177 & 9865 & 0.633346 & $5.710\times10^{6}$ \\
\hline
\end{tabular}
\caption{\textbf{Country-level statistics of the International Trade Network used in the empirical analysis}. $N^{(0)}$ is the number of countries, $L^{(0)}$ is the number of positive bilateral trade links, $d^{(0)}$ is the fraction of realized undirected country dyads, and $W$ is the total strength.}
\label{tab:itn_country_yearly_data}
\end{minipage}
\hfill
\begin{minipage}[t]{0.48\textwidth}
\centering
\scriptsize
\renewcommand{\arraystretch}{1.12}
\setlength{\tabcolsep}{5pt}
\begin{tabular}{|c|c|c|c|c|}
\hline
Year & $N^{(1)}$ & $L^{(1)}$ & $d^{(1)}$ & $W$ \\
\hline
1991 & 53 & 1079 & 0.754018 & $3.002\times10^{6}$ \\
1992 & 53 & 1113 & 0.777778 & $3.182\times10^{6}$ \\
1993 & 53 & 1095 & 0.765199 & $3.106\times10^{6}$ \\
1994 & 53 & 1100 & 0.768693 & $3.534\times10^{6}$ \\
1995 & 53 & 1098 & 0.767296 & $4.226\times10^{6}$ \\
1996 & 53 & 1120 & 0.782669 & $4.444\times10^{6}$ \\
1997 & 53 & 1133 & 0.791754 & $4.798\times10^{6}$ \\
1998 & 53 & 1133 & 0.791754 & $4.724\times10^{6}$ \\
1999 & 53 & 1131 & 0.790356 & $4.974\times10^{6}$ \\
2000 & 53 & 1132 & 0.791055 & $5.710\times10^{6}$ \\
\hline
\end{tabular}
\caption{\textbf{Macro-regional statistics at the main aggregation scale $d_c=800\,\mathrm{km}$}. $N^{(1)}$ is the number of macro-regions, $L^{(1)}$ is the number of macro-regional links including self-loops, $d^{(1)}$ is the fraction of realized undirected dyads, and $W$ is the total strength.}
\label{tab:itn_macro_yearly_data}
\end{minipage}
\end{table*}

The empirical analysis is based on country-level bilateral trade flows from the Expanded Trade and GDP Data by Gleditsch~\cite{Gleditsch2002ExpandedTradeGDP}, a standard representation of the International Trade Network as a weighted network of countries~\cite{FagioloReyesSchiavo2010,DeBenedictisTajoli2011}. We perform our analysis on the period 1991--2000 and, for each year, construct an undirected weighted representation of the ITN, where nodes are countries and edge weights are bilateral trade volumes.

For each year $t$, the country-level weighted network is represented by the matrix
\begin{equation}
\mathbf{W}^{(0)}(t)=\{w_{ij}^{(0)}(t)\}_{i,j=1}^{N^{(0)}(t)},
\end{equation}
where $w_{ij}^{(0)}(t)$ denotes the observed bilateral trade volume between countries $i$ and $j$. The corresponding binary projection is defined as
\begin{equation}
a_{ij}^{(0)}(t)=\mathbbm{1}\!\left(w_{ij}^{(0)}(t)>0\right).
\end{equation}
{Country-level self-loops are absent from the original data.} Therefore, link counts and densities at the country scale are computed on the strict upper triangular support. The observed number of country-level links is
\begin{equation}
L^{(0)}(t)=\sum_{i<j}a_{ij}^{(0)}(t),
\end{equation}
while the country-level density is
\begin{equation}
d^{(0)}(t)=\frac{2L^{(0)}(t)}{N^{(0)}(t)\left[N^{(0)}(t)-1\right]}.
\end{equation}
The strength of country $i$ is computed as
\begin{equation}
s_i^{(0)}(t)
=
\sum_{j\ne i}w_{ij}^{(0)}(t),
\end{equation}
and the observed total strength, in the convention adopted throughout the paper, is
\begin{equation}
W(t)
=
\sum_i s_i^{(0)}(t).
\end{equation}

Table~\ref{tab:itn_country_yearly_data} reports the basic yearly statistics of the country-level networks used in the empirical analysis. At the country level, the number of active countries is $N^{(0)}=173$ in 1991, $N^{(0)}=175$ in 1992 and $N^{(0)}=177$ from 1993 onward. The country-level density $d^{(0)}$ increases from about $0.55$ in 1991 to about $0.63$ at the end of the sample, while the observed total strength $W$ almost doubles over the same period.

The coarse-grained calibration layer is obtained by aggregating countries into geographically defined macro-regions. Country-to-country distances are taken from the CEPII GeoDist database~\cite{MayerZignago2011GeoDist}, using population-weighted bilateral distances. These distances are used exclusively to define the aggregation scale.

The macro-regional partition is obtained by applying single-linkage hierarchical clustering to the geographical distance matrix. Cutting the resulting dendrogram at height $d_c$ defines a set of macro-regions. The resulting dendrogram distance is ultrametric, while the input distances are ordinary geographical distances. In the main text we use $d_c=800\,\mathrm{km}$. This value yields a coarse layer with $N^{(1)}=53$ macro-regions in every year of the analysis and provides an intermediate aggregation scale, neither too close to the original country layer nor so coarse that the aggregate network becomes almost complete.

Given a partition $\mathcal{P}=\{I_1,\dots,I_{N^{(1)}}\}$, the macro-regional weighted matrix is constructed as
\begin{equation}
w_{IJ}^{(1)}(t)=\sum_{i\in I}\sum_{j\in J}w_{ij}^{(0)}(t).
\end{equation}
The corresponding binary projection is
\begin{equation}
a_{IJ}^{(1)}(t)=\mathbbm{1}\!\left(w_{IJ}^{(1)}(t)>0\right).
\end{equation}
Macro-regional diagonal entries are retained in the aggregated weighted matrix. They have a direct empirical meaning, since $w_{II}^{(1)}(t)$ represents trade whose endpoints both belong to macro-region $I$. Therefore, macro-regional link counts and densities are evaluated on the triangular support $I\le J$, rather than only on links between distinct macro-regions.

At the macro-regional level, the number of links is
\begin{equation}
L^{(1)}(t)=\sum_{I\le J}a_{IJ}^{(1)}(t),
\end{equation}
and the corresponding density is
\begin{equation}
d^{(1)}(t)=\frac{2L^{(1)}(t)}{N^{(1)}(t)\left[N^{(1)}(t)+1\right]}.
\end{equation}
The strength of macro-region $I$ is computed as
\begin{equation}
s_I^{(1)}(t)=\sum_J w_{IJ}^{(1)}(t),
\end{equation}
and, by additivity, the macro-regional strengths preserve the same observed total strength,
\begin{equation}
\sum_I s_I^{(1)}(t)
=
W(t).
\end{equation}

Table~\ref{tab:itn_macro_yearly_data} shows that, at the main aggregation scale $d_c=800\,\mathrm{km}$, the number of macro-regions is constant and equal to $N^{(1)}=53$. The macro-regional density $d^{(1)}$ is substantially larger than the country-level density, as expected after aggregation, and ranges from about $0.75$ to about $0.79$ over the period considered. By construction, the observed total strength is preserved by aggregation, so the same $W(t)$ is obtained at both the country and macro-regional layers.

For the visual diagnostics shown in the main text we use the 1993 snapshot. Table~\ref{tab:itn_cut_summary_1993} reports how the macro-regional layer changes with the dendrogram cut height in that year, for the four aggregation scales considered in the robustness analysis. As $d_c$ increases from $400\,\mathrm{km}$ to $1000\,\mathrm{km}$, the number of macro-regions $N^{(1)}$ decreases from $118$ to $32$ and the aggregate network becomes progressively denser. Aggregating too aggressively makes the macro-regional network very close to being fully connected, reducing the amount of topological information available for calibration. The main value $d_c=800\,\mathrm{km}$ provides an intermediate resolution, with $N^{(1)}=53$ macro-regions, $L^{(1)}=1095$ aggregate links and $d^{(1)}=0.765199$.

\begin{table}[!t]
\centering
\scriptsize
\renewcommand{\arraystretch}{1.08}
\setlength{\tabcolsep}{5pt}
\begin{tabular}{|c|c|c|c|}
\hline
$d_c$ (km) & $N^{(1)}$ & $L^{(1)}$ & $d^{(1)}$ \\
\hline
400  & 118 & 4384 & 0.624412 \\
600  & 85  & 2523 & 0.690287 \\
800  & 53  & 1095 & 0.765199 \\
1000 & 32  & 446  & 0.844697 \\
\hline
\end{tabular}
\caption{\textbf{Macro-regional aggregation of the 1993 International Trade Network for different dendrogram cut heights}. $N^{(1)}$ is the number of macro-regions induced by the cut, $L^{(1)}$ is the number of positive macro-regional links including self-loops, and $d^{(1)}$ is the fraction of realized undirected macro-regional dyads on the support $I\le J$.}
\label{tab:itn_cut_summary_1993}
\end{table}

\clearpage

\counterwithin*{figure}{section}
\stepcounter{section}
\renewcommand{\thefigure}{D.\arabic{figure}}

\section*{APPENDIX B.\\Dutch sectoral production data}
\hypertarget{AppB}{}

The second empirical application is based on a restricted Dutch sectoral production dataset for the year 2021, made available through a scientific collaboration with Centraal Bureau voor de Statistiek (CBS). The dataset belongs to the broader family of DPN data developed at CBS to represent domestic production relations below the standard industry level, with a commodity-group breakdown and consistency with national accounts information~\cite{MethodologyEstimatingDutchInterfirmTradeNetwork2019,Buiten2021Reconstruction}. Similar production-network data have been used in recent empirical work on Dutch value chains, export networks and productivity spillovers~\cite{Freeman2024ExportNetworks}. The dataset describes indirect production relations between Dutch economic sectors, resolved by broad product categories. In contrast with the International Trade Network considered in Appendix~\hyperlink{AppA}{A}, where nodes are countries and weights are bilateral trade flows, here nodes are economic sectors and weights quantify product-mediated production relations between sectors.

The sectoral classification is based on the Dutch Standard Industrial Classification, or \textit{Standaard Bedrijfsindeling} (SBI). The SBI classification is hierarchical. Two-digit SBI codes identify broad economic sectors, three-digit SBI codes refine them into more detailed sectoral categories, and four-digit SBI codes provide the finest sectoral resolution used in our analysis. We denote the corresponding network layers by $\Omega_0$, $\Omega_1$ and $\Omega_2$. The finest layer $\Omega_0$ corresponds to four-digit SBI sectors, the intermediate layer $\Omega_1$ corresponds to three-digit SBI sectors, and the coarsest layer $\Omega_2$ corresponds to two-digit SBI sectors. Thus, moving from $\Omega_0$ to $\Omega_2$ progressively aggregates sectoral units and yields coarser representations of the same Dutch production system.

The product dimension is represented through one-digit good groups. This level contains ten broad product layers, indexed by $g=0,\dots,9$. Each product layer defines a sectoral weighted network, so the DPN data used in the main analysis consist of ten product-specific sectoral networks at each resolution $\Omega_\ell$. The visual diagnostics reported in the main text focus on the finest sectoral layer $\Omega_0$, while the coarser layers $\Omega_1$ and $\Omega_2$ provide the aggregate levels used to test the multiscale reconstruction.

For each sectoral layer $\ell\in\{0,1,2\}$ and one-digit good group $g$, the data are represented by an undirected weighted matrix
\begin{equation}
\mathbf{W}^{(\ell,g)}
=
\left\{
w_{ij}^{(\ell,g)}
\right\}_{i,j=1}^{N^{(\ell,g)}} ,
\end{equation}
where $w_{ij}^{(\ell,g)}$ denotes the observed indirect production weight between sectors $i$ and $j$ in product layer $g$. The corresponding binary projection is
\begin{equation}
a_{ij}^{(\ell,g)}
=
\mathbbm{1}\!\left(w_{ij}^{(\ell,g)}>0\right).
\end{equation}

Sectoral diagonal entries are retained. They represent within-sector production relations and are therefore meaningful at the sectoral scale. Consequently, link counts and densities are computed on the triangular support $i\le j$. For each product layer $g$ and sectoral layer $\ell$, the number of links is
\begin{equation}
L^{(\ell,g)}
=
\sum_{i\le j}a_{ij}^{(\ell,g)},
\end{equation}
and the corresponding density is
\begin{equation}
d^{(\ell,g)}
=
\frac{2L^{(\ell,g)}}{N^{(\ell,g)}\left[N^{(\ell,g)}+1\right]}.
\end{equation}
The strength of sector $i$ in product layer $g$ is computed as
\begin{equation}
s_i^{(\ell,g)}
=
\sum_j w_{ij}^{(\ell,g)},
\end{equation}
while the observed total strength of good group $g$ is
\begin{equation}
W(g)
=
\sum_i s_i^{(\ell,g)}.
\end{equation}
By construction, $W(g)$ is preserved across sectoral resolutions.

Tables~\ref{tab:dutch_layer0}, \ref{tab:dutch_layer1} and \ref{tab:dutch_layer2} report the network statistics of the ten one-digit good-group layers used in the DPN analysis, separately for the three sectoral resolutions. The quantities $N^{(\ell)}$, $L^{(\ell)}$, $d^{(\ell)}$ and $W$ denote, respectively, the number of active sectoral nodes, the number of positive sectoral links, the density and the observed total strength of a product layer. Moving from $\Omega_0$ to $\Omega_2$ reduces the number of active sectoral nodes and links, as expected when the production system is represented at a coarser sectoral resolution. At the same time, densities increase, since the same indirect production relations are projected onto fewer sectoral units. For each product layer, $W(g)$ is preserved across sectoral resolutions, and the total strength summed over the ten one-digit product layers equals $3.126504\times10^6$.

\begin{table*}[!t]
\centering
\scriptsize
\renewcommand{\arraystretch}{1.08}
\setlength{\tabcolsep}{6pt}

\begin{minipage}[t]{0.48\textwidth}
\centering
\begin{tabular}{|c|c|c|c|c|}
\hline
$g$ & $N^{(0)}$ & $L^{(0)}$ & $d^{(0)}$ & $W$ \\
\hline
0 & 100 & 248  & 0.0491 & $1.162\times10^{5}$ \\
1 & 131 & 928  & 0.1073 & $2.220\times10^{5}$ \\
2 & 124 & 860  & 0.1110 & $3.557\times10^{5}$ \\
3 & 132 & 1319 & 0.1503 & $2.068\times10^{5}$ \\
4 & 130 & 1122 & 0.1318 & $4.626\times10^{5}$ \\
5 & 131 & 1116 & 0.1291 & $1.618\times10^{5}$ \\
6 & 131 & 1683 & 0.1947 & $6.982\times10^{5}$ \\
7 & 131 & 2183 & 0.2525 & $6.773\times10^{5}$ \\
8 & 130 & 939  & 0.1103 & $1.729\times10^{5}$ \\
9 & 130 & 784  & 0.0921 & $5.296\times10^{4}$ \\
\hline
Mean & 127.0 & 1118.2 & 0.1328 & $3.127\times10^{5}$ \\
\hline
\end{tabular}
\caption{\textbf{Fine sectoral layer of the Dutch production data}. The layer $\Omega_0$ corresponds to four-digit SBI sectors. The variable $g$ denotes the one-digit good group. Diagonal entries are retained, so link counts and densities are computed on the support $i\le j$.}
\label{tab:dutch_layer0}
\end{minipage}
\hfill
\begin{minipage}[t]{0.48\textwidth}
\centering
\begin{tabular}{|c|c|c|c|c|}
\hline
$g$ & $N^{(1)}$ & $L^{(1)}$ & $d^{(1)}$ & $W$ \\
\hline
0 & 91  & 229  & 0.0547 & $1.162\times10^{5}$ \\
1 & 119 & 805  & 0.1127 & $2.220\times10^{5}$ \\
2 & 114 & 771  & 0.1176 & $3.557\times10^{5}$ \\
3 & 120 & 1119 & 0.1541 & $2.068\times10^{5}$ \\
4 & 118 & 952  & 0.1356 & $4.626\times10^{5}$ \\
5 & 119 & 1038 & 0.1454 & $1.618\times10^{5}$ \\
6 & 119 & 1553 & 0.2175 & $6.982\times10^{5}$ \\
7 & 119 & 2005 & 0.2808 & $6.773\times10^{5}$ \\
8 & 118 & 881  & 0.1255 & $1.729\times10^{5}$ \\
9 & 118 & 735  & 0.1047 & $5.296\times10^{4}$ \\
\hline
Mean & 115.5 & 1008.8 & 0.1449 & $3.127\times10^{5}$ \\
\hline
\end{tabular}
\caption{\textbf{Intermediate sectoral layer of the Dutch production data}. The layer $\Omega_1$ corresponds to three-digit SBI sectors. The variable $g$ denotes the one-digit good group. Diagonal entries are retained, so link counts and densities are computed on the support $i\le j$.}
\label{tab:dutch_layer1}
\end{minipage}

\vspace{0.8em}

\begin{minipage}[t]{0.48\textwidth}
\centering
\begin{tabular}{|c|c|c|c|c|}
\hline
$g$ & $N^{(2)}$ & $L^{(2)}$ & $d^{(2)}$ & $W$ \\
\hline
0 & 64 & 122  & 0.0587 & $1.162\times10^{5}$ \\
1 & 81 & 529  & 0.1593 & $2.220\times10^{5}$ \\
2 & 79 & 513  & 0.1623 & $3.557\times10^{5}$ \\
3 & 81 & 775  & 0.2334 & $2.068\times10^{5}$ \\
4 & 80 & 469  & 0.1448 & $4.626\times10^{5}$ \\
5 & 81 & 760  & 0.2288 & $1.618\times10^{5}$ \\
6 & 80 & 855  & 0.2639 & $6.982\times10^{5}$ \\
7 & 80 & 1153 & 0.3559 & $6.773\times10^{5}$ \\
8 & 80 & 596  & 0.1840 & $1.729\times10^{5}$ \\
9 & 80 & 557  & 0.1719 & $5.296\times10^{4}$ \\
\hline
Mean & 78.6 & 632.9 & 0.1963 & $3.127\times10^{5}$ \\
\hline
\end{tabular}
\caption{\textbf{Coarse sectoral layer of the Dutch production data}. The layer $\Omega_2$ corresponds to two-digit SBI sectors. The variable $g$ denotes the one-digit good group. Diagonal entries are retained, so link counts and densities are computed on the support $i\le j$.}
\label{tab:dutch_layer2}
\end{minipage}

\end{table*}

At the finest sectoral resolution, $\Omega_0$, the ten one-digit product layers contain on average $127$ active sectoral nodes, about $1118$ positive links and mean density $0.1328$. The corresponding layers $\Omega_1$ and $\Omega_2$ contain, respectively, $115.5$ and $78.6$ active sectoral nodes on average. The densest product layer is $g=7$ at all sectoral resolutions, with densities $0.2525$, $0.2808$ and $0.3559$ for $\Omega_0$, $\Omega_1$ and $\Omega_2$, respectively. This confirms that the coarser sectoral descriptions preserve the main layer-level structure while compressing the number of active nodes.

\clearpage

\counterwithin*{figure}{section}
\stepcounter{section}
\renewcommand{\thefigure}{E.\arabic{figure}}

\section*{APPENDIX C.\\Robustness across ultrametric aggregation scales}
\hypertarget{AppC}{}

In the main text we used the cut $d_c=800\,\mathrm{km}$ as the reference aggregation scale. Here we check that the main conclusions are not specific to this choice by repeating the full calibration and reconstruction procedure for four additional cuts of the same single-linkage dendrogram, namely $d_c=200,400,600,1000\,\mathrm{km}$.

We restrict the robustness analysis to informative aggregation scales. Very fine cuts approach the original country layer and therefore provide little genuine coarse graining. Conversely, very large cuts produce a small number of macro-regions and a nearly complete aggregate network, so that matching the aggregate density becomes progressively less informative for the microscopic binary structure. The cuts considered below span a broad but still meaningful range between these two limiting cases.

For each value of $d_c$, the wMSM is calibrated at the corresponding macro-regional layer and then transferred to the country layer exactly as in the main analysis. The benchmark CReMB is unchanged across cuts, since it is always calibrated directly on the country-level network. All quantities in Table~\ref{tab:robustness_wmsm_absolute} are averages over the yearly snapshots 1991--2000. No year is skipped for any of the alternative cuts considered here.

\begin{table*}[!t]
\centering
\scriptsize
\renewcommand{\arraystretch}{1.12}
\setlength{\tabcolsep}{6pt}
\begin{tabular}{|c|c|c|c|c|c|c|c|c|}
\hline
$d_c$ (km)
& $\overline{N}^{(1)}$
& $\overline{d}^{(1)}$
& $\mathrm{RE}_L^{(1)}$
& $\mathrm{ARE}_k^{(1)}$
& $\mathrm{ARE}_s^{(1)}$
& $\mathrm{RE}_L^{(0)}$
& $\mathrm{ARE}_k^{(0)}$
& $\mathrm{ARE}_s^{(0)}$ \\
\hline
200  & 154.7 & 0.6283 & 0.0000 & 0.1671 & 0.0000 & 0.0316 & 0.1942 & 0.0057 \\
400  & 117.8 & 0.6482 & 0.0000 & 0.1665 & 0.0000 & 0.0548 & 0.2005 & 0.0057 \\
600  & 85.0  & 0.7110 & 0.0000 & 0.1370 & 0.0000 & 0.0628 & 0.2031 & 0.0057 \\
1000 & 32.2  & 0.8498 & 0.0000 & 0.0785 & 0.0000 & 0.1543 & 0.2521 & 0.0057 \\
\hline
\end{tabular}
\caption{\textbf{Robustness of the macro-calibrated wMSM reconstruction across alternative ultrametric cuts}. The quantities are averaged over 1991--2000. Superscript $(1)$ denotes the macro-regional calibration layer and superscript $(0)$ denotes the country reconstruction layer. The macro-regional link count is reproduced by construction for every cut. The country-level strength error remains almost unchanged because it is mainly controlled by the country-level diagonal convention, while the binary reconstruction is affected by the amount of topological information retained at the aggregate scale.}
\label{tab:robustness_wmsm_absolute}
\end{table*}

Table~\ref{tab:robustness_wmsm_absolute} shows that the reconstruction is stable over a wide range of aggregation scales. As expected, the macro-regional density increases with $d_c$, because larger cuts merge more countries into fewer blocks and make the aggregate network denser. The country-level error on the expected number of links remains small for $d_c=200,400,600\,\mathrm{km}$ and increases for the coarsest cut $d_c=1000\,\mathrm{km}$, where the macro-regional layer is substantially more compressed and less informative. The strength error remains close to the value observed in the main analysis for all cuts. This confirms that the weighted layer is controlled by the same gravity expectation, whereas the main effect of changing $d_c$ is on the transferred binary density parameter.

We also repeat the country-level comparison with the CReMB. We use the same signed relative improvement $\Delta_m(t)$ defined in the main text, so that positive values always favor the wMSM. The detailed comparison is reported in Table~\ref{tab:robustness_wmsm_cremb}.

The comparison confirms the qualitative picture observed at the main cut. Across all alternative aggregation scales, the wMSM systematically improves precision, specificity and the maximum degree error. Accuracy also remains higher in almost every year. Sensitivity is the only metric for which the behavior changes with the cut: at the finest robustness cut, $d_c=200\,\mathrm{km}$, the wMSM is essentially equivalent to the CReMB and slightly better on average, while for coarser cuts the CReMB recovers a larger fraction of observed links. This is consistent with the interpretation given in the main text. Increasing $d_c$ removes microscopic topological information from the calibration layer and makes the transferred wMSM probabilities more selective. As a consequence, false positives decrease, improving $\mathrm{PPV}$ and $\mathrm{TNR}$, while some true links are assigned lower probability, reducing $\mathrm{TPR}$.

Overall, the robustness analysis shows that the performance of wMSM does not depend on a finely tuned aggregation scale. The coarsest cut naturally leads to a stronger loss of information, but the country-level reconstruction remains competitive with the fine-scale benchmark. The main conclusion is therefore stable: a density parameter calibrated only on a coarse-grained layer can be transferred to the country scale and still recover a high-quality microscopic binary support.

\begin{table*}[!b]
\centering
\scriptsize
\renewcommand{\arraystretch}{1.12}
\setlength{\tabcolsep}{7pt}
\begin{tabular}{|c|c|c|c|c|c|}
\hline
$d_c$ (km)
& $\Delta_{\mathrm{TPR}}$
& $\Delta_{\mathrm{PPV}}$
& $\Delta_{\mathrm{TNR}}$
& $\Delta_{\mathrm{ACC}}$
& $\Delta_{\mathrm{MRE}_k}$ \\
\hline
200
& $+0.21$ $(7/10)$
& $+3.48$ $(10/10)$
& $+7.43$ $(10/10)$
& $+2.80$ $(10/10)$
& $+7.22$ $(10/10)$ \\
400
& $-1.42$ $(1/10)$
& $+4.29$ $(10/10)$
& $+9.80$ $(10/10)$
& $+2.57$ $(10/10)$
& $+12.98$ $(10/10)$ \\
600
& $-2.00$ $(1/10)$
& $+4.57$ $(10/10)$
& $+10.50$ $(10/10)$
& $+2.50$ $(10/10)$
& $+15.01$ $(10/10)$ \\
1000
& $-9.04$ $(0/10)$
& $+7.61$ $(10/10)$
& $+18.62$ $(10/10)$
& $+0.92$ $(9/10)$
& $+18.57$ $(10/10)$ \\
\hline
\end{tabular}
\caption{\textbf{Country-level comparison between the macro-calibrated wMSM and the country-calibrated CReMB across alternative ultrametric cuts}. Entries report the mean signed relative improvement $\Delta_m$, in percentage, with positive values favoring the wMSM. Parentheses indicate the number of yearly snapshots, out of ten, in which the wMSM outperforms the CReMB.}
\label{tab:robustness_wmsm_cremb}
\end{table*}

\clearpage

\counterwithin*{figure}{section}
\stepcounter{section}
\renewcommand{\thefigure}{F.\arabic{figure}}

\section*{APPENDIX D.\\CReMB as a fine-scale benchmark}
\hypertarget{AppD}{}

The Conditional Reconstruction Method for weighted networks was introduced to reconstruct weighted networks by separating the inference of the binary topology from the inference of link weights conditional on link existence~\cite{CReM2020}. This construction builds on maximum-entropy network reconstruction methods for economic and financial systems~\cite{Mastrandrea2014EnhancedReconstruction,EstimatingTopologicalProperties2015,CimiModel2015,NetworkReconstructionDensitySampling2017,ReconstructionMethods2018,cimini2021reconstructing}, where the available information is typically limited to node-level aggregate quantities.

Let $\mathbf{W}=\{w_{ij}\}$ be an undirected weighted network {on the layer-specific support $\mathcal D$ (strict upper triangle for countries, upper triangle including the diagonal for sectors)}, and let
\begin{equation}
a_{ij}=\mathbbm{1}(w_{ij}>0)
\end{equation}
be its binary projection. In the conditional reconstruction framework, the probability distribution of weights is factorized into two parts. First, a binary model assigns to each dyad $(i,j)$ a link probability
\begin{equation}
P(a_{ij}=1)=p_{ij},
\qquad
P(a_{ij}=0)=1-p_{ij}.
\end{equation}
Second, conditional on $a_{ij}=1$, a positive weight is drawn from a distribution with prescribed conditional mean. If $a_{ij}=0$, the weight is identically zero. Thus the unconditional expected weight satisfies
\begin{equation}
\mathbb{E}[w_{ij}]
=
p_{ij}\,\mathbb{E}[w_{ij}\mid a_{ij}=1].
\end{equation}

Different variants of the method correspond to different choices of the binary layer. In the CReMA, the adjacency matrix is assumed to be known and the binary probabilities are effectively fixed by the observed topology. In the CReMB, instead, the topology is not fixed: it is described by a matrix of link probabilities $\mathbf{P}=\{p_{ij}\}$. This makes the CReMB suitable for reconstruction settings in which the binary network itself must be inferred.

Here the binary layer is specified through the density-corrected Gravity Model (dcGM), a one-parameter fitness model in which connection probabilities depend on the product of node strengths~\cite{EnhancedGravityModelTrade2019,DiVece2022GravityMER}. In the undirected case, one sets
\begin{equation}
p_{ij}^{\mathrm{CReMB}}
=
\frac{z\,s_i s_j}{1+z\,s_i s_j},
\end{equation}
where $s_i$ is the observed strength of node $i$ and $z>0$ is a global density parameter. The parameter $z$ is fixed by matching the expected number of links to the observed one,
\begin{equation}
\sum_{{(i,j)\in\mathcal D}}p_{ij}^{\mathrm{CReMB}}=L.
\end{equation}
The expected degree of node $i$ is therefore
\begin{equation}
\left\langle k_i\right\rangle_{\mathrm{CReMB}}
=
\sum_{{j}}p_{ij}^{\mathrm{CReMB}}.
\end{equation}

Once the binary probabilities have been fixed, the weighted layer is reconstructed conditionally on link existence. Since we work with discrete positive weights, the conditional maximum-entropy distribution with prescribed mean is the geometric distribution on $\{1,2,\dots\}$,
\begin{equation}
q_{ij}(w\mid a_{ij}=1)
=
\left(\frac{1}{\mu_{ij}}\right)
\left(1-\frac{1}{\mu_{ij}}\right)^{w-1},
\qquad w\in\{1,2,\dots\},
\end{equation}
where
\begin{equation}
\mu_{ij}
=
\mathbb{E}[w_{ij}\mid a_{ij}=1].
\end{equation}
At the level of ensemble averages, the model is fully characterized by the probability $p_{ij}^{\mathrm{CReMB}}$ and the conditional mean $\mu_{ij}$.

The unconditional expected weight is assigned through the gravity closure
\begin{equation}
\left\langle w_{ij}\right\rangle_{\mathrm{CReMB}}
=
\mathbb{E}_{\mathrm{CReMB}}[w_{ij}]
=
\frac{s_i s_j}{W},
\end{equation}
where
\begin{equation}
W=\sum_i s_i
\end{equation}
is the total strength. Since
\begin{equation}
\mathbb{E}_{\mathrm{CReMB}}[w_{ij}]
=
p_{ij}^{\mathrm{CReMB}}
\mathbb{E}_{\mathrm{CReMB}}[w_{ij}\mid a_{ij}=1],
\end{equation}
the corresponding conditional expected weight is
\begin{equation}
\mu_{ij}^{\mathrm{CReMB}}
=
\frac{s_i s_j}{W\,p_{ij}^{\mathrm{CReMB}}},
\qquad p_{ij}^{\mathrm{CReMB}}>0.
\end{equation}
The conditional construction therefore separates the role of topology and weights: the binary layer determines how likely a dyad is to be connected, while the conditional layer determines how much weight a realized link is expected to carry.

Combining the two layers gives
\begin{equation}
p_{ij}^{\mathrm{CReMB}}\mu_{ij}^{\mathrm{CReMB}}
=
\frac{s_i s_j}{W},
\end{equation}
so that the expected weighted matrix has the same gravity form as in standard strength-driven reconstruction methods. On the full dyadic support, including diagonal entries, this prescription reproduces strengths exactly:
\begin{equation}
\sum_j \left\langle w_{ij}\right\rangle_{\mathrm{CReMB}}
=
\frac{s_i}{W}\sum_j s_j
=
s_i.
\end{equation}
If self-loops are removed, the expected strength instead becomes
\begin{equation}
\sum_{j\ne i}\left\langle w_{ij}\right\rangle_{\mathrm{CReMB}}
=
s_i\left(1-\frac{s_i}{W}\right),
\end{equation}
so the deviation from exact strength matching is entirely due to the exclusion of diagonal dyads.

\end{document}